\documentclass[a4paper,11pt]{article}
\pdfoutput=1 

\usepackage{jinstpub} 
\usepackage{ptdr-definitions}
                     
\usepackage{amssymb}

\newcommand{\cmsAuthorMark}[1]
{\hbox{\textsuperscript{\normalfont#1}}}
\newskip{\cmsinstskip} \cmsinstskip=0pt plus 4pt
\newskip{\cmsauthskip} \cmsauthskip=16pt

\title{\boldmath The CMS Phase-1 Pixel Detector Upgrade}

\author{The Tracker Group of the CMS Collaboration}

\abstract{The CMS detector at the CERN LHC features a silicon pixel detector as its innermost subdetector. The original CMS pixel detector has been replaced with an upgraded pixel system (CMS Phase-1 pixel detector) in the extended year-end technical stop of the LHC in 2016/2017. The upgraded CMS pixel detector is designed to cope with the higher instantaneous luminosities that have been achieved by the LHC after the upgrades to the accelerator during the first long shutdown in 2013-2014. Compared to the original pixel detector, the upgraded detector has a better tracking performance and lower mass with four barrel layers and three endcap disks on each side to provide hit coverage up to an absolute value of pseudorapidity of 2.5. This paper describes the design and construction of the CMS Phase-1 pixel detector as well as its performance from commissioning to early operation in collision data-taking.}


\begin{document}

\maketitle
\flushbottom

\newpage
\hspace{0pt}
\vfill
\vspace{-8cm}
\begin{center}
In memory of Gino Bolla, the physicist, the friend.
\end{center}
\begin{center}
$\ast$ 25 September 1968
\end{center}
\vspace{-1cm}
\begin{center}
$\dag$ 4 September 2016 
\end{center}
\vfill
\hspace{0pt}

\cleardoublepage

\section{Introduction}

The CMS experiment~\cite{cmspaper} at the CERN Large Hadron Collider (LHC) includes a silicon pixel detector as the innermost part of the tracking system. The pixel detector provides 3-dimensional space points in the region closest to the interaction point that allow for high-precision, charged-particle tracking and for vertex reconstruction~\cite{Chatrchyan:2009aa, Chatrchyan:2014fea}. The pixel detector is located in a particularly harsh radiation environment characterized by a high track density. The original pixel detector~\cite{cmspaper} consisted of three barrel layers at radii of 44, 73, and 102\,\mm and two endcap disks on each end at distances of 345 and 465\,\mm from the interaction point. It was designed for a maximum instantaneous luminosity of $\rm 1\times10^{34}$\,\percms and a maximum average pileup (number of inelastic interactions per bunch crossing) of 25 in LHC operation with 25\,ns bunch spacing. With the upgrade of the accelerators during the first long shutdown (LS1, 2013-2014), these parameters have been exceeded and the luminosity and pileup have more than doubled compared to the design values. In order to maintain efficient and robust tracking at CMS under these conditions, the original pixel detector has been replaced by a new system, referred to as the \cmsph~\cite{Dominguez:1481838}. The installation of the \cmsph took place during the extended year-end technical stop of the LHC in 2016/2017\footnote{The LHC year-end technical stop in 2016/2017 was extended by two months and lasted from December 2016 to April 2017.}.

The \cmsph constitutes an evolutionary upgrade, keeping the well-tested key features of the original detector and improving the performance toward higher rate capability, improved radiation tolerance, and more robust tracking. It is expected to deliver high-quality data until the end of LHC Run 3 (currently expected for 2024), after which the whole CMS tracker detector will be replaced in preparation of the High-Luminosity LHC~\cite{Collaboration:2272264}.

In this paper, the design and construction of the \cmsph are described and its performance from commissioning to early operation in collision data-taking is presented. Issues experienced during the first data-taking period are discussed and improvements and modifications that have been implemented during the 2017/2018 LHC year-end technical stop, or will be implemented during the second long shutdown (LS2, 2019-2021), are explained. 

The outline of the paper is as follows. The overall system aspects, main design parameters, and performance goals are described in Sec.~\ref{s:design}. The design, assembly, and qualification of the detector modules is discussed in Sec.~\ref{s:modules}. Sections~\ref{s:mechanics},~\ref{s:services},~\ref{s:power}, and ~\ref{s:cooling} discuss the detector mechanics, readout electronics and data acquisition system, as well as the power system and cooling. In Sec.~\ref{s:pilot} the commissioning of a pilot system is reviewed and in Sec.~\ref{s:integration} the integration, testing, and installation of the final detector system is described. Results from detector calibration and operations are discussed in Sec.~\ref{s:calib} and~\ref{s:ops}. A summary and conclusions are presented in Sec.~\ref{s:summary}. A glossary of special terms and acronyms is given in Sec.~\ref{chap:glossary}.

\section{Design of the \cmsph}
\label{s:design}

\begin{figure}[tb!]
  \centering
    \includegraphics[trim=0 0 0 200,clip,width=1.0\textwidth]{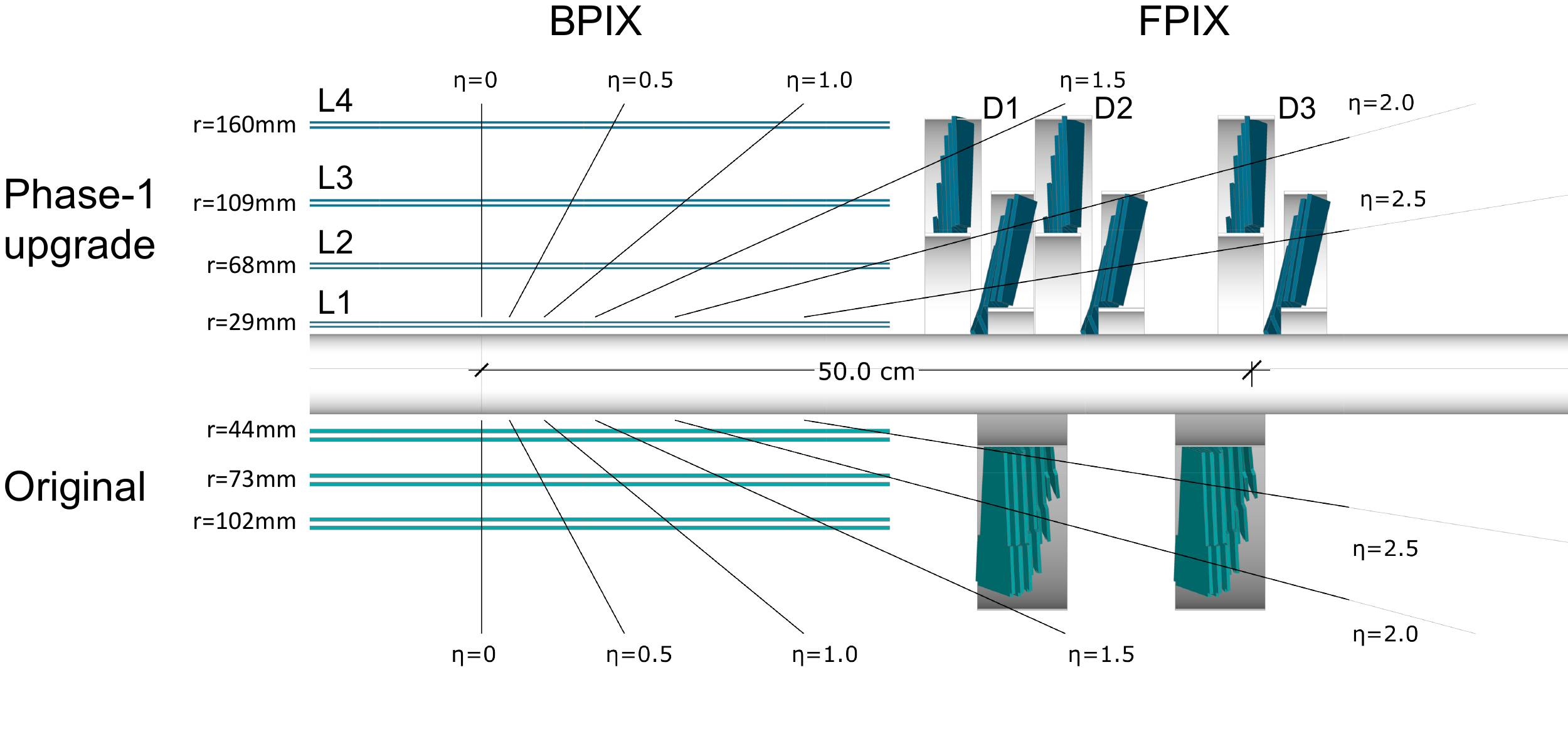}
    \caption{Layout of the \cmsph compared to the original detector layout, in longitudinal view.}
    \label{fig:layout}
\end{figure}

The layout of the \cmsph is optimized to have four-hit coverage over the pseudorapidity range $|\eta|<2.5$\footnote{CMS uses a right-handed coordinate system. The $x$-axis points to the center of the LHC ring, the $y$-axis points up vertically and the $z$-axis points along the beam direction. The azimuthal angle $\phi$ is measured in the $xy$-plane and the radial coordinate is denoted by $r$. The polar angle $\theta$ is defined in the $rz$-plane and the pseudorapidity is $\eta=-\ln\tan(\frac{\theta}{2})$.}, improved pattern recognition and track reconstruction, and added redundancy to cope with hit losses. 
During LS1, a new beam pipe with a smaller radius of 23\,\mm, compared to a radius of 30\,\mm of the original beam pipe, was installed in CMS. This allowed for placement of the innermost layer of the \cmsph closer to the interaction point compared to the original pixel detector. The \cmsph consists of four concentric barrel layers (L1-L4) at radii of 29, 68, 109, and 160\,\mm, and three disks (D1-D3) on each end at distances of 291, 396, and 516\,\mm from the center of the detector. The layout of the \cmsph is compared to the one of the original pixel detector in Fig.~\ref{fig:layout}. The total silicon area of the \cmsph is 1.9\,m$^2$, while the total silicon area of the original pixel detector was 1.1\,m$^2$.

The \cmsph is built from 1856 segmented silicon sensor modules, where 1184 modules are used in the barrel pixel detector (\bpix) and 672 modules are used for the forward disks (\fpix). Each module consists of a sensor with $160\times416$ pixels connected to 16 readout chips (ROCs). In total there are 124 million readout channels. The design of the detector modules is discussed in more detail in Sec.~\ref{s:modules}. 

\begin{figure}[b!]
  \centering
    \includegraphics[trim=0 0 0 0,clip,width=0.8\textwidth]{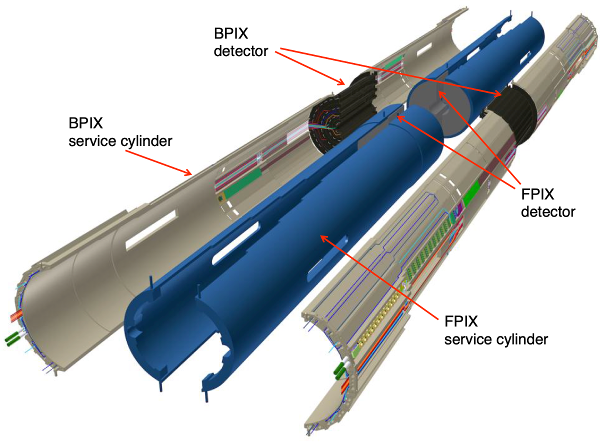}
    \caption{Drawing of the Phase-1 \bpix and \fpix detectors together with the service half-cylinders that hold the readout and control circuits as well as power and cooling lines.}
    \label{fig:ST}
\end{figure}

The main dimensional parameters of the \cmsph are reviewed in Tab.~\ref{t:design}. The \bpix and \fpix detectors are independent components, both mechanically and electrically. The \bpix detector consists of two half-barrels (Fig.~\ref{fig:ST}) with a total length of 540\,\mm, each divided into four layers (called half-shells). 
Similarly the \fpix detector is assembled from twelve half-disks (six half-disks on each side) with a radial coverage from 45 to 161\,\mm. The half-disks are further divided into inner and outer half-rings supporting 22 and  34 modules, respectively. The division of the detector into mechanically independent halves makes it possible to install the pixel detector inside the CMS detector with the beam pipe in place. This scheme allowed the \cmsph to be installed within the limited period of time during the 2016/2017 LHC extended year-end technical stop. Furthermore, it permits access to the detector for maintenance work and refurbishment also during the short periods of regular LHC year-end technical stops. 

\begin{table}[!htb]
  \begin{center}
    \caption{Summary of average r, $z$ positions and number of modules for the four \bpix layers and the six \fpix rings.}
    \label{t:design}
   \vspace{0.1in}
   \begin{tabular}{l | c | c | c}
      \hline
       \multicolumn{4}{c}{\bpix} \\ 
        \hline
        \hline
     Layer & Radius [\mm] & $z$ position [\mm] & Number of modules  \\ 
    \hline
    \hline
    
     L1 & 29   &$-270$\,to\,$+270$  & 96 \\ \hline
    L2 & 68   & $-270$\,to\,$+270$  & 224 \\ \hline
    L3 & 109 & $-270$\,to\,$+270$  & 352 \\ \hline
    L4 & 160 & $-270$\,to\,$+270$ & 512 \\
    \hline
  \hline
    \multicolumn{4}{c}{\fpix} \\ \hline
    \hline
      Disk & Radius [\mm] & $z$ position [\mm] & Number of modules  \\ \hline
    \hline
    D1 inner ring & 45--110 & $\pm$338 &    88 \\ \hline
    D1 outer ring & 96--161& $\pm$309 &    136 \\ \hline
    D2 inner ring & 45--110 & $\pm$413 &    88 \\ \hline
    D2 outer ring & 96--161 & $\pm$384 &    136 \\ \hline
    D3 inner ring & 45--110 & $\pm$508 &    88 \\ \hline
    D3 outer ring & 96--161 & $\pm$479 &    136 \\ \hline
      \hline
   \end{tabular}
  \end{center}
\end{table}

The \bpix and \fpix detectors are each supplied by four service half-cylinders that hold the readout and control circuits and guide the power lines and cooling tubes of the detector, as shown in Fig.~\ref{fig:ST}. 
The \bpix detector is divided into two mechanically independent halves, both composed of one half detector and two service half-cylinders. The \fpix detector is divided into four mechanically independent quadrants, each formed by three half-disks installed in a service half-cylinder. 

The \cmsph is required to fit into the same mechanical envelope as the original system and to partly reuse existing services. This has put strong constraints on the design of the new system. In particular, higher bandwidth electronics are needed to transmit the increased data volume from the \cmsph through the existing optical fibers to the data acquisition (DAQ) system. Since the \cmsph has 1.9 times more channels than the original pixel detector, the power consumption increases accordingly. The \cmsph uses \dcdc power converters to supply the necessary current to the modules while reusing the existing cables from the power supply racks to the tracker detector patch panel inside the CMS magnet bore. Sections~\ref{s:services} and~\ref{s:power} give more information about the readout and power systems of the \cmsph. 

\begin{figure}[tb!]
  \centering
    \includegraphics[trim=0 0 0 0,clip,width=0.45\textwidth]{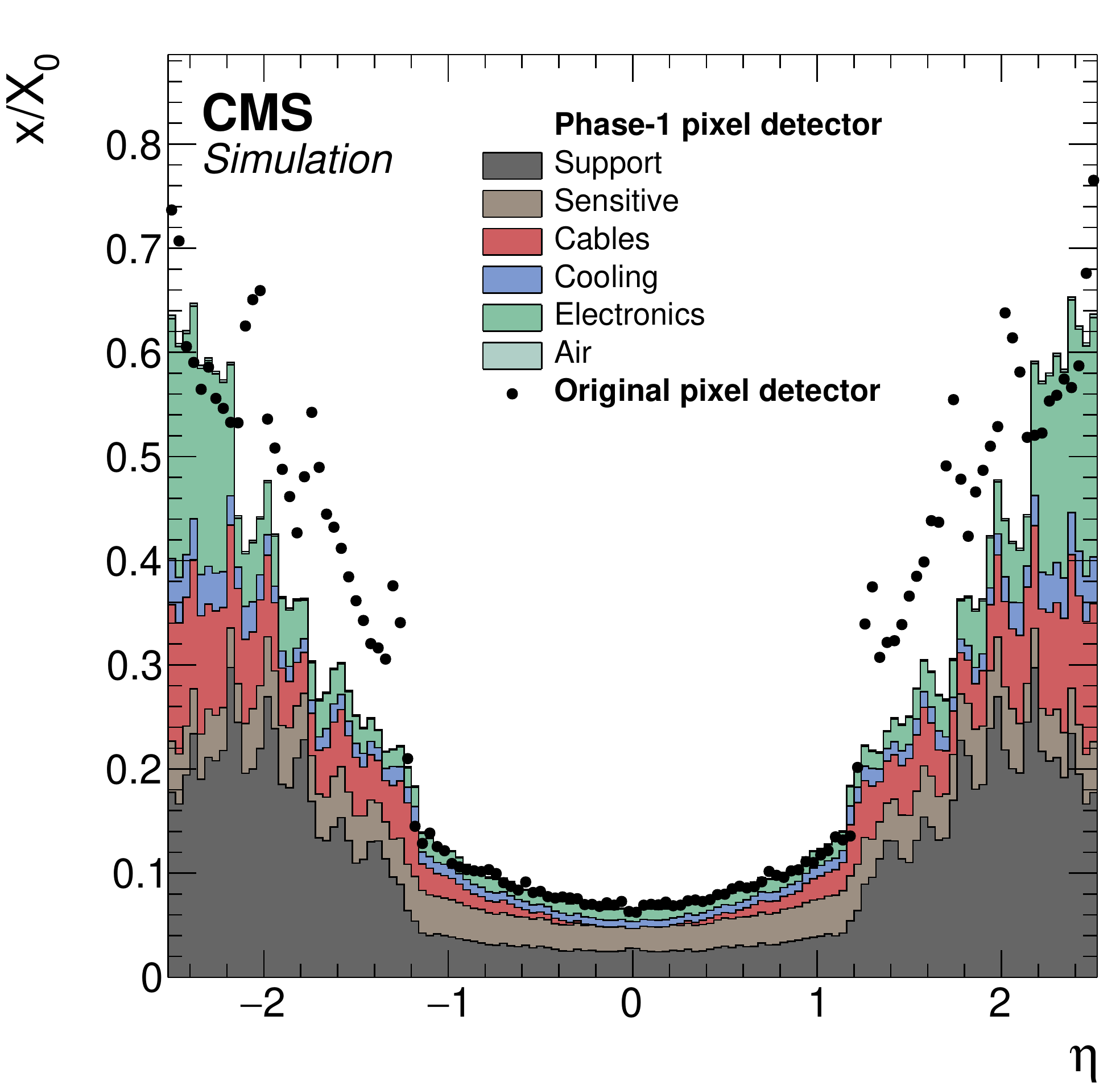}
    \includegraphics[trim=0 0 0 0,clip,width=0.45\textwidth]{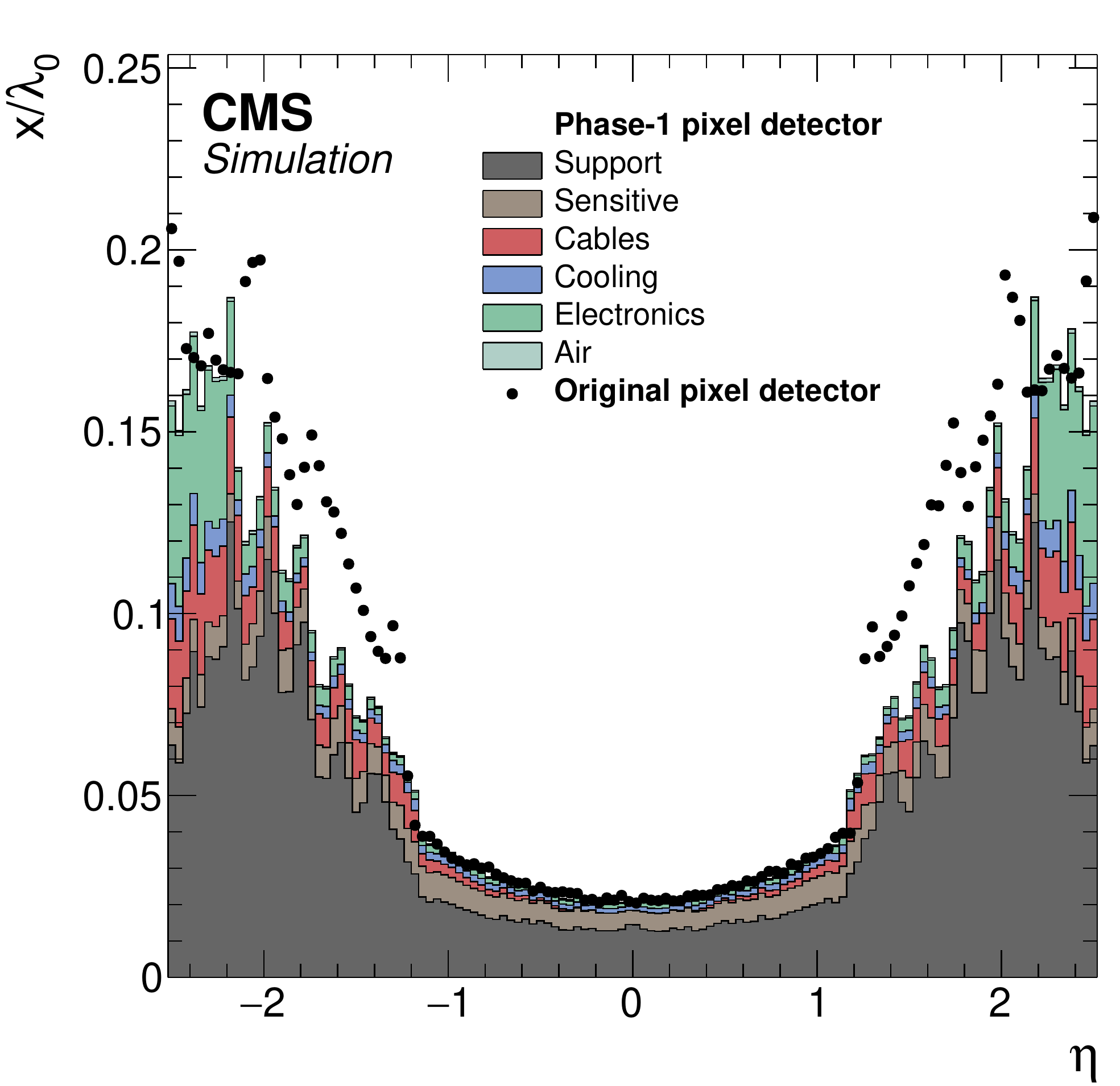}\\
    \caption{Material budget in units of (left) radiation lengths, $X_0$, and (right) hadronic interaction lengths, $\lambda_0$, as a function of pseudorapidity, $\eta$, as obtained from simulation. The material budget of the original pixel detector is compared to the \cmsph within the tracking acceptance. The material budget of the \cmsph is split into the contributions of the different categories. The peaks in the distribution in the forward region reflect the disk structure. The largest values for radiation length and hadronic interaction length lie outside the tracking acceptance at around $|\eta|=3.5$ and amount to 1.9\,$\rm{x/X_0}$ and 0.36\,$\rm{x/\lambda_0}$, respectively.}
    \label{fig:material}
\end{figure}

In order to optimize the tracking and vertexing resolution, it is crucial to minimize the material used in the detector. Despite the additional sensor layers, the material budget of the \cmsph in the central region is almost unchanged compared to the original detector, while it is significantly reduced in the forward region at $|\eta|>1$. This is achieved by using advanced carbon-fiber materials for the mechanical structure and adopting the use of a lower mass, two-phase \coo cooling system. Furthermore, the electronic boards on the service half-cylinders are placed in higher pseudorapidity regions, outside of the tracking acceptance. Figure~\ref{fig:material} shows the material budget of the \cmsph compared to the original pixel detector within the tracking acceptance in terms of radiation lengths and hadronic interaction lengths. The material budget is obtained from \texttt{GEANT4}-based simulation models~\cite{geant} of the CMS pixel detectors. 

With the innermost layer placed at a radius of 29\,\mm from the beam, the modules in this region have to withstand very high radiation doses and hit rates, as shown in Tab.~\ref{t:requirements}. A hadron fluence of $3.6\times10^{15}\Neq$ (fluence measured in units of 1\,MeV neutron equivalents) is expected to be accumulated in the innermost layer after collecting an integrated luminosity of 500\,\fbinv. This fluence is about twice as high as the operational limits of the proposed system, as defined by the charge collection efficiency of the sensor~\cite{Dominguez:1481838}. Therefore, the innermost \bpix layer will be replaced during LS2. The fluence in the second layer of the \bpix detector is about four times less, and hence the outer \bpix layers will stay operational during the entire period. The same is true for the modules in the \fpix detector. 

The expected hit rates in the outer \bpix layers and the \fpix detector are two to three times higher compared to the original detector and increase to almost 600\,\mhzcm for \bpix L1. The ability of the \cmsph to cope with these hit rates is achieved by the design of new ROCs, as discussed in Sec.~\ref{s:roc}. Because of these improvements, the \cmsph has the same, or even better, performance compared to the original detector at twice the instantaneous luminosity, as discussed in Sec.~\ref{s:ops}.

\begin{table}
  \begin{center}
    \caption{Expected hit rate, fluence, and radiation dose for the \bpix layers and \fpix rings~\cite{ref:fluka}. The hit rate corresponds to an instantaneous luminosity of $\rm 2.0\times10^{34}$\,\percms~\cite{Dominguez:1481838}. The fluence and radiation dose are shown for integrated luminosities of 300\,\fbinv for \bpix L1 and 500\,\fbinv for the other \bpix layers and \fpix disks.}
    \label{t:requirements}
   \vspace{0.1in}
   \begin{tabular}{l | c |   c |   c }
     \hline
   & Pixel hit rate  & Fluence & Dose  \\ 
    & [\mhzcmnospace]  & [$10^{15}\Neq$] & [\mradnospace]  \\ 
      
    \hline
    \hline
    \bpix L1    & 580 & 2.2 & 100  \\ \hline
    \bpix L2    & 120 &  0.9 &  47\\ \hline
    \bpix L3    & 58 &  0.4 &  22\\ \hline
    \bpix L4    & 32 &  0.3  &  13\\
    \hline
    \hline
     \fpix inner rings  & 56-260 &  0.4-2.0   & 21-106\\
     \fpix outer rings  & 30-75 &  0.3-0.5  & 13-28\\ 
      \hline
    \hline
   \end{tabular}
  \end{center}
\end{table}

\section{Silicon sensor modules}
\label{s:modules}

\begin{figure}[b!]
	\centering
	\includegraphics[trim=0 0 0 350,clip,width=1\textwidth]{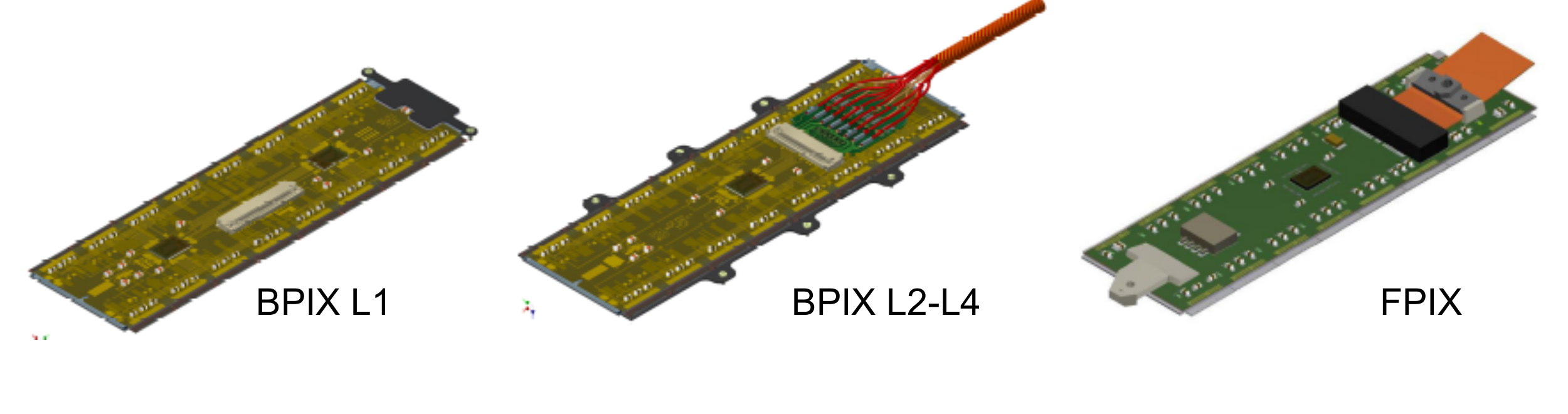}\\
	 \caption{Drawings of the pixel detector modules for \bpix L1 (left), \bpix L2-4 (middle), and the \fpix detector (right).}
	 \label{fig:modulebpixfpix}
\end{figure}

The \cmsph uses a similar module design as the \bpix modules of the original detector. A pixel detector module is built from a planar silicon sensor with a size of $\rm{18.6\times66.6\,\mm^2}$ (active area of $\rm{16.2\times64.8\,\mm^2}$), bump-bonded to an array of $2\times8$ ROCs. Each ROC is segmented into 4160 readout channels and reads out the pulse height information for each pixel. The standard pixel size is $100\times 150\,\mum^{2}$ (as in the original pixel detector). Since two ROCs can only be placed at some minimum distance from each other, pixels along the ROC boundaries have twice the area and those at the corners have four times the area of a standard pixel. On the other side of the silicon sensor, a high-density interconnect (HDI) flex printed circuit is glued and wire-bonded to the ROCs. A token bit manager chip (TBM) controls the readout of a group of ROCs and is mounted on top of the HDI (two TBMs in the case of L1 modules). In order to simplify module production and maintenance, the same rectangular module geometry is used for the \bpix and \fpix detectors. Drawings of the \cmsph modules are shown in Fig.~\ref{fig:modulebpixfpix}.

In the \bpix detector, the orientation of the sensor surface of the modules is parallel to the magnetic field, as in the original pixel detector. The pixels are oriented with the long side parallel to the beam line. In the \fpix detector, the modules in the outer rings are rotated by $20^\circ$ in a turbine-like geometry, similar to the original detector. However, to obtain optimal resolution in both the azimuthal and radial directions for the inner ring, the modules in the inner ring are arranged in an inverted cone array tilted by $12^\circ$ with respect to the beam line, combined with the $20^\circ$ rotation (also shown in Figs.~\ref{fig:layout} and~\ref{fig:disk}). The sensor orientation in the \fpix detector is such that the long side of the pixel is in the radial direction, and thus different with respect to the original detector. 


\subsection{Sensors}
\label{s:sensor}

The sensor design of the original pixel detector was
the result of an extensive R\&D program described in Ref.~\cite{ref:pixPh0}. Studies with irradiated sensors have continued and have shown 
that the sensors also fulfill the requirements of the \cmsph~\cite{ref:pixPh1}. 



The sensors of the \bpix and \fpix detectors were produced by different companies in order not to depend on a single source. The \bpix sensors were produced by CiS Forschungsinstitut f\"ur Mikrosensorik in Erfurt, Germany, while the \fpix sensors were manufactured by SINTEF Micro-systems and Sensors in Oslo, Norway. To achieve optimal yield, the sensor concept and design was tailored to each vendor's production process, which led to two quite different sensor types.



Both types of sensors are made of silicon and follow the n-in-n approach~\cite{ref:pixPh0}, with strongly n-doped ($\rm{n^+}$) pixelated implants on an n-doped silicon bulk and a p-doped back side. In a reverse-bias configuration, the $\rm{n^+}$ implants collect electrons. This is advantageous since the electrons have a higher mobility compared to holes and therefore are less affected by charge trapping caused by radiation damage in the silicon after high irradiation~\cite{ref:rose, Rohe:807069}. This leads to a high signal charge even after a high fluence of charged particles.  After irradiation-induced space charge sign inversion, the highest electric field in the sensor is located close to the n-electrodes used to 
collect the charge, which is also advantageous as it allows the sensors to be operated under-depleted. A further consequence of the higher mobility of the electrons is the larger 
Lorentz drift of the signal charges. This drift leads to increased charge sharing between neighboring pixels and 
is exploited to improve the spatial resolution. 



In n-in-n sensors, the junction that depletes the sensitive volume is realized as a large-area implant
on the back side of the sensor. In order to guarantee a controlled termination of the junction towards the 
edge of the device, a series of guard rings are implemented, meaning that both sides of the
sensor need photolithographic steps. The guard-ring scheme allows all sensor edges to be at ground potential, which greatly simplifies the construction of detector modules, because no high-voltage protection is needed to adjacent components
like ROCs or neighboring modules.


The interface between the silicon substrate and the silicon oxide carries a slight positive charge which increases by orders of magnitudes after ionizing radiation. This causes a conducting electron accumulation layer, which may short the electron-collecting electrodes. Therefore an n-side isolation is required. The technical implementation of
this isolation has a large impact on the pixel cell layout and was chosen to best match the techniques offered by the two vendors.


%

\begin{figure}[tb!]
\centering
  \includegraphics[height=5cm]{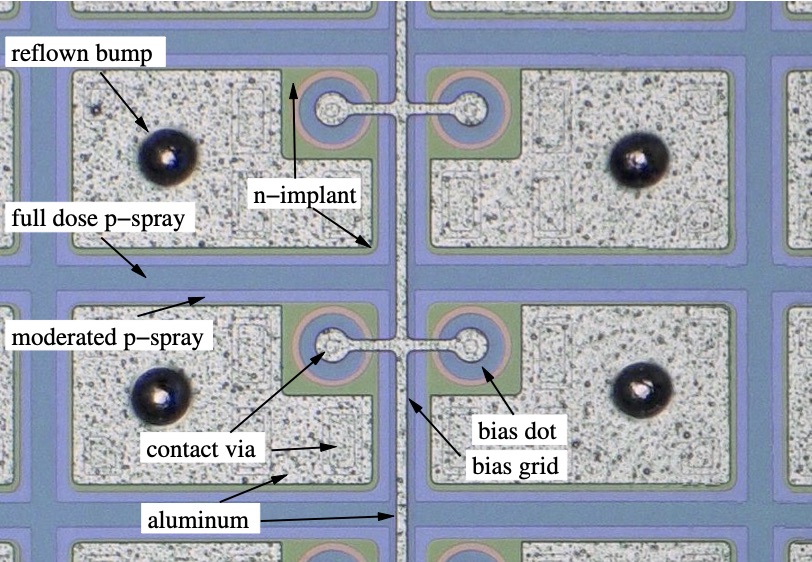}
  \includegraphics[height=5cm]{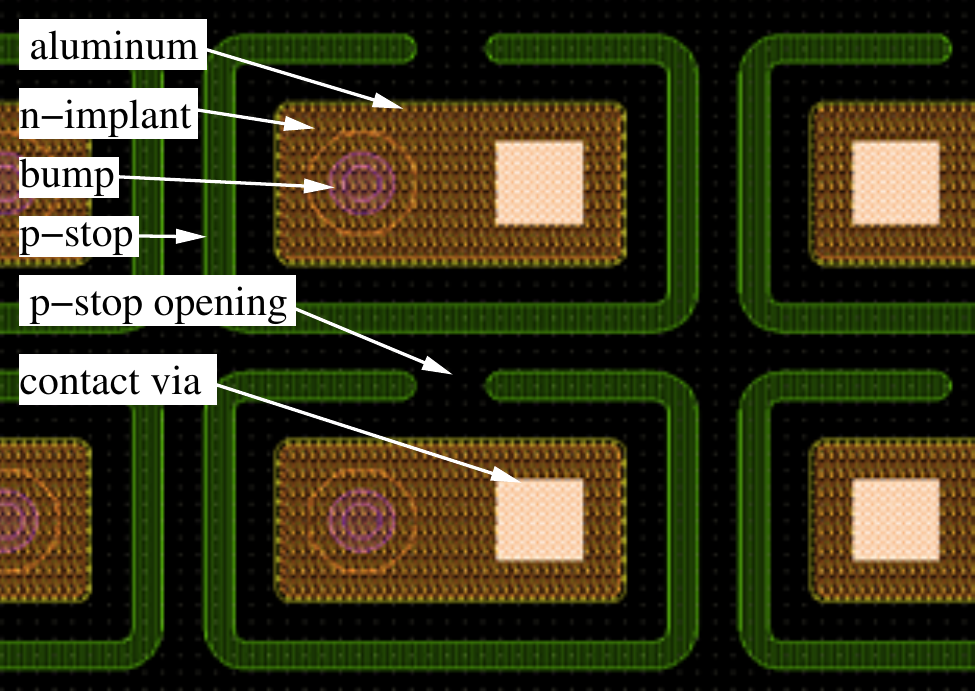}
  \caption{Photograph of four pixel cells on a \bpix sensor (left) and schematic of two pixel cells on an \fpix sensor (right).}
  \label{fig:sensor-pic}
\end{figure}

In the case of the \bpix sensors, the n-side isolation was
implemented through the moderated p-spray technique~\cite{ref:mod}
with a punch-through biasing grid. The moderated p-spray technique allows for
small distances between the pixel implants. Such a layout leads to a homogeneous 
electric field inside the sensor. The small gaps between pixel implants also facilitate
the implementation of punch-through bias structures, the bias dots. The bias dots provide a highly resistive connection to each pixel. This can be used to apply bias voltage to the sensor prior to any further assembly. This in turn allows sensor quality assurance measurements prior to bump bonding, such as the current-voltage (IV) characteristic. A photograph of four pixel cells in a \bpix sensor is shown in Fig.~\ref{fig:sensor-pic}~(left).

The \bpix sensors were produced on approximately 285\,\mum thick phosphorous-doped 
4-inch wafers from silicon mono-crystals produced in the float-zone (FZ) process. The resistivity of 3.7\,k$\Omega$\cm 
leads to an initial full depletion voltage of 55\,V.
The crystal orientation is $\langle 111 \rangle$. The first processing step
is an oxidation according to the recommendation of the 
ROSE collaboration~\cite{ref:rose} (Diffusion Oxygenated FZ material).
All \bpix sensors were processed using silicon originating from the same
ingot. Therefore, the variations of the full depletion voltage are small.

Three sensors were placed on each 4-inch wafer. A wafer was accepted when at least
two of the sensors fulfilled the specifications. Most critical 
was the requirement of a maximum current of $2\,\mu$A at 150\,V reverse bias voltage, measured at a temperature of +17\,\cel.





The \fpix sensors use open p-stops for n-side isolation. Each pixel is surrounded by an individual p-stop which has an opening on one side, as shown in Fig.~\ref{fig:sensor-pic}~(right). In between the p-stops, electrons will accumulate close to the surface and form a resistive grid covering the
whole sensor. The resistance of the grid depends strongly on the back side voltage. The openings connect each pixel to this grid providing the same functionality as the bias
dots in the \bpix layout. Owing to the presence of the p-stops, the distance between the charge-collecting
pixel electrodes is larger compared to the \bpix sensors, which leads to a smaller capacitance.
The disadvantage -- a less homogeneous drift field -- is of less importance in the forward region
as the charge sharing between pixels is caused by the geometric tilt of the modules and not by the magnetic field.

The \fpix sensors have been produced on 300\,\mum thick 6-inch FZ-wafers with $\langle 100 \rangle$-orientation. Eight sensors were placed on one wafer. A wafer was accepted if at least six of the sensors fulfilled the specifications~\cite{Arndt:2003ck, Dilsiz:2019rsr}. 

By design, the position resolution of the \cmsph depends strongly on the charge sharing
between pixels. The pixel shape of $100\times150\,\mum^2$ means that in order to obtain an optimal
position measurement in the azimuthal direction (the bending plane for charged particle tracks within the CMS magnet) the charge width\footnote{The charge width is defined as the projection on the module coordinates of the area where the charge is collected on the detector surface.} has to be
of the order of the pixel pitch, that is 100\,\mum. In the strong magnetic field of 3.8\,T provided by the CMS
magnet, the Lorentz angle (LA) for the drifting electrons has a value of about $27^\circ$.
With the sensor thickness of 285\,\mum, this produces a charge width of 145\,\mum, which is sufficient to share the charge between at least two pixels.
The LA depends strongly on the bias voltage of the sensor and weakly on the temperature. It is
also affected by the radiation damage in the sensor.  This means that in order to obtain the optimal position
resolution the LA has to be regularly monitored (Sec.~\ref{s:la}).

The durability of the modules is, to a large extent, defined by the possibility of increasing the sensor bias voltage to obtain a 
sufficiently high signal charge. During operation in CMS, modules in the innermost layer of the \cmsph have been run with high 
efficiency at a bias voltage of 450\,V up to an integrated luminosity of almost 120\,\fbinv. After the replacement of the innermost \bpix layer during LS2, the new innermost layer must withstand a fluence that is expected to be about twice as high until the end of Run~3. In order to maintain a high enough signal charge, the pixel detector power supplies have been upgraded to deliver a maximum voltage of 800\,V during LS2.

\subsection{Readout chip}
\label{s:roc}


The upgraded ROCs used in the \cmsph (\psidig~\cite{KaStli:2013vja}, \proc~\cite{Starodumov_2017}) are manufactured in the same 250\,nm CMOS technology as the ROC used in the original pixel detector (\psiana~\cite{Kastli:2005jj}). The design requirements for \psidig and \proc are summarized in Tab.~\ref{t:roc}. 

\begin{table}
  \begin{center}
    \caption{Parameters and design requirements for \psidig and \proc.}
    \label{t:roc}
   \vspace{0.1in}
   \begin{tabular}{l | c | c }
     \hline
      & \psidig & \proc \\
      \hline
      \hline
      Detector layer & \bpix L2-L4 and \fpix & \bpix L1 \\ 
   	 ROC size & $10.2\times7.9\,\mm^2$ & $10.6\times7.9\,\mm^2$ \\ 
   	 Pixel size  & $100\times150\,\mum^2$ & $100\times150\,\mum^2$ \\
         Number of pixels & $80\times52$ & $80\times52$ \\
    	 In-time threshold  & $<2000$\,\e & $<2000$\,\e \\  
    	 Pixel hit loss & $<2\%$ at 150\,\mhzcm & $<3\%$ at 580\,\mhzcm \\
     	 Readout speed & 160\,\mbs & 160\,\mbs\\
	Maximum trigger latency & 6.4\,$\mu$s & 6.4\,$\mu$s\\
        Radiation tolerance & 120\,\mrad & 120\,\mrad\\
      \hline
      \hline
   \end{tabular}
  \end{center}
\end{table}


The \psidig is used in the outer \bpix layers (L2-4) and in the \fpix detector. It maintains the well-tested and reliable core of the original ROC and its readout architecture based on the column-drain mechanism~\cite{HORISBERGER2001148}.  

The pixel matrix of the \psidig consists of an array of pixel unit cells (PUC) arranged in 26 double columns of 2x80 pixels each, which are controlled by the double-column periphery. The double columns, the double-column periphery, and the chip periphery are the three main functional units of a ROC. They fulfill the task of recording the position and charge of all hit pixels with a time resolution of 25\,ns, and store the information on-chip during the Level-1 trigger latency of the CMS experiment (currently 4.15\,$\mu$s). The behavior of the \psidig is controlled by means of 19 digital-to-analog converter (DAC) registers which can be programmed using a 40\,MHz serial bus. The design of the pixel matrix for the \psidig remains essentially unchanged compared to the \psiana, except for the implementation of an improved charge discriminator. The main modifications made in the chip periphery are to overcome the limitations of the \psiana at high rate. The ROCs need two different power supplies, namely +2.5\,V and +1.5\,V. These supply the digital and analog circuits, respectively, through internal linear voltage regulators. The power consumption of the \psidig is about 41\,mW for the analog part. The power consumption of the digital part has a static contribution of 70\,mW and a dynamic contribution that amounts to about 31\,mW per 100\,\mhzcm hit rate.

The PUC can receive a signal either through a charge deposition in the sensor or by injecting a calibration signal. Within the PUC, the signal is passed through a two-stage pre-amplifier and shaper system to a comparator, where zero-suppression is applied. The comparator threshold is set by a DAC for the whole ROC, but can be adjusted via a 4-bit DAC (trim bits) for each pixel individually. Furthermore, the comparator of a pixel can be disabled by setting a mask bit. If a signal exceeds the comparator threshold the analog pulse-height information is stored, the corresponding pixel becomes insensitive, and the column periphery is notified. The column periphery writes the value of the bunch crossing counter into a time-stamp buffer and issues a readout token. A column-drain mechanism is initialized to read out the pixel hit information. Hit pixels send the registered analog pulse-height information together with the pixel address to the column periphery for storage in the data buffers, before being set again into data taking mode. The communication between the PUC and the periphery allows for three pending column drains, meaning that the double columns are capable of recording new hits while still copying information from the previous hits to the buffers in the periphery. Upon arrival of the Level-1 trigger-accept (L1A) signal, the double-column periphery verifies the pixel hit information by comparing the time stamp with a counter delayed with respect to the bunch crossing counter by the trigger latency. In case of agreement the double column is set into readout mode and is not ready to accept any new data, otherwise the data in the corresponding buffer are discarded. When a readout token issued by the TBM arrives at the double-column periphery the validated data are sent to the chip periphery and the double column is reset. 

The main changes for the \psidig compared to the \psiana include the increase of the size of the data (from 32 to 80) and time-stamp (from 12 to 24) buffers to store the hit information during the trigger latency, the implementation of an additional readout buffer stage to reduce dead time during the column readout, and the adoption of 160\,\mbs digital readout. The readout speed of the ROC itself is unchanged, but the transition from 40\,MHz analog coded data to 160\,\mbs digital data allows faster readout of the modules. Consequently an 8-bit successive approximation analog-to-digital converter (ADC) running at 80\,MHz has been implemented in the \psidig. Digitized data are stored in a $64\times23$ bit first-in-first-out register, which is read out serially at 160\,MHz. A phase-locked loop (PLL) circuit has been added to derive the 80 and 160\,MHz clock frequencies from the LHC clock. 

The improvements in the design of the charge discriminator reduce cross talk between pixels and time walk of the signal~\cite{Kastli:2005jj} and thus lead to lower threshold operation (below 1500\,\e with noise less than 100\,\e in a module). Time walk is caused by the fact that the rise time of the amplified signal cannot be infinitely fast. Therefore, signals with different amplitudes cross the threshold at different times, with the low amplitude signals crossing the threshold later than the high amplitude signals. Also the decision speed of the comparator increases for small signals just slightly above the threshold. If the low amplitude signals are delayed beyond the 25\,ns time window between LHC collisions, they will appear in the next bunch crossing and will be lost for hit reconstruction. The threshold needed for the signal charge to be recorded in the correct bunch crossing is higher than the pixel threshold, and is referred to as ``in-time'' threshold. The effect of the time walk on the threshold was reduced significantly, from about 1000\,\e in \psiana to about 300\,\e in \psidig. 

Furthermore, higher radiation tolerance is achieved. 
The radiation tolerance of the \psidig has been tested after irradiation to up to 150\,\mrad using a 23\,MeV proton beam at ZAG Zyklotron AG in Karlsruhe, Germany. The \psidig shows excellent performance, and threshold and noise characteristics remain basically unchanged after irradiation~\cite{Hoss_2016}. In addition, comprehensive test beam studies have been conducted to verify the design and to quantify the performance of detector assemblies with the new ROCs in terms of tracking efficiency and spatial resolution~\cite{Dragicevic_2017}. Leakage currents from sensors will increase significantly after irradiation and the pixel input circuit has to be able to absorb it. There is no dedicated circuitry for current compensation in the pixels, however currents up to 50\,nA/pixel, can be absorbed, after feedback adjustments, without any observable gain changes.  At 100\,nA/pixel there is only a small (about 10\%) gain degradation.

\begin{figure}[tb!]
	\centering
	\includegraphics[trim=0 0 0 0,clip,width=1\textwidth]{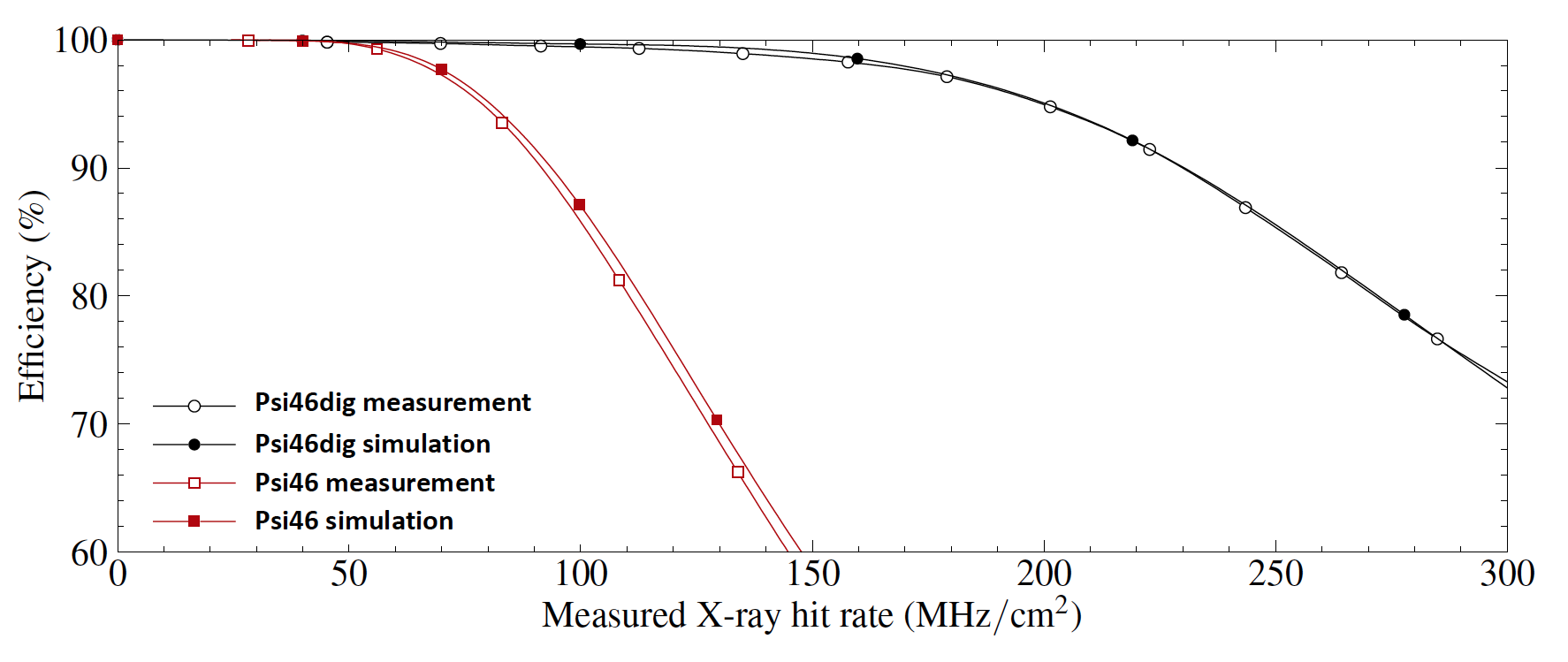}
	 \caption{Measured and simulated efficiencies for \psiana (used in the original pixel detector) and \psidig (used in the \cmsph in the outer \bpix layers and the \fpix detector) as a function of X-ray hit rates~\cite{marcorossini}.}
	 \label{fig:psi46digeff}
\end{figure}

The single-pixel hit efficiency at high rates has been measured using the internal calibration signal while exposing the ROC to high-rate X-rays~\cite{marcorossini}. The efficiency has been measured for pixel hit rates up to 300\,\mhzcm and was found to be in excellent agreement with expectations based on detailed architecture simulations, as shown in Fig.~\ref{fig:psi46digeff}. Based on the same architecture simulation but now using simulated proton-proton collision events, the data losses in the \fpix detector and in the outer \bpix layers are less than 2\% at the expected maximum hit rate of 120\,\mhzcm. The \psidig fulfills all the design requirements for the outer \bpix layers and the \fpix detector and has performed very well during the proton-proton collision data-taking in 2017 and 2018.

The \proc has been designed for the innermost layer, where hit rates of up to almost 600\,\mhzcm are expected. The main design requirements for the \proc are faster hit transfer from pixels to the periphery as well as dead-time free buffer management. This has been achieved by a complete redesign of the double column unit. In the \proc, the pixels within a double column are dynamically grouped into clusters of four and read out simultaneously to enable faster readout. Reading groups of four pixels in one step avoids the need for each hit pixel to initiate its readout sequence by the periphery. This approach speeds up the readout process significantly even if some pixels, out of the group of four, are read out despite having no hit. The readout is zero-suppressed in order to remove pixels in the clusters without measured signal amplitude. Furthermore, a new checkout mechanism has been implemented that allows the column-drain mechanism to run continuously and that does not require a buffer reset. An improved communication logic design between the PUC and the periphery allows for seven pending column drains in the \proc, compared to three pending column drains in the \psidig. The power consumption of the analog part of the \proc is the same as for the \psidig. The static digital power is 90\,mW with an increase of 20\,mW per 100\,\mhzcm hit rate.

The radiation tolerance of the \proc has been tested under proton irradiation with doses up to 480\,\mrad~\cite{Starodumov_2017}. The \proc remained fully operational after irradiation to a dose of 120\,\mrad, which is larger than the total dose expected during its operation in the innermost \bpix layer (accounting for the replacement of \bpix L1). Even at doses of up to 480\,\mrad only a slight degradation in the performance was observed. Furthermore, the high-rate performance of the \proc has been studied in laboratory tests with X-rays as well as in a 200\,MeV high-rate proton beam at the PSI Proton Irradiation Facility~\cite{Starodumov_2017}. 

The \proc has delivered high-quality physics data during operation in 2017 and 2018. Two shortcomings of the \proc have been noticed during data-taking. One is a higher-than-expected noise at high hit rates because of cross talk between pixels. The second is a lower-than-expected efficiency, also at high hit rates, because of a rare loss of data synchronization in double columns. Both problems were mitigated by operational procedures, avoiding any compromise in the data quality. Nevertheless, a revised design of the \proc, to be used in the replacement of the innermost \bpix layer in Run~3, has been developed. The higher-than-expected noise was traced to an inappropriate shielding of the circuitry for calibration pulse injection connected to the pre-amplifier input node. This issue has been addressed in the revised version. In addition, the routing and shielding of power and address lines was improved. Both changes lead to lower noise and lower cross talk between pixels.

\begin{figure}[tb!]
	\centering
	\includegraphics[trim=0 0 0 0,clip,width=0.65\textwidth]{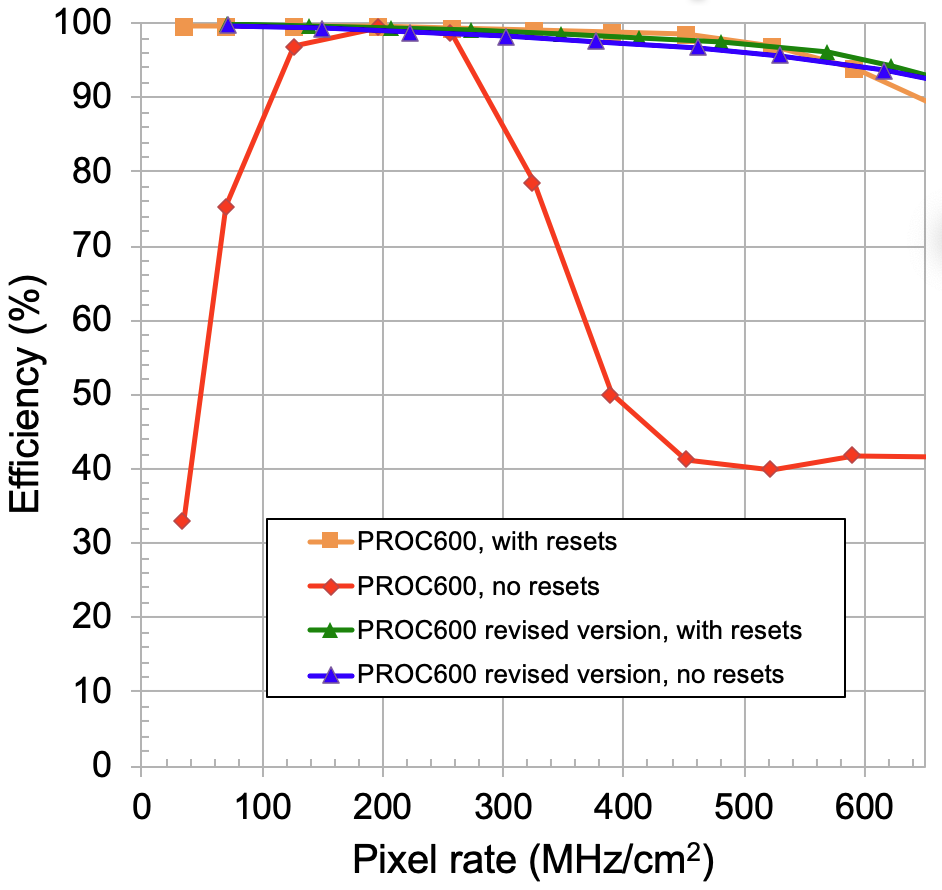}
	 \caption{Measured single-pixel hit efficiency for the \proc as a function of pixel hit rate in the 200\,MeV proton beam at PSI. The version of the \proc used in the CMS experiment in operation in 2017 and 2018 is compared to the revised version of the chip that will be used in the replacement of \bpix L1 in Run 3. The yellow and green points labeled "with resets" correspond to the data-taking mode in which every trigger signal was preceded by a reset signal.}
	 \label{fig:effproc}
\end{figure}

The main change in the revised version of the \proc addresses the rare events of data synchronization loss in the double columns. The issue has been tracked to a timing error in the time-stamp buffer of the double column, which leads to inefficiencies at low and high hit rates. When the time-stamp buffer is full, the double column no longer acquires new hits until the content of the buffer cell with the oldest hit information is cleared. If a coincidence of a return to acquisition mode and a new hit in a pixel occurs, it can generate a spurious column drain and thus loss of synchronization of the double column such that subsequent hits are not assigned to the correct event. Since the time stamp buffer is filled more frequently when the occupancy is high, the effect of the spurious signal affects the efficiency when running at high luminosity. At low rates, a very specific sequence of events generates a wrong time-stamp buffer-full signal. This in itself would not be a problem, but in combination with the issue of spurious column drains described above can again lead to a loss of synchronization. In both cases synchronization is restored by sending a reset signal to the double column. During operation, reset signals were sent at a frequency of 70\,Hz to mitigate the inefficiencies at low instantaneous luminosity~\cite{ref:bora}. Both timing issues have been corrected in the buffer logic of the revised \proc. The high-rate performance of the revised version of the \proc has been studied by measuring the single-pixel hit efficiency during operation in the high-rate proton beam at the PSI Proton Irradiation Facility. The result is shown in Fig.~\ref{fig:effproc}. The revised version of the \proc maintains a single-pixel hit efficiency above 95\% at rates up to 600\,\mhzcm. Furthermore, the time-walk behavior has been optimized by adjusting the speed of the comparator.

\label{sec:roc}

\subsection{Token bit manager}
%
\label{s:tbm}

The main functionality of the TBM is to synchronize the module data transmission. The TBM issues a readout token upon arrival of an L1A signal from the CMS back-end trigger electronics~\cite{Khachatryan:2016bia}. The
token is passed to each ROC in turn and the readout is initialized. The last ROC in the chain sends the token back to the TBM. The TBM multiplexes the signals from the ROCs, adds a header and a trailer to the data stream, and drives the signal through the readout link. Before issuing the next readout token, the TBM awaits the return of the previous token. Trigger signals that arrive during the readout of a previous event are placed on a stack (up to 32 trigger signals).  

The TBMs for the \cmsph are a digital
evolution of the respective analog chip used in the original CMS pixel detector \cite{Bartz:2005qc}. To increase the data bandwidth sent from a module, two 160\,\mbs ROC signal paths, with one path inverted, are multiplexed into a 320\,\mbs signal, then encoded into a 4-bit/5-bit Non-Return to Zero Inverted (NRZI) 400\,\mbs data stream. This is suitable
for use in transmitting the data optically to the downstream DAQ system. By adopting a digital readout at 400\,\mbs and using four links per module in \bpix L1, the readout bandwidth is increased by a factor of four compared to the original innermost layer with two analog links at 40\,MHz. The
digital TBM has single output (TBM08) and dual output (TBM09, TBM10)
versions, which are used in different parts of the detector. TBM09 and TBM10 models differ only in the delay between the trigger and when
tokens are sent to initiate ROC readout. The TBM08 version has two independent 160\,\mbs ROC 
readout paths, and the TBM09 and TBM10 versions have four separate, semi-independent, 
160\,\mbs ROC readout paths. The headers and trailers corresponding to the two 160\,\mbs ROC readout paths that share a 400\,\mbs readout link are sent synchronously. 
The time when one of the 160\,\mbs readout paths in TBM09 and TBM10 is idle is filled with digital 
zeros.

Each TBM has a 5-bit hub address. The address is defined by the voltage levels applied to the corresponding pads on the TBM. The pads are either connected through wire bonds to the HDI or to internal pull-down resistors. The hub address is used to uniquely identify each module served by a given control link to send the configuration information to the correct TBM for subsequent loading into the ROCs. In the \bpix detector, up to 28 modules are served by the same control link, while in the \fpix detector 14 modules share a common link. 

In addition to the increased output bandwidth, several features were added to the TBMs for the \cmsph. A token timeout was added that resets the ROCs and drains buffered data of the triggered events on the stack if a token does not return within an adjustable amount of time.
The adjustment is 6.4\,$\mu$s times an 8-bit setting and can be disabled. The setting used during operation corresponded to
147\,$\mu$s, making sure that very long readouts did not block the DAQ system.
The functionality of issuing an automatic reset was added in order to send periodic ROC resets every $N$ triggers, where $N$ is a 
multiple of 256 times an 8-bit setting. However, this functionality has not been used. Instead it was decided
that the periodic reset signals are issued centrally, by the CMS trigger control and distribution system (TCDS), to recover from the loss of synchronization of the
\proc. The periodic reset signals are sent after 3000 bunch crossings without L1A signals in order to drain the data from the ROC buffers and
avoid losing the data. As a result of adopting a digital readout, delay adjustments could be added for the ROC readouts, the token outputs, the data headers and the data trailers.
Relative phase adjustments between the 40\,MHz incoming clock, the 160\,MHz clock
and the 400\,MHz clock were also added. The size of the data header was increased
to allow additional TBM status information to be transmitted (currently an unused feature). 
The size of the data trailer was increased to indicate whether a token timeout 
and/or automatic reset occurred and to transmit the number of buffered trigger signals 
on the stack waiting to be processed.


Operation of the TBM during collision data taking revealed a vulnerability to single-event upsets (SEUs). SEUs are interactions in which a highly ionizing particle deposits a significant amount of energy in silicon that affects the functioning of a transistor and changes its state. The transistor design can be modified to make it more robust, however the effect cannot be completely eliminated. Therefore, with some probability, every transistor in the readout chain of the pixel detector will be affected by SEUs. For the CMOS technology used in the design of the ROC and TBM chips the SEU probability was measured using pion beams and was found to be $\rm 2.4\times10^{-14}\,\cm^2$ per storage cell for unprotected transistors and $\rm 2.6\times10^{-16}\,\cm^2$ per storage cell for protected transistors~\cite{Kastli:2005jj}. Because of a design issue in the TBM, a single transistor in the circuit responsible for event synchronization is not protected against SEUs and cannot be reset by a reset signal. Whenever it is blocked, the readout chain is interrupted and can only be recovered by a power cycle. This means
that parts of the pixel detector have to be periodically power-cycled. Since the flux of particles is highest in the region closest to the interaction point, \bpix L1 is most affected by this issue. The observed fraction of TBM cores (two cores per TBM) which would be blocked without intervention in \bpix L1 is about 0.7\% per 100\,\pbinv of integrated luminosity, which translates into an overall inefficiency of about 0.2\%. For the outer layers the flux of charged particles is significantly lower compared to L1 and therefore SEU effects in the TBMs are much reduced.

An additional iteration of the TBM chips was designed in the spring of 2018 to address the TBM SEU issue and to add an adjustable delay of up to 32 ns to the 40\,MHz clock. The delay was added to allow finer adjustment of the relative timing between modules (Section~\ref{s:phase}). The new chips will be used in new modules in \bpix L1 that will be incorporated during the consolidation work in LS2.


\subsection{BPIX module construction}

The 1184 \bpix modules installed in the detector all have the same geometry, but three module designs were needed to meet the requirements of the different layers. \bpix L3 and L4 modules have an identical design, based on the \psidig ROC and the TBM08, with one readout link per module. The \bpix L2 modules also feature the \psidig ROC, but use a TBM09 chip that has two readout links to match the higher data volume at smaller radii. The hit rates in \bpix L1 requires not only two TBM10 chips with a total of four readout links, but also the \proc, different cables, and, because of severe space constraints, a different mounting scheme. The different module types are summarized in Tab.~\ref{t:mod} (and displayed in Fig.~\ref{fig:modulebpixfpix}).

\begin{table}
  \begin{center}
    \caption{Overview of module types used in the \cmsph.}
    \label{t:mod}
   \vspace{0.1in}
\resizebox{\textwidth}{!}{   \begin{tabular}{l | c | c | c | c | c  }
     \hline
      & ROC & Number of  & TBM & Number of  & Number of  400\,\mbs  \\   
      & & ROCs & &  TBMs & readout links per module \\
      \hline
      \hline
      \bpix L1 & \proc & 16 & TBM10 & 2 & 4  \\
      \bpix L2 & \psidig & 16 & TBM09 & 1 & 2  \\
      \bpix L3, L4 \& \fpix & \psidig & 16 & TBM08 & 1 & 1  \\
      \hline
      \hline
   	   
   \end{tabular}}
  \end{center}
\end{table}


The production of the \bpix modules was shared by five consortia, including institutions from Germany, Switzerland, Italy, Finland, and Taiwan, in five different module-assembly centers. ROCs and TBMs were probed on wafers before dicing. The yield was 93\% for ROCs and 53\% for TBMs\footnote{A batch of TBMs produced for the new L1 of BPIX was found to have a yield close to 90\% after a new wafer probe cleaning method was used before probing.}. While the assembly tools and procedures were mostly standardized~\cite{Konig:2007pd}, each center used different bump-bonding techniques. Two module-assembly centers worked with industrial vendors for the bump bonding (IZM~\cite{r:izm, ref:Fritzsch} for INFN and ADVACAM~\cite{r:advacam} for CERN/National Taiwan University/University of Helsinki). Bump bonding for the Swiss institutions was done in cooperation with Dectris~\cite{r:dectris}, based on the indium process developed at PSI~\cite{Broennimann:2005qv}. The KIT/RWTH Aachen consortium developed a cost-effective combination of an in-house flip-chip step at KIT with SnPb bumps deposited on ROC wafers by RTI~\cite{r:rti, ref:huffman, Caselle:2016hjf}.
The DESY/University of Hamburg production was done completely in-house except for the under-bump metallization (UBM) of sensors. The sensor UBM consists of electroless nickel plating by PacTech~\cite{r:pactech, Hansen:2017nsr}, which is also used in the KIT process. Solder balls were placed sequentially with a laser-assisted solder-sphere jetting technique, at a rate of 6-8 balls per second~\cite{Hansen:2017nsr}. The quality of the bump bonding was in each case verified as soon as the bump-bonded assemblies ('bare modules') were produced, or received by the module-assembly center. Thereby, single ROCs were contacted individually with a probe-card while supplying bias voltage to the sensor. This was performed to verify the functionality of ROCs and TBMs, to measure the leakage current of the sensor, and to check the quality of the bump-bonding process. The replacement of single ROCs that failed in an otherwise good module was practiced by four of the five centers, and about  10-20\% of the modules were reworked. The total bump-bonding yield, including rework, varied from center to center, and ranged between 84 and 96\%.

To turn the bare module into a full module, a thin four-layer HDI was glued onto the back of the sensor and wire-bonded to the ROCs. The gluing stations were operated manually with alignment pins guaranteeing sufficient mounting precision. TBMs had already been mounted, wired-bonded and tested on the HDI before joining the HDI and the bare module. In contrast to the original pixel detector modules, the upgraded modules feature a detachable cable. A temporary short cable stayed connected during module assembly and testing and was only replaced by the long final cable for mounting on the detector mechanics.

\begin{figure}[tb!]
	\centering
      \includegraphics[trim=0 0 0  0,clip,width=0.8\textwidth]{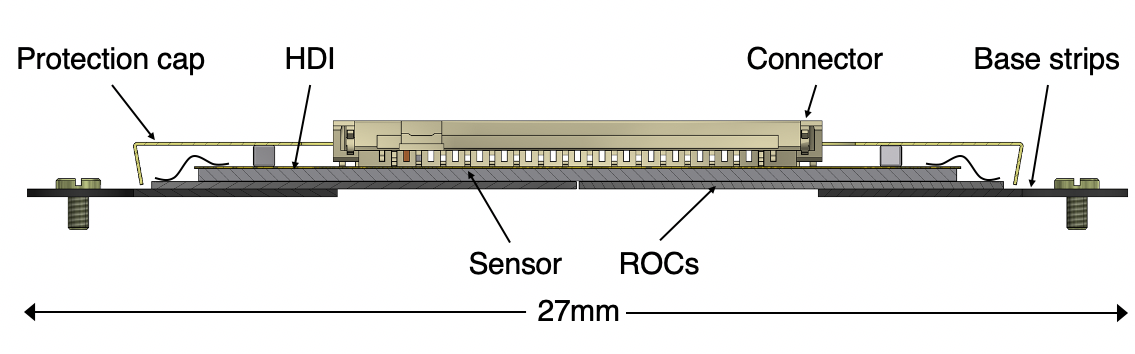}
	 \caption{Cross section of a pixel detector module for \bpix L2-4 cut along the short side of the module.}
	 \label{fig:modxsec}
\end{figure}

\bpix L2-4 modules are mounted using base strips, that were glued under the ROCs on the two long sides of the module (Fig.~\ref{fig:modulebpixfpix}, center). Silicon nitride was chosen as the material for the base strips, as it is an insulator with a similar coefficient of thermal expansion as silicon. The base strips have extensions with holes for mounting screws that match the mounting points on the \bpix mechanics. A drawing of the cross section of a detector module for \bpix L2-4 is shown in Fig.~\ref{fig:modxsec}. The tight space requirements in the innermost \bpix layer required a different scheme with clamps between two modules (Fig.~\ref{fig:modulebpixfpix}, left). 

\begin{figure}[tb!]
	\centering
	\includegraphics[trim=0 100 0 120,clip,width=0.7\textwidth]{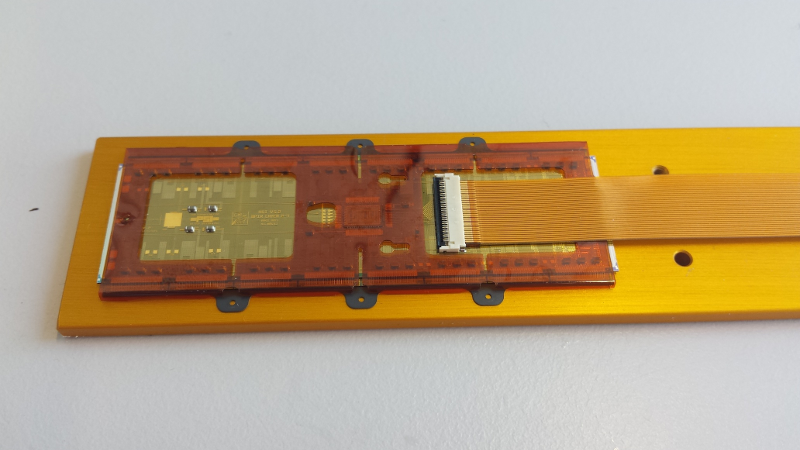}
	 \caption{Picture of a \bpix L2 module after assembly of the protection cap. The amber-colored protection cap covers the wire bonds of ROCs and TBMs. In the picture, the flat polyimide flex cable used for module testing is connected to the module. }
	 \label{fig:modulewithcap}
\end{figure}

The last assembly step, after the successful completion of all quality assurance tests, was the mounting of a protection cap made from a 75\,\mum thick polyimide foil. The cap protects the wire bonds of ROCs and TBMs against mechanical damage from the cables of other modules when mounted on the \bpix mechanics. A picture of a \bpix L2 module after assembly of the protection cap is shown in Fig.~\ref{fig:modulewithcap}. A single L2 module, excluding the cable, has a mass of about 2.4\,g and represents a thickness corresponding to 0.8\% of a radiation length at normal incidence angle.  


\subsection{FPIX module construction}

The 672 \fpix modules installed in the detector all have the same design and are instrumented with \psidig ROCs and a TBM08 with one readout link per module (Tab.~\ref{t:mod}). The \fpix modules (Fig.~\ref{fig:modulebpixfpix}) use a different sensor, HDI and aluminum/polyimide flat flex cable and are not interchangeable with \bpix modules. 

The bump-bonding procedure relied on a more automated version (Datacon APM2200 bump bonder) of the process used by the vendor RTI for the original \fpix detector (FC150 bump bonder)~\cite{Merkel:2007zz}. The yield for pre-production modules was 87\% and therefore it was felt that no bare-module testing was needed for the production. However, during early production a poor yield of 30\% was seen. After disassembling and inspecting a malfunctioning module it was found that abrasive blade dicing debris was damaging ROCs. Though a visual inspection of ROCs prior to bump bonding helped to identify damaged ROCs, higher yields were recovered only when the photoresist mask was left in place during dicing as was done during the pre-production. After this change the yield improved to 85\%. Because of the poor yield in the early production it was decided to perform testing of bare modules for the last two batches of bump-bonded modules. After failing modules had been reworked, the yield rose to above 90\%. 

\begin{figure}[tb!]
	\centering
	\includegraphics[trim=0 0 0 0,clip,width=0.8\textwidth]{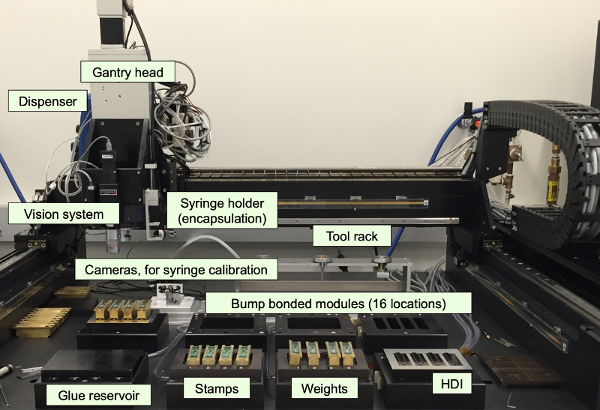}
	 \caption{A picture of the robotic gantry as used for the assembly of the \fpix modules. Labels indicate the main tools used for the assembly.}
	 \label{fig:gantry}
\end{figure}

The bare modules were sent from RTI to two \fpix module assembly and testing sites in the USA. Surface mount components were soldered to the HDI, and a TBM08 chip was glued and wire-bonded to the HDI. The HDI was then glued and wire-bonded to the bare module. Modules were assembled using an automated gantry (Fig.~\ref{fig:gantry}) with 10\,\mum precision using a vacuum chuck and pick and placement equipment. The vision system labeled in Fig.~\ref{fig:gantry} used the fiducial markers on the HDI and bare module to assemble parts with sufficient precision. The wire bonds were encapsulated with Dow Corning Sylgard~186 to protect against humidity and to mechanically support the wire bonds. The gantry used for the module assembly steps is software controlled for both the gluing and encapsulation steps, leading to high reproducibility. 




\subsection{Module qualification and grading}
\label{s:testing}


Modules have been tested in similar test setups in all production centers in Europe and the USA. Up to four modules could be tested in parallel on each setup. The modules were connected via copper cables to adapter cards, which were in turn connected to a digital test board (DTB) via a SCSI cable. The DTB is a compact DAQ system and contains an FPGA and a NIOS processor. The module testing software was running on a PC, connected via USB to the DTB. Modules were placed in a cold-box that provided the desired temperature and relative humidity ($<10\%$) during testing. All modules were tested at two temperatures, once at +17\,\cel  and twice at $-20$\,\cel, which corresponds to the temperature range expected during operation. Additionally, each module was exposed to ten thermal cycles between +17\,\cel  and $-20$\,\cel in order to verify that thermal stress does not create any damage in the module.

The testing procedure includes verification of the basic functionality of the module, the measurement of the efficiency of the pixel unit cells, and equalization and calibration of the pixel response. In addition to the functionality tests of ROCs and TBMs, the IV characteristic of each module was determined by measuring the leakage current as a function of the reverse bias voltage. Module qualification was performed using the \pxar testing framework~\cite{Spannagel:297247}. After completion of the module evaluation process, modules suitable for detector installation were selected based on predefined grading criteria. Details of the module grading criteria and production yield are discussed below. 


At the beginning of the testing procedure all modules were tested for basic functionality and initial parameters for the DAC settings were obtained. First, the \vana DAC setting, which regulates the analog supply voltage of the amplifiers, was adjusted. It was checked that the ROC can be programmed by changing the \vana DAC setting and verifying the corresponding change in the analog current. The \vana DAC was then set such that each ROC drew the nominal analog current of 24\,mA. In the next step, the delay settings for the readout of the ROCs and TBM(s) were adjusted by performing a scan over the available phase space. The delay settings were scanned for all TBM cores simultaneously and set within the center of the valid region. Then, the setting of the comparator threshold (\vthrcomp) was adjusted by using the charge injection mechanism of the ROC. The amplitude of the injected signal is controlled by the \vcal DAC, and its time delay by the \caldel DAC. A two-dimensional scan over the \vthrcomp and \caldel DAC settings was performed for a set of pixels. The working point was chosen in the center of the valid range for \caldel and at a value of the \vthrcomp that is well above the noise level, where the hit detection efficiency is maximal. 

%




In the following step, the functionality of every pixel was verified. Several internal calibration signals were sent to each pixel and its response was analyzed. If the number of signals sent to the pixel and the number of signals read out from the pixel was the same, the pixel was considered functional. At the same time, the address of the pixel was verified by comparing the address to which the internal calibration signal was sent with the decoded address. Finally, the functionality of the pixel mask mechanism was tested. Calibration signals were sent to masked pixels and it was verified that there was no response from the corresponding pixels.



\begin{figure}[tb!]
\centering
  \includegraphics[width=0.45\textwidth]{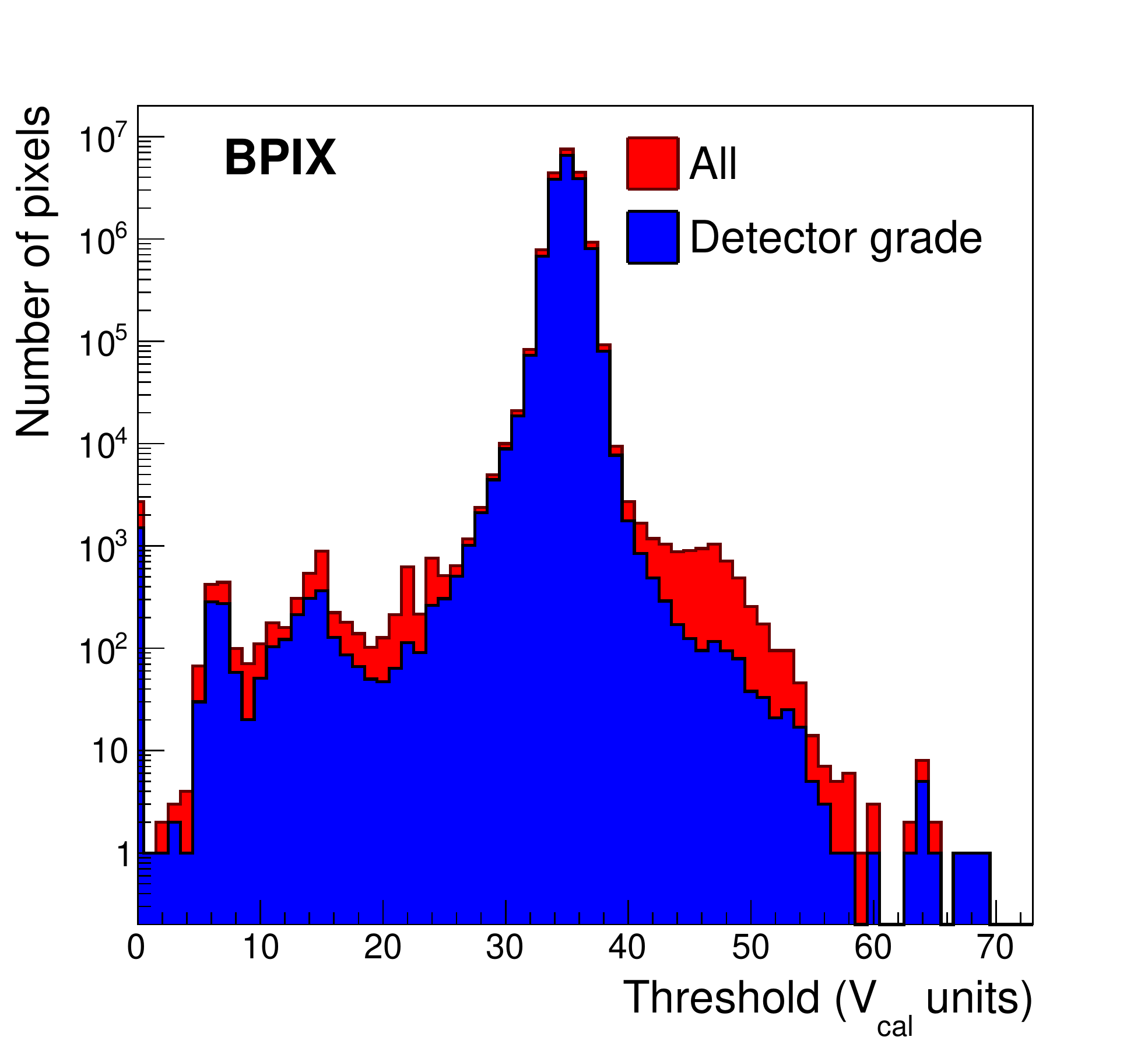}
  \includegraphics[width=0.45\textwidth]{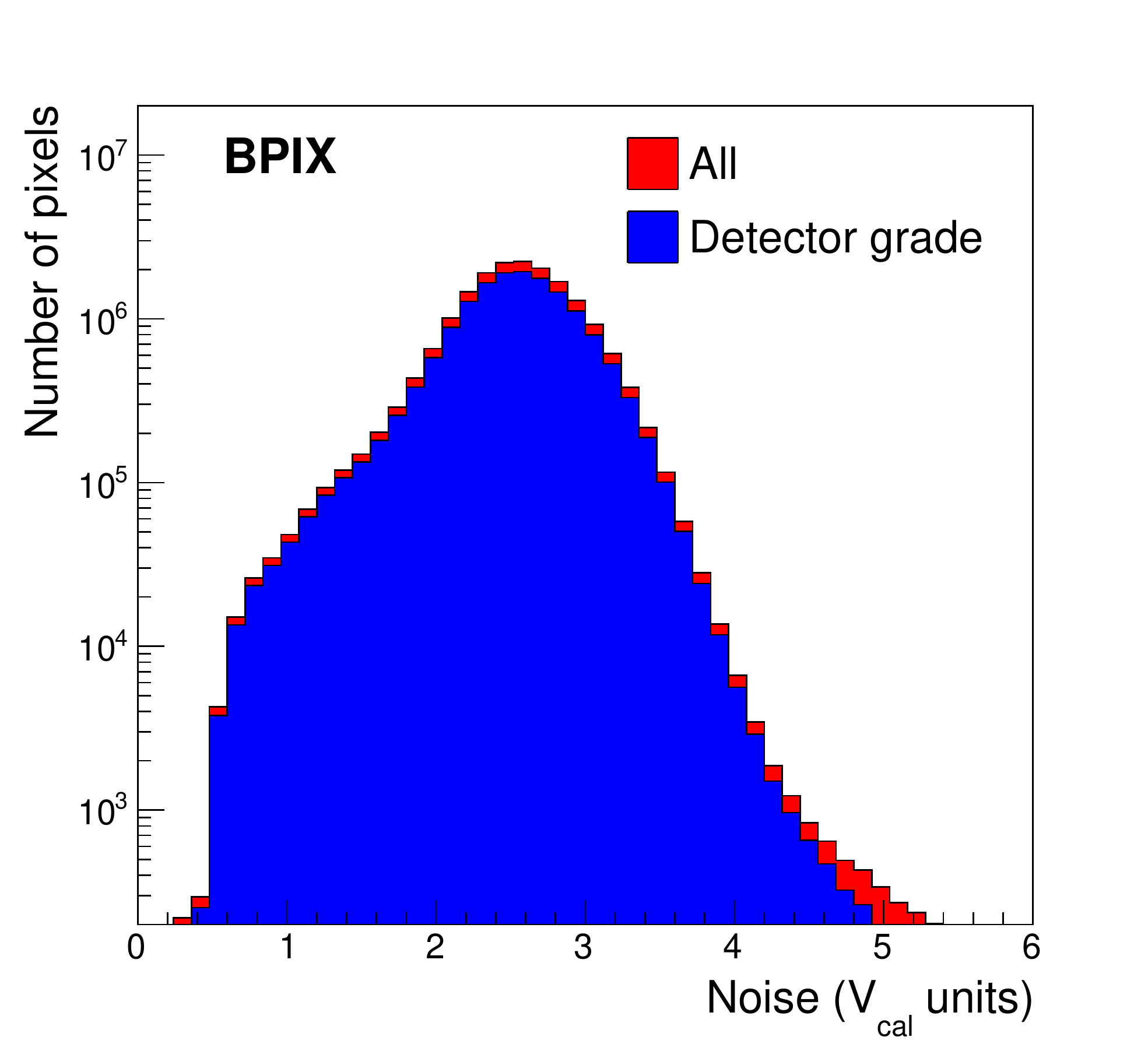}\\
  \includegraphics[width=0.45\textwidth]{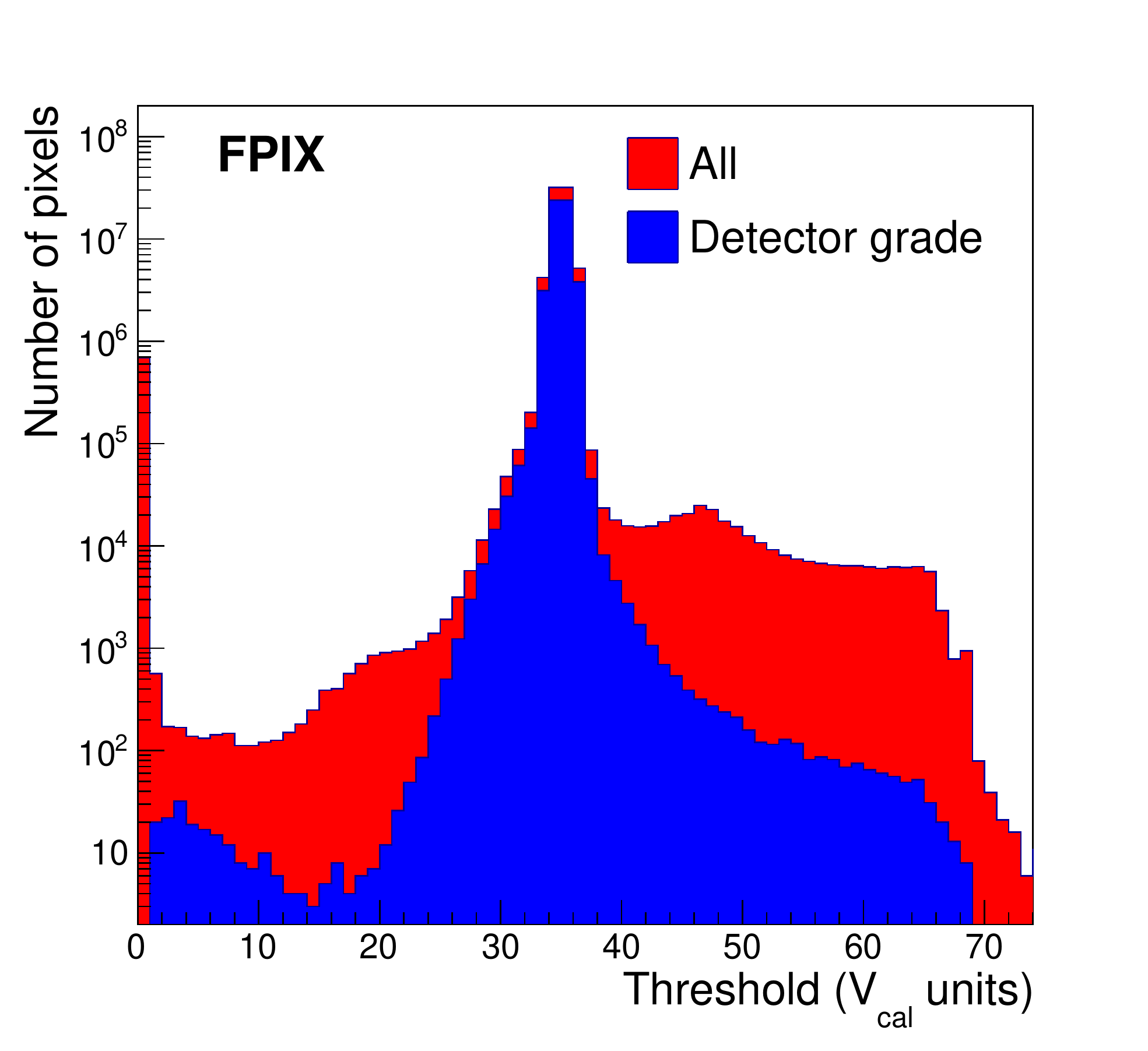}
  \includegraphics[width=0.45\textwidth]{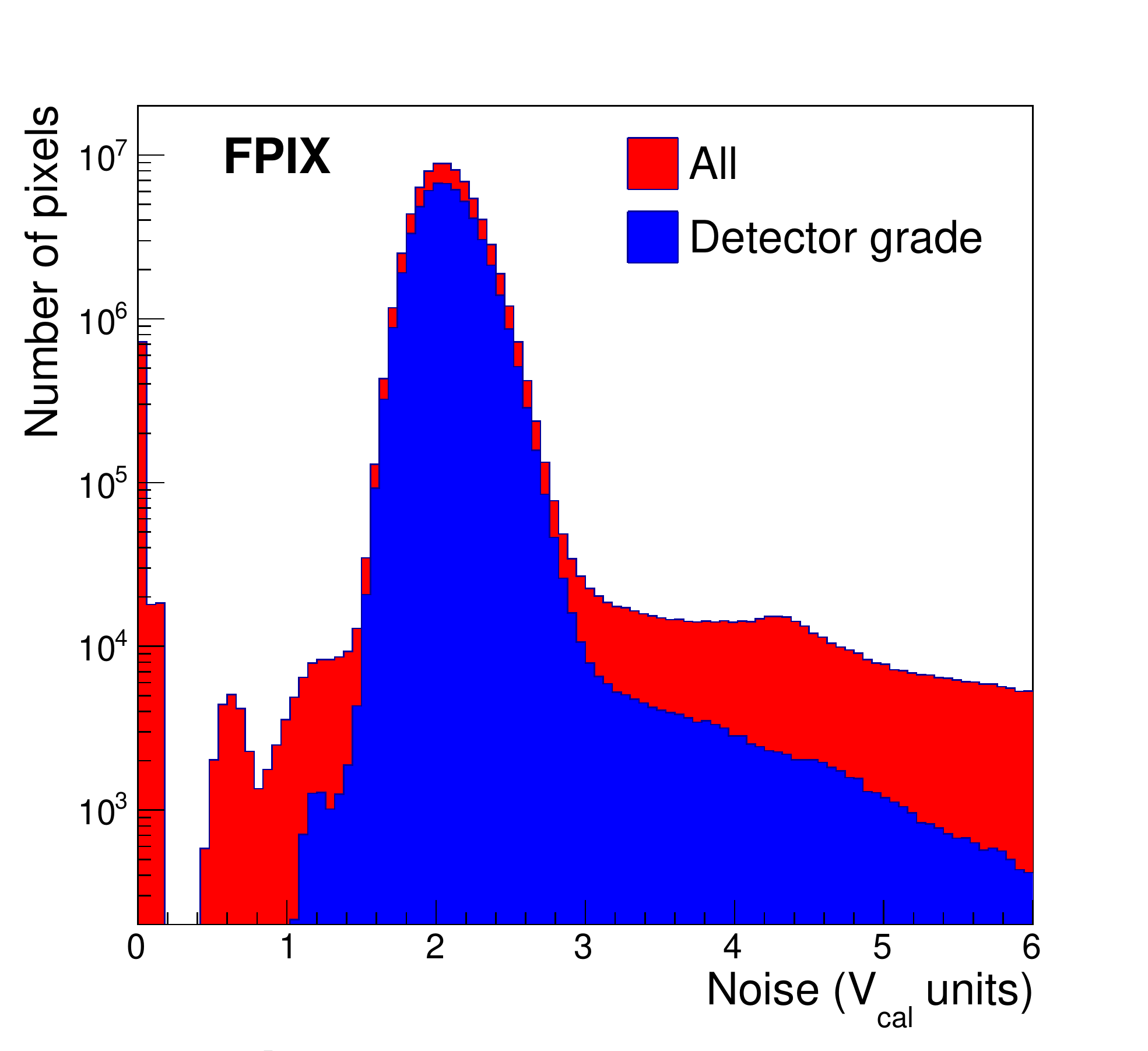}
  \caption{Distribution of threshold (left) and noise (right) per pixel for all modules that completed the qualification process. The upper plots are for \bpix modules, the lower plots for \fpix modules. The module grading criteria are discussed in the main text.}
  \label{fig:pixel_trim_noise}
\end{figure}

The trimming procedure was then performed. This procedure aims to equalize the thresholds of all 4160 pixels within a ROC. While the global threshold can be adjusted per ROC, the individual pixel thresholds are fine-tuned by the trim bits, whose range is set by the \vtrim~DAC.
The trim bits and \vtrim~DAC were iteratively adjusted to a target threshold, corresponding to 2000\,\e for module testing. Furthermore, the functionality of the trim bits was verified by sequentially enabling each bit and checking its effect on the pixel threshold distribution. 
The threshold and noise after each iteration of the trimming procedure were determined by measuring the S-curve~\cite{Chatrchyan:2009aa}. The S-curve measures the hit detection efficiency as a function of the amplitude of the injected calibration signal, adjusted by the \vcal DAC. It gets its name from the shape of the error function used to fit the measured curve of efficiency vs. \vcal for each pixel. The value corresponding to a response efficiency of 50\% determines the threshold, while the noise is proportional to the width of the error function. The threshold and noise distributions for \bpix and \fpix modules that completed the evaluation process are shown in Fig.~\ref{fig:pixel_trim_noise}. Besides determining the average threshold and noise per ROC, the measurement was also used to flag noisy pixels. Some differences in the noise and threshold distributions of the \bpix and \fpix detector modules can be
attributed to the different sensor technologies and the fact that the sensors are operated at different bias voltages. The tails at high values of the
noise distributions are mostly due to the larger area pixels placed at the edges of the ROCs. 


\begin{figure}[tb!]
\centering
  \includegraphics[width=0.5\textwidth]{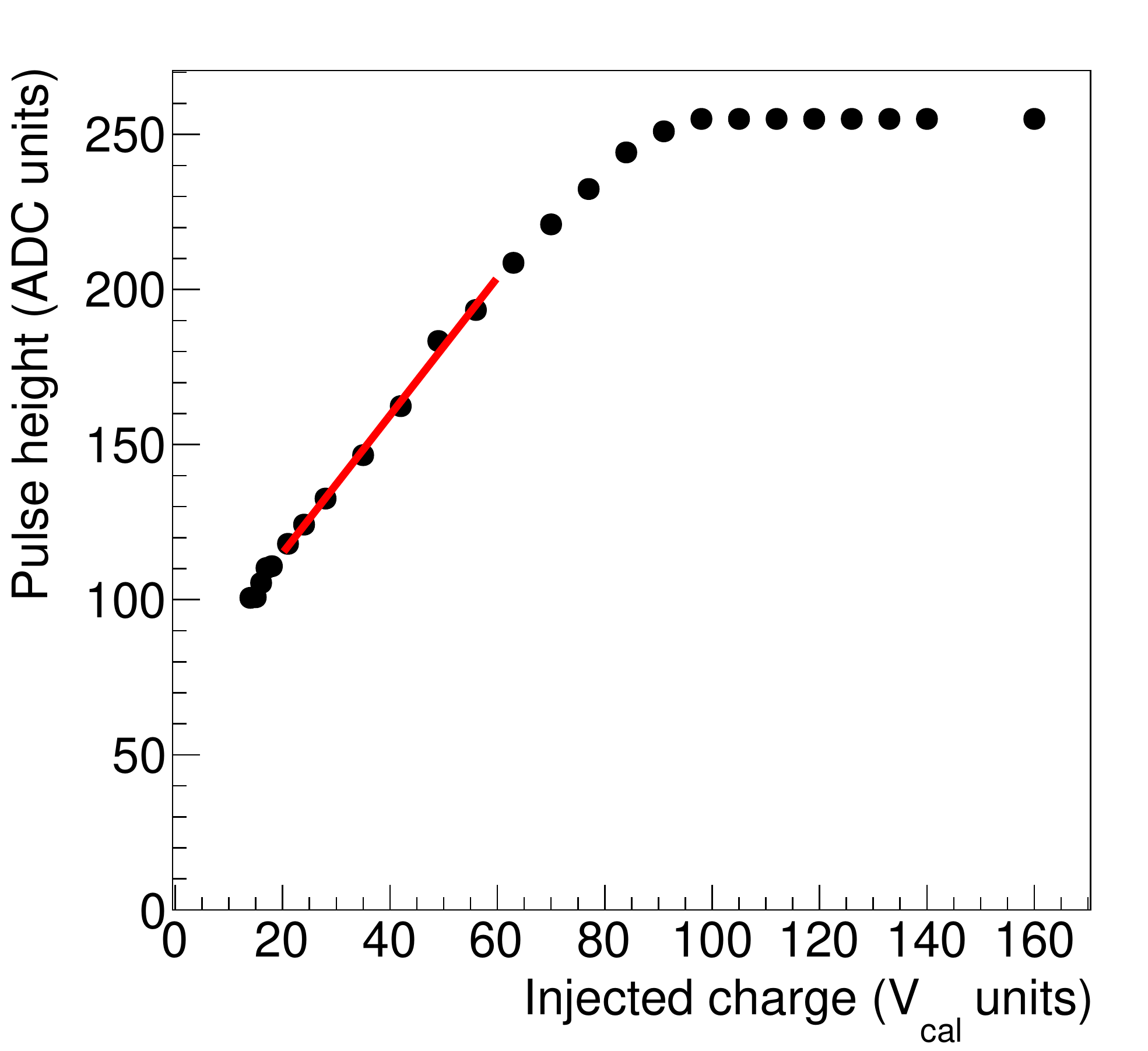}
  \caption{Pulse height measured in one pixel as a function of amplitude of the internal calibration signal in units of \vcal. The slope of the curve below saturation is fitted with a linear function.}
  \label{fig:PH_curve}
\end{figure}

In the next step, the pulse height response was adjusted and calibrated. The appropriate dynamic range for the 8-bit ADC that digitizes the recorded pulse height is set by two DACs, called \phoffset and \phscale. \phoffset adds a constant offset to the pulse height measurement, while \phscale effectively sets the gain of the ADC. To use the ADC most effectively, the pulse height response is optimized by injecting signals with different charge amplitude. The calibration evaluates and records the most probable value of the pulse height distribution as a function of the \vcal~DAC setting for each pixel, as shown in Fig.~\ref{fig:PH_curve}. These curves were fit with a linear function and the fit parameters were stored for offline reconstruction of the hit position (discussed in Sec.~\ref{gaincalibration}). It is sufficient to only consider the linear range of the distribution since it covers the range of charges most relevant for the position measurement. Details about this calibration procedure can be found in Ref.~\cite{Chatrchyan:2009aa}. 

A test was then performed to check the quality of the bump bonding by identifying missing or bad connections. The test made use of a feature in the ROC that allows a calibration signal to be sent to a top metal pad in the ROC~\cite{Kastli:2005jj}. The pad and the sensor are coupled capacitively through an air gap. The generated signal was read out by the ROC, and in this way the pixel capacitance was measured. Any pixel that had a capacitance five standard deviations below the mean was flagged as faulty.
The different bump-bonding technologies used at the various module production sites generated differences in the
bump heights, which strongly influenced the capacitive couplings. As a result the parameters of the bump-bonding
test had to be adapted for each production site. The fraction of disconnected bumps varied between 0.01\% and 0.05\%, depending on the technology used. The average fraction of disconnected bumps was 0.025\%, randomly distributed with some tendency to be higher at module corners.
The last step in the testing procedure was the measurement of the leakage current of the sensor as a function of the applied bias voltage. The IV characteristic of the modules was determined to assess whether the module was damaged by handling and to ensure that there was no intrinsic sensor defect that manifests itself only at low or high temperature.

\begin{figure}[tb!]
\centering
  \includegraphics[width=0.5\textwidth]{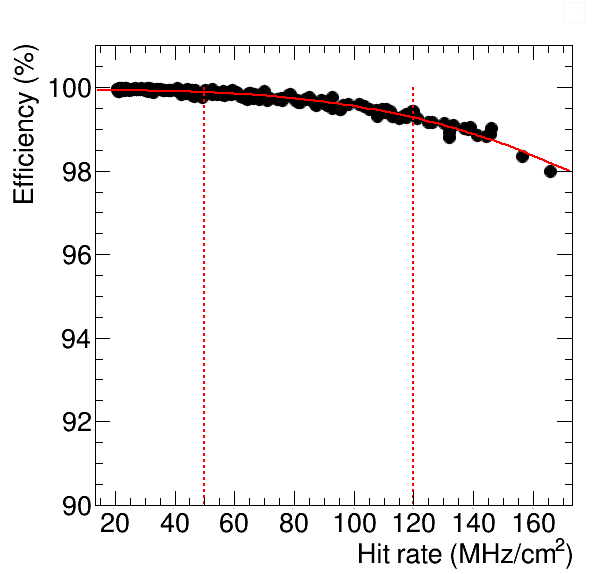}
  \caption{Single-pixel hit efficiency versus X-ray hit rate measured for double columns in a ROC of a \bpix L2 module. The dashed lines indicate the values at which the single-pixel hit efficiency per double column is determined during the module qualification and cover the range of expected hit rates in L2-L4. The solid red line is a fit to the data points using a polynomial function.}
  \label{fig:Hit_efficiency_L2}
\end{figure}

All modules were tested with X-rays in order to measure the hit-detection efficiency at high occupancy and to calibrate the absolute energy response of every ROC~\cite{Chatrchyan:2009aa}. First, each module was exposed to a direct high-rate X-ray beam. The number of hits per pixel was recorded in several runs with different hit rates ranging from 20 up to 160\,\mhzcm. For \bpix L1 modules the high-rate performance was tested up to 600\,\mhzcm. The hit efficiency was measured using the internal calibration signal while exposing the module to high-rate X-rays to emulate the occupancy expected during detector operation. As an example, the result of an X-ray test of a \bpix L2 module is shown in Fig.~\ref{fig:Hit_efficiency_L2}. 

\begin{figure}[tb!]
\centering
  \includegraphics[width=0.5\textwidth]{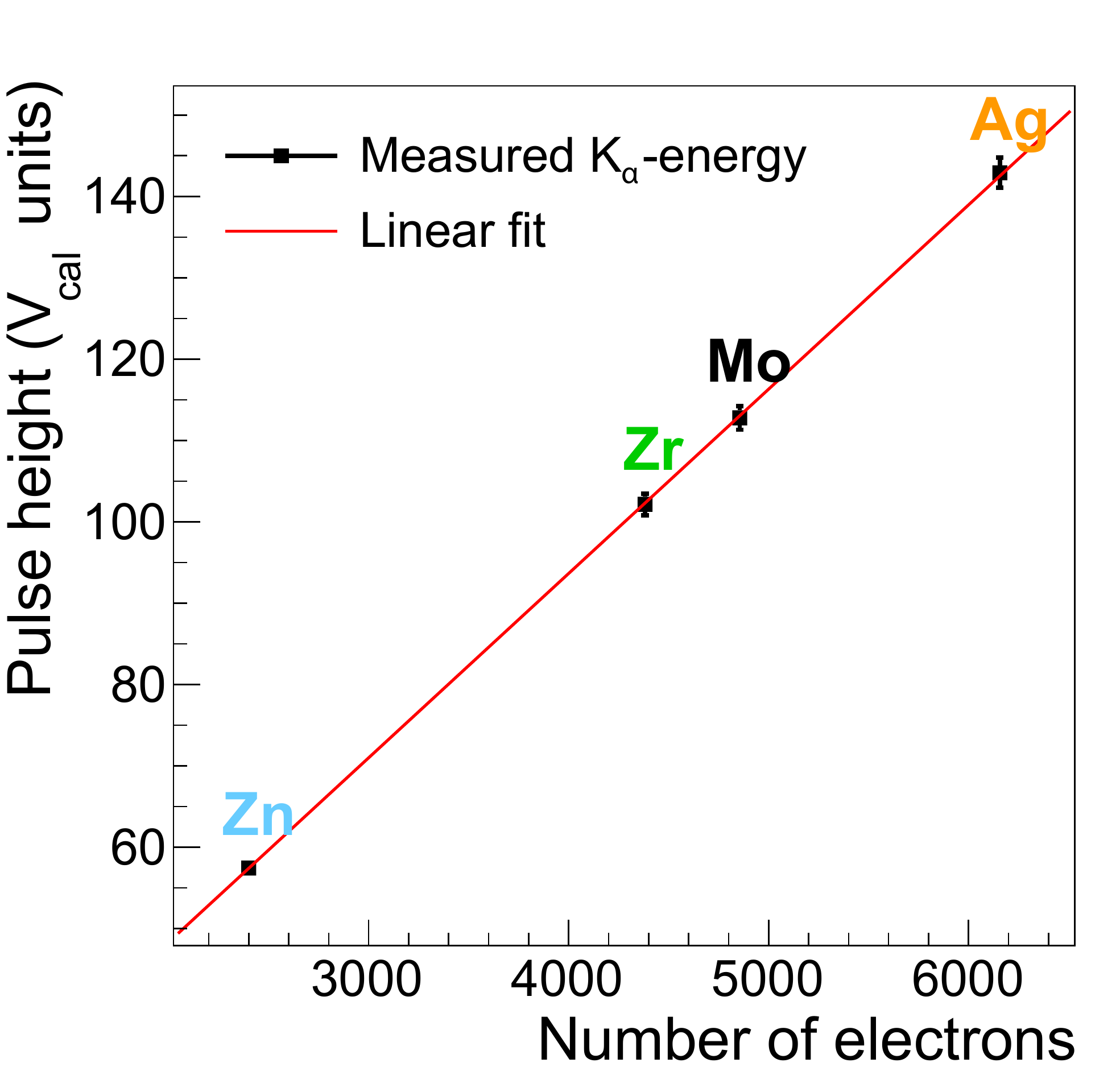}
  \caption{Calibration of one ROC on a pixel detector module with monochromatic X-rays. The measured pulse height is plotted against the expected number of electrons from the K$_{\alpha}$-transitions in the different target materials, as given in Ref.~\cite{r:nist}. The pulse height in units of the internal calibration signal \vcal has a linear dependence on the number of electrons expected to be collected from different target materials. The red line shows a fit with a linear function to the data points.}
  \label{fig:XrayCalibrationChipResults}
\end{figure}

To determine the absolute energy calibration of the module, tests with different X-ray energies were performed. Here, the direct X-rays illuminate different metallic foils (targets), which then produce monochromatic X-rays at a known energy depending on the material of the foil. Four different foil types were used: tin, zinc, molybdenum, and silver, and the pixel pulse height was measured for each target. The most probable value of each pulse height distribution (in units of \vcal, according to the calibration described before) was calculated and plotted versus the expected number of electrons produced by monochromatic X-rays (assuming 3.6\,eV per electron-hole-pair), as illustrated in Fig.~\ref{fig:XrayCalibrationChipResults}. This gives the pulse height calibration in units of electrons. The calibration varies by about 15\% from pixel to pixel within a ROC as well as among ROCs. The average charge calibration for \psidig was measured to have a slope of $47\pm5$ electrons per \vcal unit and an offset of $-60\pm130$ electrons. For the \proc the average calibration was measured on a subset of modules to have a slope of $50\pm3$ and an offset of $-670\pm220$. The quoted uncertainties are the standard deviations of the ROC-to-ROC spread.

All modules were graded once they completed the module testing procedure. The module grades were based on the functionality, X-ray, and IV test results. In order to qualify for detector installation a module had to fulfill the following main selection criteria:

\begin{itemize}
	\item all ROCs and TBMs programmable, valid timing settings found and no decoding errors;
      \item less than 4\% of all pixels defective;
      \item pixel mask mechanism functional for all pixels; 
      \item mean noise per ROC less than 600 electrons and spread of pixel thresholds less than 400 electrons;
      \item no double column with single-pixel hit efficiency less than 95\% at 120\,\mhzcm (600\,\mhzcm for \bpix L1);
      \item sensor leakage current below $10\,\mu$A at 150\,V bias voltage and +17\,\cel.
\end{itemize}

\begin{figure}[tb!]
\centering
  \includegraphics[width=0.47\textwidth]{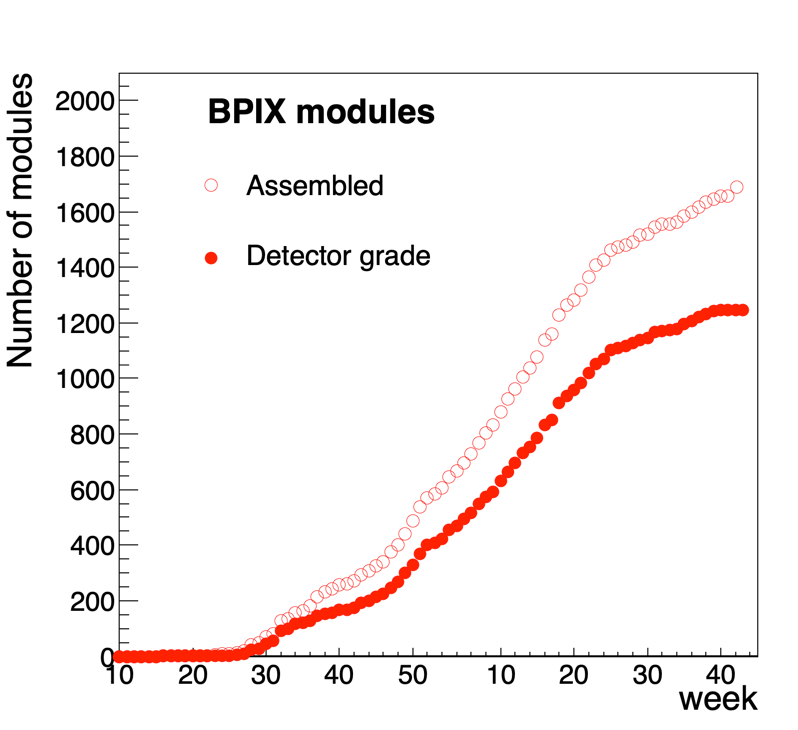}
   \includegraphics[width=0.47\textwidth]{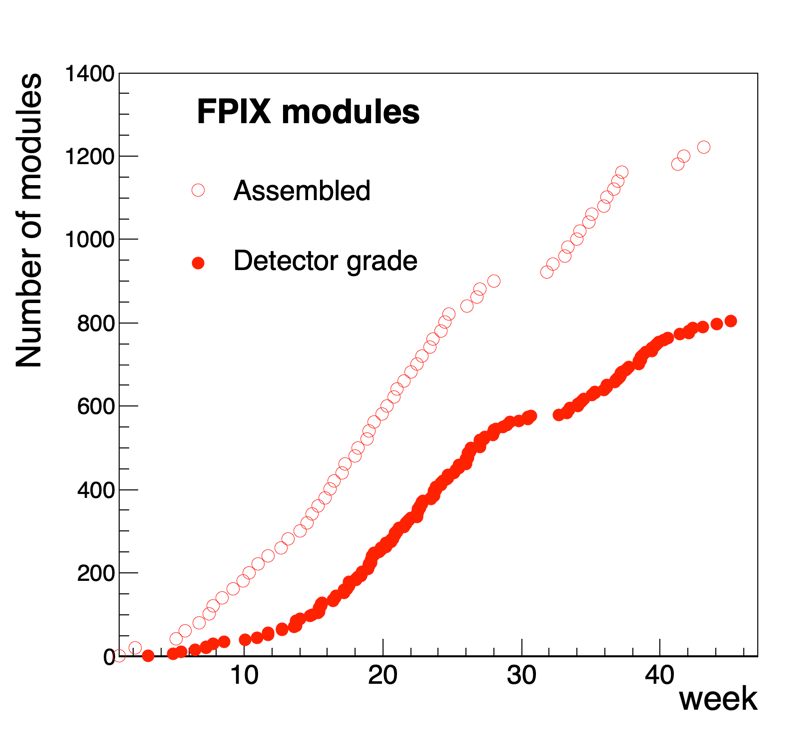}
  \caption{Module production trend versus time for \bpix L2-L4 modules (left) and \fpix modules (right). The L2-L4 \bpix modules were produced from March 2015 to September 2016, while the \fpix modules were produced between January and November 2016. Detector-grade modules pass the selection criteria to qualify for installation in the detector.}
  \label{fig:module_yield}
\end{figure}

In total, 1634 (141) modules were built for \bpix L2-L4 (L1), out of which 1246 (117) were accepted for installation and 1088 (96) were installed. The overall yield of the \bpix module assembly, starting from a good bare module, ranged from 65\% to 85\% in the different production centers. This number includes some low-yield phases during  production start-up. The main causes of module loss were sensor leakage currents that were unacceptably high or failure of individual double columns during high-rate X-ray tests. Other sources of loss were a high number of pixel defects in one ROC, HDI defects, other types of ROC failures, handling mistakes, and various accidents. The number of \bpix L2-L4 modules produced over time is presented in Fig.~\ref{fig:module_yield} (left). The relatively small number (96) of installed L1 modules allowed module construction to start almost a year later than the production for the outer layers, giving more time for the development of the \proc. The L1 modules were built by the Swiss consortium at a time when the production of L2-4 modules had almost finished.

Out of 1223 \fpix modules produced, 816 were accepted for detector installation. 
The module production as a function of time is shown in Fig.~\ref{fig:module_yield} (right). The primary cause for low quality modules during production was the damage that occurred during the dicing of the ROC wafers described earlier. 
Furthermore, the tests identified an issue that occurred during the sintering of the sensor wafers by the manufacturer. This caused charge traps to form between the bulk silicon and the surface $\textrm{SiO}_{2}$ layer. 
The issue was mitigated by increasing the sensor bias voltage for operation to 300\,V. 

\section{Mechanics}
\label{s:mechanics}



In both the \bpix and \fpix detectors, the sensor modules are mounted on light-weight mechanical structures with thin-walled stainless steel tubes used for the \coo cooling. Since the modules are not in direct contact with the cooling loops, carbon fiber and graphite materials with high thermal conductivity are used in the detector mechanical structures. The most stringent requirement regarding the thermal performance of the detector mechanics comes from keeping a safe margin to thermal runaway of the sensor power in the innermost layer. A thermal resistance of 1.8\,K/W has been achieved between the coolant and the sensor modules in the innermost layer, which is well below the requirement of 2.5\,K/W obtained from simulations assuming a power dissipation per module of 6\,W as expected after collecting an integrated luminosity of 300\,\fbinv~\cite{maxrauchmaster}. 

The \bpix and \fpix detectors are each connected to four service half-cylinders, which host the auxiliary electronics for readout and powering. The auxiliary electronics on the service half-cylinders act as pre-heaters that stimulate the gas phase of the \coo cooling. In this section the main design parameters and the construction of the detector mechanics, cooling loops and service half-cylinders are presented. 

\subsection{\bpix mechanics}
\label{s:bpixmech}

\begin{figure}[tb!]
  \centering
  	 \includegraphics[trim=50 100 50 20,clip,width=0.4\textwidth]{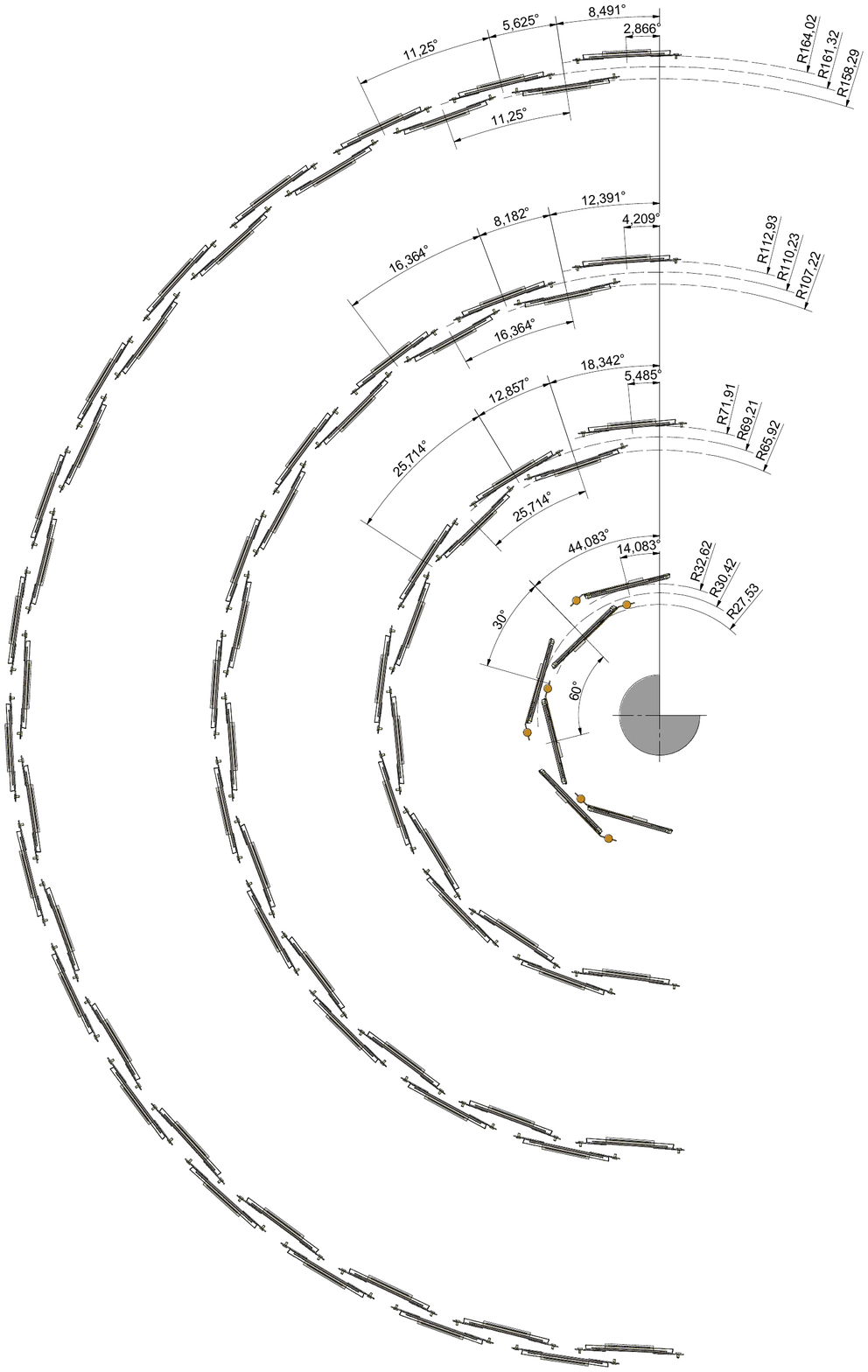}
  	 \caption{Drawing of one \bpix detector half in the $r\phi$ view, showing the arrangement of ladders in the four detector layers.}
    \label{fig:ladderarrangement}
\end{figure}

The \bpix modules are mounted on four concentric, cylindrical layers each formed by an alternating arrangement of ladders at smaller and larger radii, as shown in Fig.~\ref{fig:ladderarrangement}. Each ladder supports eight detector modules. Modules are mounted on the inward and outward facing sides of the inner and outer ladders, respectively. The ladder arrangement provides between 0.5\,mm and 1.0\,mm of overlap in the $r\phi$ direction in the active area of the sensors. The cylindrical layers are divided into half-shells in the longitudinal direction, such that overlap is provided in the boundary region. This is achieved by radially displacing the ladders of one half-shell in the overlap region. In the innermost layer the radial positions of the modules are 27.5, 30.4, and 32.6\,mm, leading to non-negligible differences in hit rates and radiation levels (up to 20\%). 

\begin{figure}[tb!]
  \centering
  	 \includegraphics[trim=50 20 50 20,clip,width=0.6\textwidth]{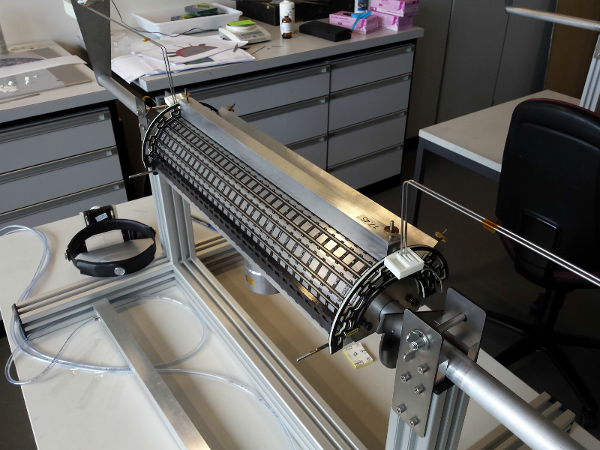}
  	 \caption{Picture of the \bpix L2 mechanics during production. }
    \label{fig:halfshell}
\end{figure}

The ladders have a length of 540\,mm and a thickness of 500\,\mum and are made from 8-ply unidirectional carbon-fiber layups made from K13D2U pre-pregs. The ladders are suspended between end rings. The end rings for L1, L2, and L3 consist of a CFK/Airex/CFK sandwich structure, while the end rings for L4 are entirely built from CFK to provide mechanical stability. The detector mechanics consists of 148 ladders with 25 variants in shape. Each ladder is custom made and machined from a CFK sheet using water jet cutting. Nuts are glued into the ladders in order to fix the detector modules by screws. The ladders are glued from both sides onto the cooling loops, as shown in Fig.~\ref{fig:halfshell}, using a rotatable jig. Each \bpix half-shell underwent thermal cycling between $+25$\,\cel and $-18$\,\cel, to make sure that there is no delamination of the glued structures. 

\begin{figure}[tb!]
  \centering
    \includegraphics[trim=0 0 0 0,clip,width=0.7\textwidth]{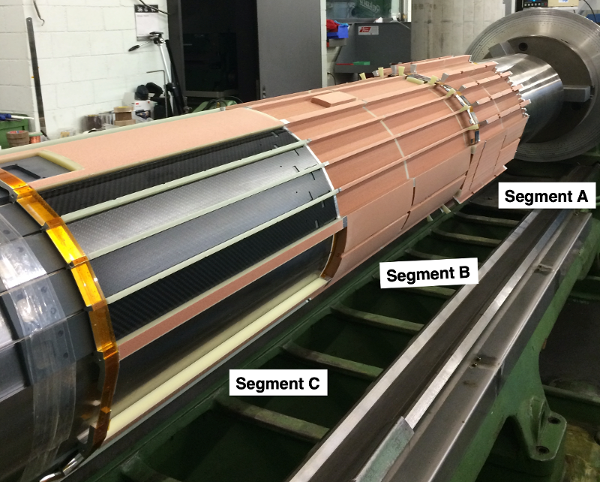}
    \caption{Picture of a \bpix service half-cylinder during construction. The service half-cylinder is placed on a rotatable mandrel. Segments A and B are made from Airex foam, while Segment C is made from carbon fiber plates.}
    \label{bpixstmandrel}
\end{figure}

The \bpix service half-cylinders are composed of three segments, labeled A, B, and C in Fig.~\ref{bpixstmandrel}, which shows one of the four \bpix service half-cylinders during construction. Segment A is farthest from the interaction point and houses the \dcdc converters. The optoelectronics components are placed in Segment B, while Segment C provides the space for the module connections (Fig.~\ref{fig:bpixst}). Segments A, B, and C form a mechanical unit with a length of 1.7\,m. The inner radius of the \bpix service half-cylinder is 175\,mm, while the outer radius is 190\,mm up until Segment B and increases to 215\,mm in Segment A. The radii are chosen within the boundary conditions of ensuring a clearance of at least 3\,mm during installation, while providing sufficient space to route the services to and from the detector. An additional mechanical support structure (Segment D) is placed in between the \bpix detector mechanics and the service half-cylinder to guide the module cables along~$z$. The \bpix module cables are micro-twisted-pair copper cables with a length of about 1\,m. Segment D is a separate mechanical unit made from a CFK envelope with a length of 470\,mm, an inner radius of 180\,mm and an outer radius of 200\,mm. 

The \bpix service half-cylinders were constructed on a rotatable mandrel. First, the inner shielding foil was placed on the mandrel. The shielding is made from an aluminum foil with a thin layer of high-pressure fiberglass laminate for electrical isolation. In the next step, four layers of Airex foam, which had been baked into a cyclindrical form beforehand, were glued together. A computer-controlled mill was used to trench the channels for the electronics boards into the foam. Threaded pins were glued to the Airex foam to fix the stack of electronics boards within the channels. Aluminum ribs were placed in between the segments to ensure mechanical rigidity. To further increase the stiffness, fiberglass ribs were placed within the segments. In the final step, carbon fiber plates were added to form Segment C. 
An outer shielding foil was screwed to the \bpix service half-cylinders after the placement and testing of the components of the detector readout, power, and cooling system, described in Sec.~\ref{s:integration}. All electronics components within a service half-cylinder are connected to a common ground at the aluminum flange between Segments B and C. This ground is connected with a single wire per service half-cylinder to the end flange, where it is clamped to the central CMS ground. The inner und outer shields of the \bpix service half-cylinders are also connected to this ground. Furthermore, shielding foils are mounted on the inside and the outside of the \bpix detector and connected with single wires to the shields of the service half-cylinders.

\begin{figure}[tb!]
  \centering
  	 \includegraphics[trim=0 -100 0 0,clip,width=0.7\textwidth]{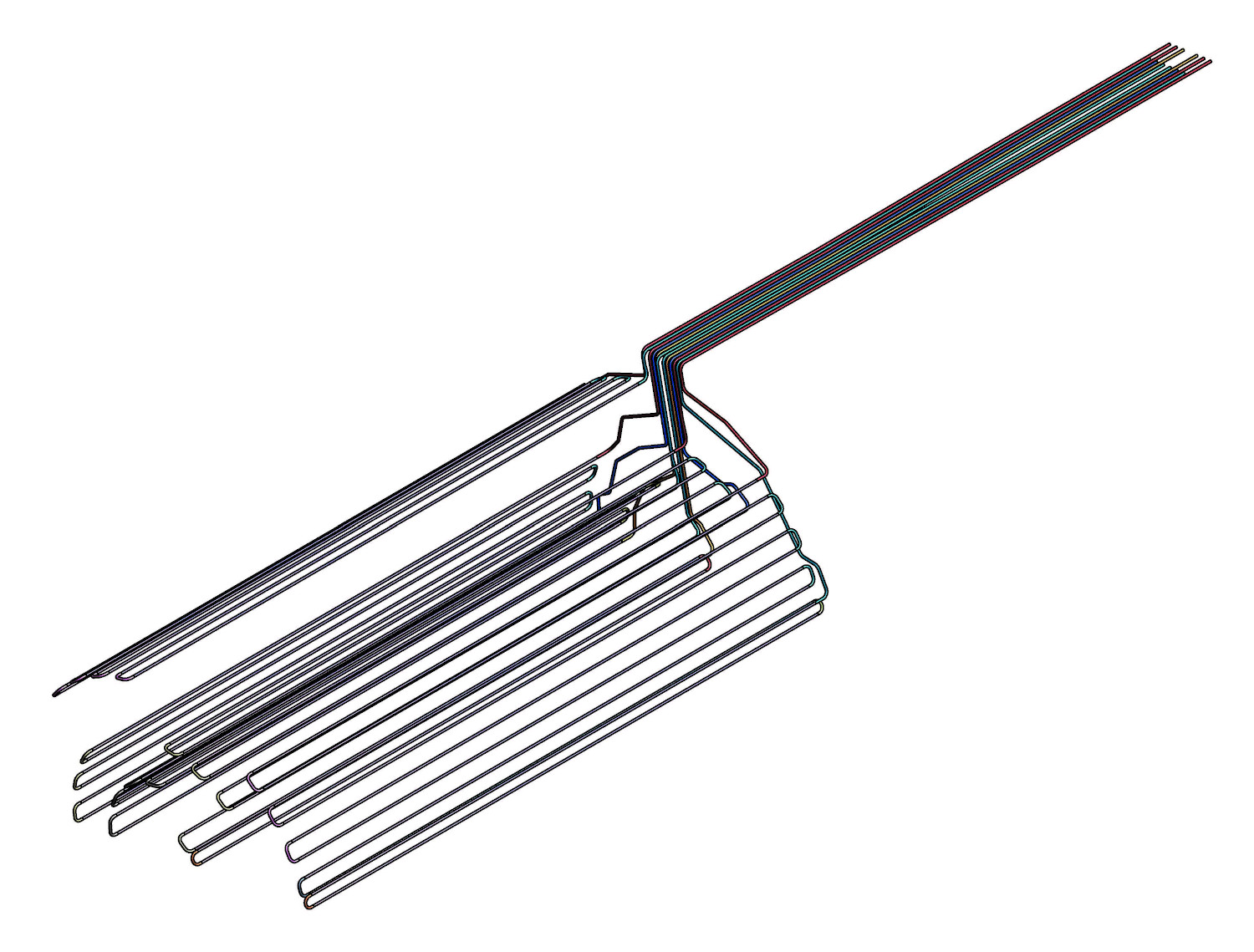}
	  \includegraphics[trim=0 0 0 0,clip,width=0.8\textwidth]{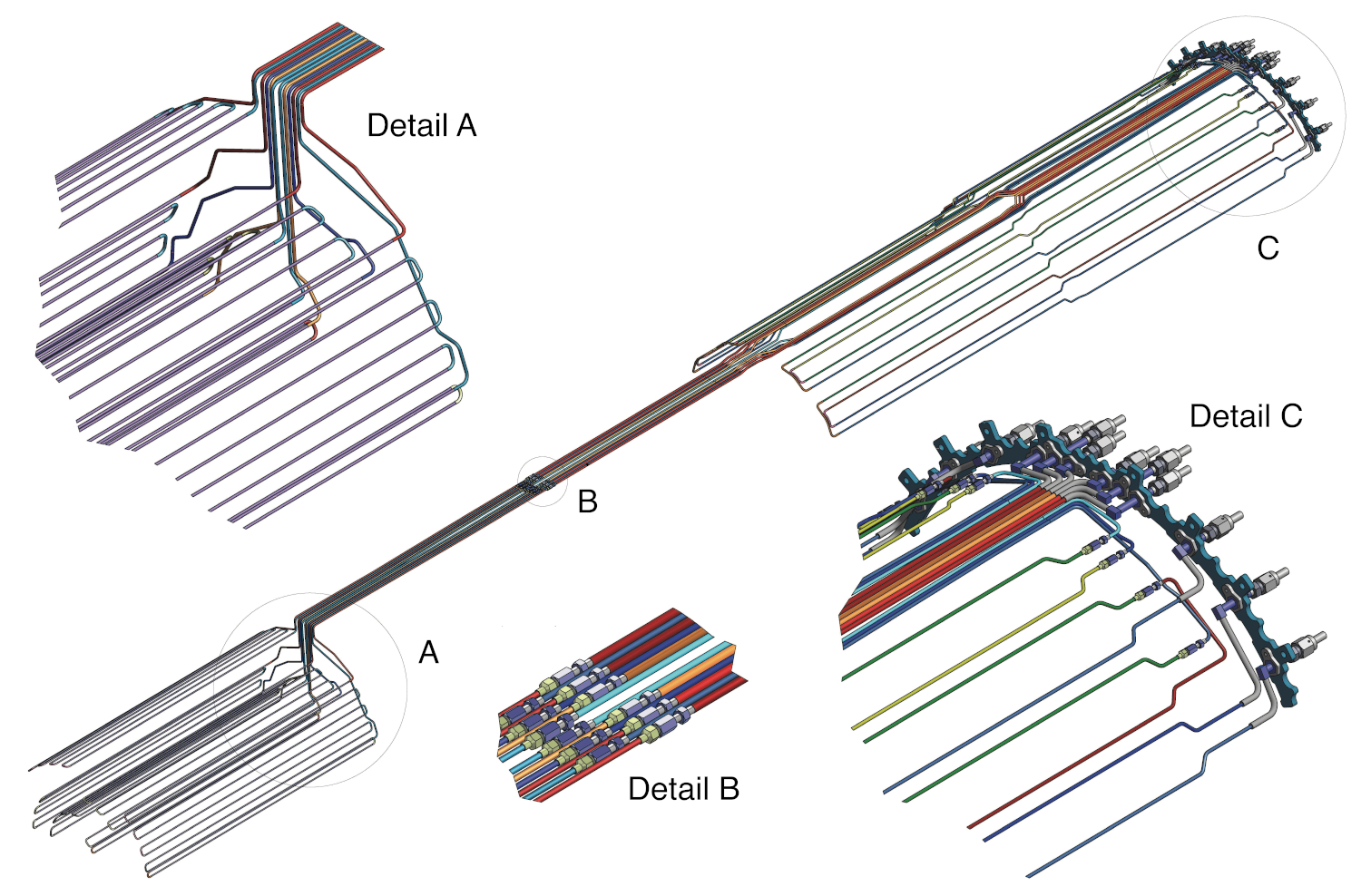} 
  	 \caption{Drawing of one quarter of the cooling loop system for the \bpix detector (top) and the service half-cylinder (bottom).}
    \label{fig:bpixcoolingmech}
\end{figure}

The \bpix cooling lines are constructed in a complex looping structure in order to cool the components on the service half-cylinders as well as the detector modules. The cooling loops have lengths between 957\,cm and 1225\,cm, depending on their position, and are designed to dissipate a power of up to 240\,W at a temperature of $-20$\,\cel, which is the power expected when running an irradiated detector at the highest instantaneous luminosities. In the active region of the \bpix detector, stainless steel tubes with an inner diameter of 1.7\,mm and a wall thickness of 50\,\mum are used. The cooling line segments of the detector are joined with those of the service half-cylinder using custom-made mini fittings, provided with a thread, and a copper ring for sealing. The cooling tubes on the service half-cylinder have a wall thickness of 200\,\mum and inner diameters of 1.8\,mm and 2.6\,mm for supply and return lines, respectively. The inlet and outlet of each cooling loop are located at the same end of the detector and the number of cooling loops is distributed evenly on the $+z$ and $-z$ end of the detector. Swagelok VCR connections are used at the end flange of the \bpix service half-cylinder to connect the cooling loops to capillaries, which in turn connect to the transfer lines from the \coo cooling plant. A total of 24 cooling loops are used in the \bpix system (four cooling loops for L1 and L2 each, and eight cooling loops for L3 and L4 each). A drawing of one quarter of the \bpix cooling loop system separated in detector and service half-cylinder parts is presented in Fig.~\ref{fig:bpixcoolingmech}. 

The cooling tubes, the mini fittings, as well as the adapter to the VCR fittings have all been produced using the same stainless steel quality (DIN 1.4441). The cooling tubes were bent in an industrial CNC bending machine, allowing for a minimal bending radius of 5\,mm. The tubes were cut to the correct lengths by electrical discharge machining. An industrial computer-controlled laser welding process was used to join the cooling tubes, mini fittings, and adapters. A total of 386 welds were needed in the detector part of the cooling system, another 288 welds in the service half-cylinder part. Because of the tight tolerances needed in the laser welding process, using a sleeve, high-precision tube alignment is achieved. A further benefit of laser welding is that the joints can be made without adding any additional material and with local heating only. 

The quality of welded samples of cooling tubes was assessed based on X-ray inspection. In order to check the quality of the laser welding process during production, all loops underwent a visual inspection assessing the geometry of the completed component and the contour of the welding seam. Furthermore, all individual components as well as the completed loops were pressure-tested (using nitrogen at 200\,bar) immediately after the laser welding.  

In addition to the quality assurance during production, a long term test was carried out using a test loop with shorter length but including all components of a complete \bpix cooling loop (detector and service half-cylinder parts with mini fittings and VCR connections). The test loop was tested under pressure (using helium at 200\,bar) for several months and no issues were observed. 

A summary of the weights of the components of the \bpix detector is given in Tab.~\ref{t:bpixweights}.

\begin{table}
  \begin{center}
    \caption{Summary of weights of the components of the \bpix detector.}
    \label{t:bpixweights}
   \vspace{0.1in}
   \begin{tabular}{l | c }
     \hline
     Component & Weight [g] \\
      \hline
      \hline
      Silicon sensor modules (1184) &  2700\\ 
      Detector mechanical structure (with cooling loops) &  2500\\ 
      Service cylinder mechanical structure (4) & 11600\\
      Service cylinder electronics and \dcdc converters & 32300 \\
      Service cylinder cooling loops & 2800 \\
      \hline
      \hline
   \end{tabular}
  \end{center}
\end{table}

\subsection{\fpix mechanics}

\begin{figure}[tb!]
  \centering
  	 \includegraphics[trim=0 0 0 0,clip,width=0.3\textwidth]{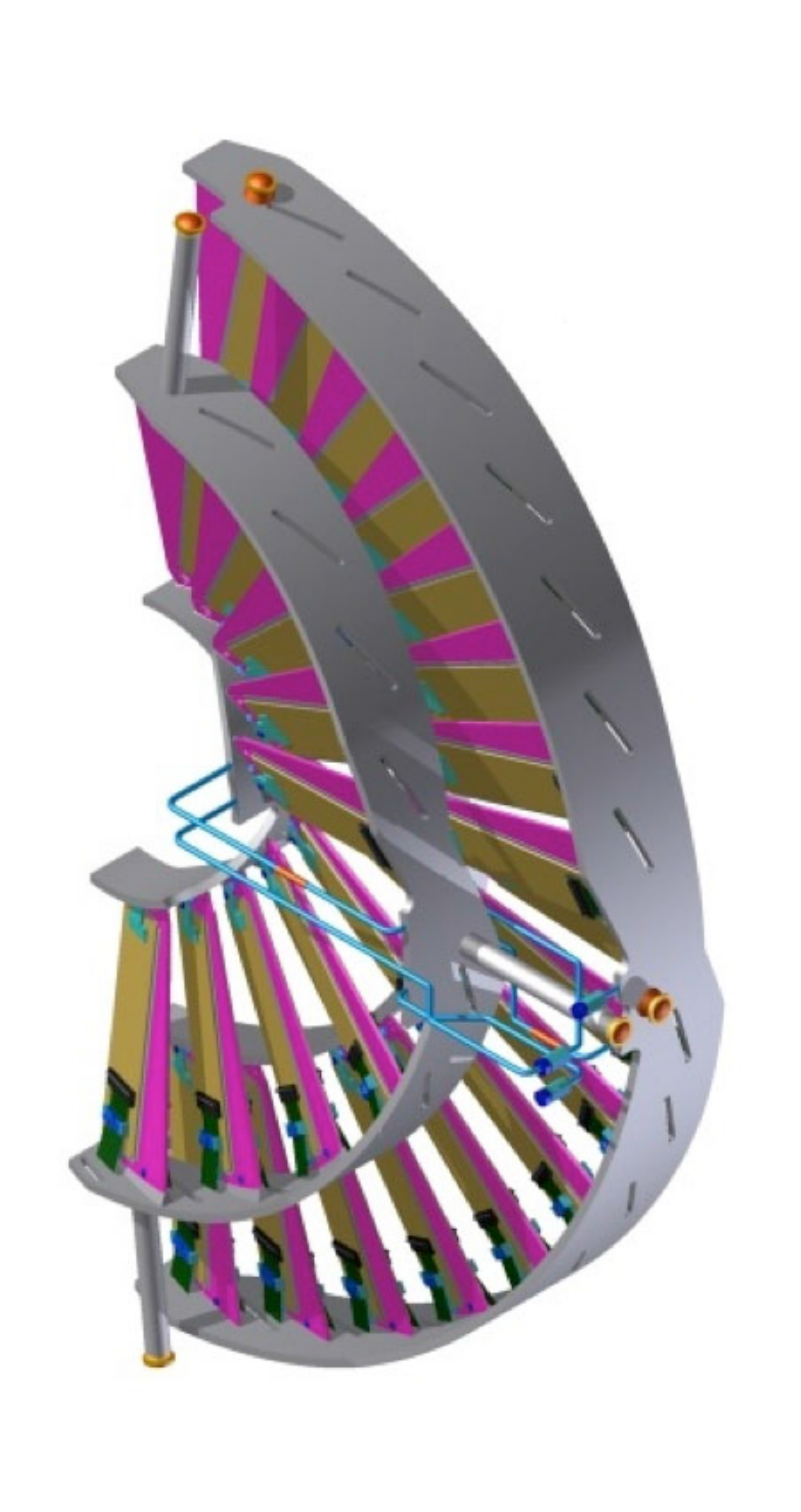}
   	 \includegraphics[trim=0 0 0 0,clip,width=0.4\textwidth]{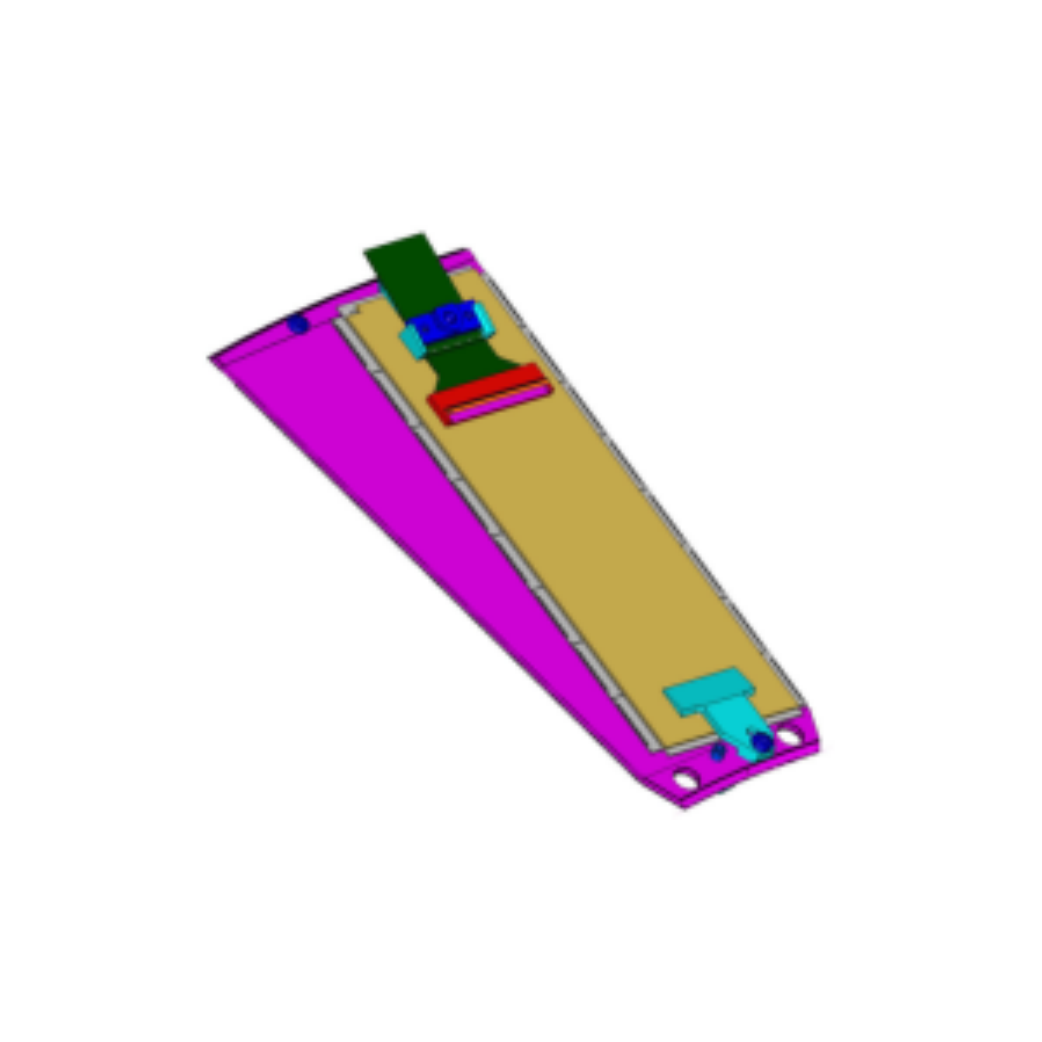}
  	 \caption{Drawing of an \fpix half-disk made from two half-rings of modules mounted on blades that are suspended between graphite rings (left), and a close up of a module mounted on a blade (right). }
    \label{fig:disk}
\end{figure}

The three half-disks forming one \fpix quadrant are supported by a carbon-fiber composite service half-cylinder. The \fpix half-disks consist of two turbine-like mechanical support structures with an inner assembly providing a sensor coverage from radii of 45\,mm to 110\,mm with 11 blades, while the outer assembly covers radii from 96\,mm to 161\,mm with 17 blades (Fig.~\ref{fig:disk}). The half-disks serve as the cooling isotherms for the sensor modules. One module is mounted on each side of the blades with a small overlap in coverage at the outer edge of the blade with adjacent modules and a larger overlap closer to the beam. The flat panel blades are made of 0.6\,mm thick sheets of thermal pyrolytic graphite (TPG) encapsulated between two 70\,\mum thick single-ply carbon fiber face sheets. 
The blades are suspended between an inner and an outer graphite ring. The graphite rings are 2.4\,mm thick and reinforced on the side facing away from the blades with a 6-ply carbon fiber skin.  The TPG/carbon fiber blades are glued into slots machined into the graphite rings; the slots are filled with TC5022 thermal compound 
before blade insertion, and the blade-ring joint is sealed with an epoxy fillet. 

\begin{figure}[bt!]
  \centering
   \includegraphics[trim=0 0 0 0,clip,height=7cm]{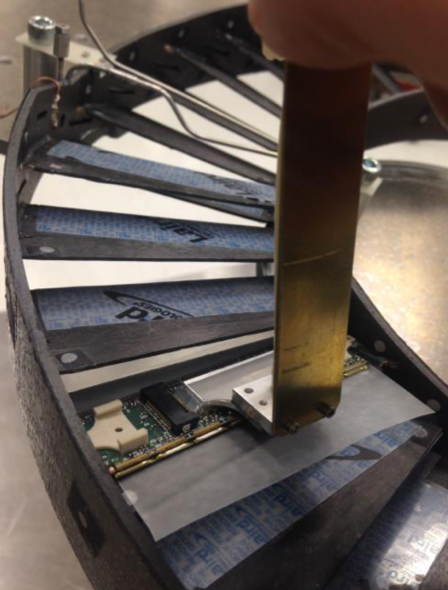}
    \includegraphics[trim=0 0 0 0,clip,height=7cm]{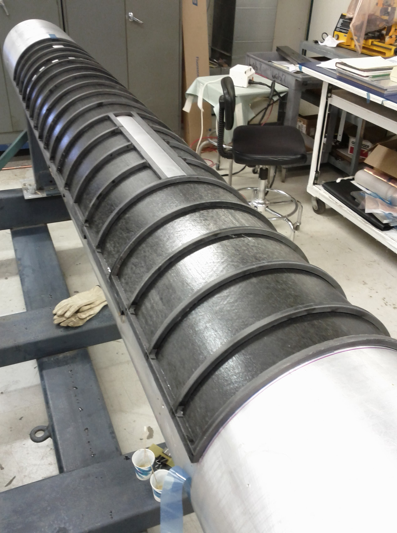}
    \caption{ (Left) \fpix module placement on a blade of a half-ring using a vacuum tool. (Right) \fpix service half-cylinder during construction. The rear section of the service half-cylinder is in the foreground of the picture and the 8\,\mm square profile ribs are clearly visible.}
    \label{fig:cyl}
\end{figure}

The graphite rings contain machined U-shaped channels for the stainless steel cooling tubes of 1.45\,mm inner diameter and 100\,\mum wall thickness. The cooling loops are embedded using TC5022 thermal compound and held by the carbon fiber skins that are glued to the outer radial surface of the outer graphite ring and the inner radial surface of the inner graphite ring. There are two cooling loops for each \fpix half-disk, one serving the inner, the other serving the outer ring assembly. The sensor modules are secured to the blades with two screws threaded into inserts glued into the blade, and full-face glued with Laird TPCM 583 
phase change film (Fig.~\ref{fig:cyl}, left). 

Given that the modules have full face contact to the blade, and that both the TPG of the blade and the graphite of the rings have high thermal conductivity ($1000$~and~$100\, \rm W/(mK)$, respectively), two critical joints dominate the thermal path resistance: a) the small cross section transition of the blade to the graphite ring, and b) the embedding of the stainless steel tubing into the grooves machined into the graphite. 

\begin{figure}[tb!]
  \centering
   	 \includegraphics[trim=0 0 0 0,clip,width=0.6\textwidth]{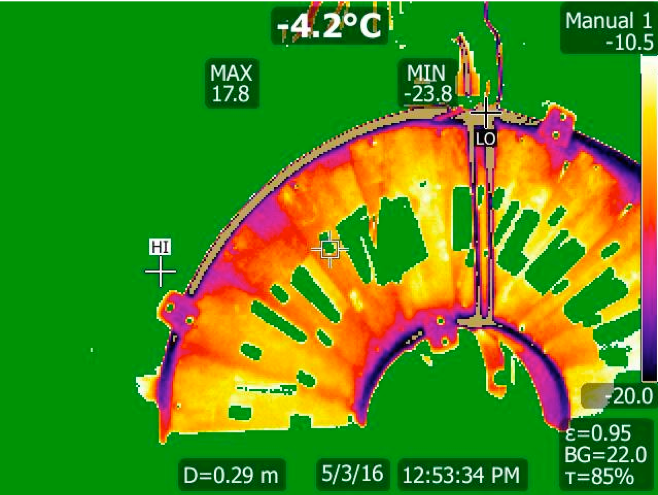}
	\caption{ Thermal image of an \fpix disk during quality control. An infrared camera observed the disk through a porthole in the test enclosure and was programmed such that the allowed temperature range, between $-20.0$\,\cel and $-10.5$\,\cel, was rendered in the color spectrum from purple to white, while temperatures outside of the allowed range were rendered green. The blade surfaces of the disk, except for the areas where the heaters and co-cured reflective pads are located (corresponding to the green large and small rectangles), appear to be within the acceptable range.}
    \label{fig:diskthermal}
\end{figure}

During construction, all half-disks were qualified using a thermal imager, dummy heaters, and a \coo cooling plant at FNAL (Fig.~\ref{fig:diskthermal}). For the thermal acceptance test, before modules were mounted, each disk was cooled with two-phase \coo in a test enclosure. Ohmic heaters were clamped to each blade, providing a power load of 3\,W per module, equivalent to the full end-of-life detector power load. A disk structure was accepted if all surface temperatures were within 10\,\cel of the coolant temperature, except for temperatures measured on the clamp-on heaters themselves and on reflective pads co-cured to the surface of the blades. No production disk structures were 
rejected due to thermal considerations.

The \fpix service half-cylinders are 2220\,mm long, with an outer radius of 175\,mm and an inner radius of 165\,mm. The service half-cylinders consist of a double-walled rear section that is 1500\,mm long, and a corrugated single-wall front section. 
The rear skins are 12-ply carbon fiber layups made from K13C2U pre-pregs, cured at 1\,bar 
on a curved mold. The skin spacing is set by hollow carbon fiber ribs of 8\,mm square profile. The ribs are fabricated from spiral wound M46J pre-pregs 
on round silicone tubing that is cured in a square cross section steel mold with the tubing pressurized to 5\,bar. The inner silicone mold is pulled from the finished rib.
The skins and ribs, including the solid carbon fiber reinforcement ring at the transition to the corrugated front section, are assembled with room temperature cure epoxy on a cylindrical mold with added precise alignment features locating the edges of the service half-cylinder (Fig.~\ref{fig:cyl}, right). The corrugated front section consists of a single 14-ply carbon fiber layup, made from axial K13C2U and $90^\circ$ M46J pre-pregs on a corrugated steel mold, a short section of 12-ply reinforcement skin, and a solid carbon fiber reinforcement ring at the very front, both made from K13C2U. The front section supports the half-disks via kinematic mounts, with a ruby ball on each of three disk support stalks sliding in aluminum bushings. The bushings are glued into the half-cylinder skin at approximately $90^\circ$ spacing around the circumference.

The mechanical design for the \fpix detector was optimized for maximum coverage of the sensor elements in the given radial space, motivating the single-shell front section. Finite element analysis showed that solid carbon fiber support hoops are needed to stiffen the corrugated single shell in order to achieve satisfactory stiffness. 


The end flanges of the service half-cylinders are machined from aluminum, with thermally insulating inserts for the cooling tubes and cover plates for the optical fiber channels machined from polyether ether ketone (PEEK). The end flange is glued to the half-cylinder on a coordinate measuring machine to control the alignment between the half-cylinder support feet in the corrugated section and the end flange.

Stainless steel lines supplying \coo coolant to the half-disks are routed along the inner surface of the \fpix service half-cylinder. The thin-walled 316L stainless steel coolant supply (2.2\,mm inner diameter and 130\,\mum wall thickness) and return (2.7\,mm inner diameter and 180\,\mum wall thickness) tubes connect to the half-disks with custom-designed metal seal couplers. The coupler glands, that are laser welded to the tubing, are machined from low impurity VIMVAR 304L steel. Tubing and coupler steel are selected for a ferrite content of the weld pool of around 4\%, which, together with the low impurity content, protects against solidification cracking~\cite{ref:Stephanie}. Extensive pressure testing, along with temperature cycling, was done after the laser welding before the pipes were integrated into the detector structures. Pressure tests were also performed after several assembly steps and prior to installation certification.


 \dcdc converters, control and readout electronics are located inside the service half-\linebreak cylinder, connected to the sensor modules with aluminum flex cables. The electronic boards are attached to the service half-cylinder via inserts glued into the double wall structure, with the \dcdc converters and the opto-hybrids mounted on cooled aluminum bridges clamped to the \coo supply and return lines. Ground connections from the electronics boards to the service half-cylinder are provided by gold-plated contact pads on copper mesh polyimide flex circuits co-cured with the inner half-cylinder carbon fiber skin~\cite{ref:co-cure}. The end flange is electrically connected via short jumpers to contact pads on the co-cured grounding mesh of the inner half-cylinder wall. The CMS central ground wire is connected at the end flange.

A summary of the weights of the components of the \fpix detector is given in Tab.~\ref{t:fpixweights}.

\begin{table}
  \begin{center}
    \caption{Summary of weights of the components of the \fpix detector.}
    \label{t:fpixweights}
   \vspace{0.1in}
   \begin{tabular}{l | c }
     \hline
     Component & Weight [g] \\
      \hline
      \hline
      Silicon sensor modules (672) &  1940\\ 
      Half-disk mechanical structure &  3800\\ 
      Service cylinder mechanical structure (4) & 14000\\
      Service cylinder electronics and \dcdc converters & 28800 \\
      Service cylinder cooling loops & 5200 \\
      \hline
      \hline
   \end{tabular}
  \end{center}
\end{table}

\section{Readout architecture and data acquisition system}
\label{s:services}



\begin{figure}[tb!]
  \centering
    \includegraphics[trim=0 0 0 0,clip,width=0.8\textwidth]{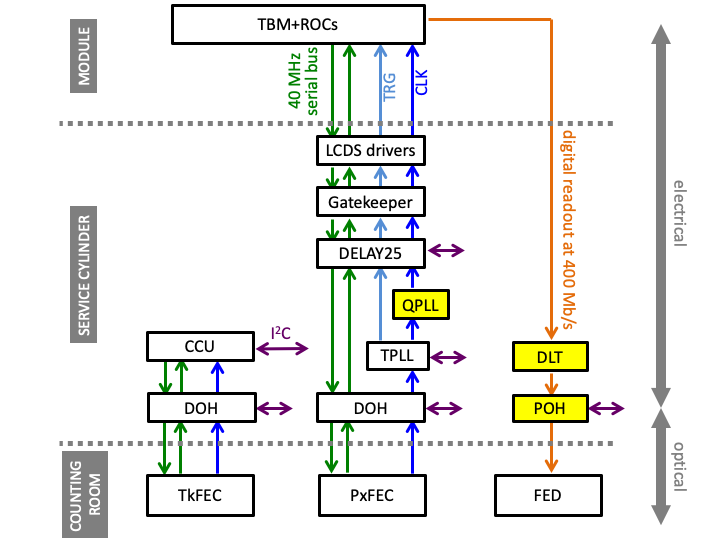}
    \caption{Overview of the readout and control system of the \cmsph. The components in the control and readout chain that have been specifically designed for the \cmsph are highlighted in yellow. All other electronics components on the service half-cylinders are identical to those used in the original pixel detector.}
    \label{fig:pixeldaq}
\end{figure}

An overview of the \cmsph readout architecture is given in Fig.~\ref{fig:pixeldaq}. The \cmsph is organized into 48 independent readout groups (known as sectors), serving up to 39 modules each. The communication between the detector and the DAQ back-end electronics boards in the underground counting room is done via optical links. A sector includes digital opto-hybrids (DOHs) as well as auxiliary chips (TPLL~\cite{ref:tpll}, QPLL~\cite{ref:qpll}, DELAY25~\cite{ref:delay25}, Gatekeeper~\cite{ref:gatekeeper}, LCDS drivers~\cite{ref:naegeli}) for the transmission of control, clock, and trigger signals. Except for the QPLL chips, which were added to reduce the jitter on the clock signal, all components used in the control and readout chain are identical to those used in the original detector system. A 40\,MHz serial bus is implemented for the control, programming, and readback of the TBM. The change, with respect to the previous system, from 40\,MHz encoded analog to 400\,\mbs digital data in the readout of the TBM required the adoption of new optical converters, called pixel opto-hybrids (POHs)~\cite{Troska:2012zz}. POHs are built from four or seven transmitter optical subassemblies (TOSAs)~\cite{ref:tosa}, linear laser-driver (LLD)~\cite{ref:lld}, and digital level-translator (DLT) chips~\cite{ref:naegeli}, and have been designed specifically for use in the \cmsph. 

The electronics for the readout and control system are integrated on the detector service half-cylinders. The slow control links are implemented as a ring architecture with communication-and-control unit ASICs (CCUs)~\cite{CCU}. Two \itwoc channels of each CCU are used to program the readout electronics on the service half-cylinders and up to 16 parallel interface adapter (PIA) channels are used to enable/disable the \dcdc converters and to generate reset signals for the readout electronics on the service half-cylinders and the detector modules.  

The off-detector VME-based DAQ system used for the original detector has been replaced by a microTCA-based system~\cite{r:mtca} with high-speed signal links for providing data rates up to 10\,\gbs for the transfer of the data from each pixel detector DAQ module to the CMS central DAQ system. The FC7 carrier board~\cite{Pesaresi:2015qfa, Auzinger:2016niy} was selected as the platform for the DAQ modules of the \cmsph. The DAQ system to control and read out the full pixel detector consists of 108 front-end driver modules (FEDs), which receive and decode the pixel hit information; three front-end controller modules (TkFECs), which serve the detector slow control; 16 pixel front-end controller modules (PxFECs) used for module programming and clock and trigger distribution; and 12 AMC13 cards~\cite{Hazen:2013rma} providing the clock and trigger signals. Application-specific mezzanine cards and firmware make an FC7 carrier board into a FED or FEC. A detailed description of the \cmsph DAQ system can be found in Ref.~\cite{ref:bora}.

The 400\,\mbs data streams from the detector modules are transmitted electrically to the POHs along the module cables. On the POH the signals received from the TBM are converted by the DLT to levels compatible with the input of the LLD.  The LLD drives the TOSAs by biasing the lasers at their working point and modulating them with a current proportional to the input signal. The modulation gain and bias current are configurable through an \itwoc interface. The POH transmits DC-balanced digital signals via optical fibers to the FEDs. The optical fibers connected to the lasers on the POHs are grouped in bundles of 12 fibers. A FED has two connectors and thus serves 24 links.



\begin{figure}[bt!]
  \centering
    \includegraphics[trim=0 0 0 0,clip,width=0.9\textwidth]{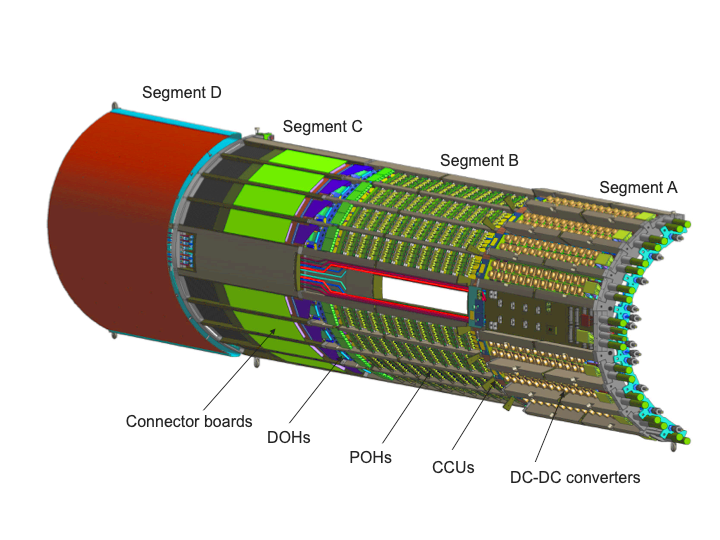}
    \caption{Drawing of a \bpix service half-cylinder.}
    \label{fig:bpixst}
\end{figure}

Figure~\ref{fig:bpixst} shows a drawing of one of the four \bpix service half-cylinders. The electronics components are organized in eight slots on each service half-cylinder. A slot consists of a stack of printed circuit boards (PCBs), POHs, DOHs (Segments B+C), and \dcdc converters (Segment A). One rigid-flex board with eight CCUs (plus one for redundancy) spans each service half-cylinder azimuthally, with one CCU serving one slot. For monitoring purposes, the \bpix service half-cylinders are equipped with a total of 132 temperature sensors and four humidity sensors, placed on the PCBs and cooling lines. 

The micro-twisted-pair cables coming from the \bpix detector modules are connected to the service half-cylinders using dedicated PCBs (called connector boards), which are located in Segment C. There is a stack of three connector boards in each slot, serving modules in L1-2, L3, and L4, respectively. The module readout signals are routed from the connector boards via adapter boards and polyimide flex cables to the motherboard on which the POHs are mounted (Segment B). Each slot holds a POH motherboard with up to 14 POHs. Optical fibers are connected to the TOSAs on the POHs and run to the end flange of the service half-cylinders. 

Boards with two DOHs per slot transmit the control signals between the PxFECs and the detector modules. One control link (one DOH) serves L1 and L2, the other one serves L3 and L4. The electrical signals from the DOHs are passed through the polyimide flex cables and adapter boards to the connector boards. An LCDS driver chip mounted next to each module connector on the connector board drives the control signals via the micro-twisted-pair cables to the modules. The LCDS driver chip features delays that can be adjusted in hardware to compensate the differences in signal propagation time occurring because of different cable lengths. 

\begin{figure}[tb!]
  \centering
    \includegraphics[trim=0 0 0 0,clip,width=0.9\textwidth]{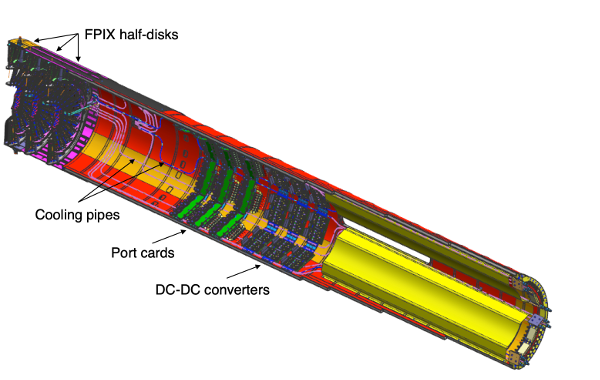}
    \caption{Drawing of an \fpix service half-cylinder and three half-disks.}
    \label{fig:fpixst}
\end{figure}

In the \fpix detector, most of these electronics is implemented in a single card, known as port card. A port card is connected to seven \fpix modules via aluminum flex cables. Each port card hosts one POH with seven TOSAs, one DOH, as well as the auxiliary chips (TPLL, QPLL, DELAY25, Gatekeeper, LCDS drivers). In total there are 24 port cards per service half-cylinder. In addition, there is a CCU board with four CCUs (plus one for redundancy) per service half-cylinder. A total of 116 temperature sensors and four humidity sensors are installed on the \fpix service half-cylinders. Furthermore, there is an auxiliary board using a CCU channel and dedicated ASICs to read out additional temperature sensors on the cooling lines and \dcdc converters. A drawing of the \fpix service half-cylinder with readout electronics, \dcdc converters and cooling loops is shown in Fig.~\ref{fig:fpixst}.

In order to test the performance of the \cmsph readout system and gain experience with its operations, test stands were set up at University of Zurich for the \bpix detector and at FNAL and CERN for the \fpix detector. The setups included a slice of the full CMS pixel detector DAQ, together with all components of the upgrade power system and control and readout chain, as well as a number of detector modules. A comprehensive testing and measurement program was carried out which demonstrated the performance of the readout system in view of operation in CMS~\cite{ref:bora}.

\section{Power system}
\label{s:power}

The power system involves the distribution of the low and high voltages to the pixel modules, as well as the distribution of the supply voltage to the components on the service cylinders. The distribution of the high voltage and the supply voltage to the service cylinders are conceptually unchanged with respect to the original detector. 

For the low voltage distribution, the original pixel detector featured a direct, parallel powering scheme, in which several pixel detector modules were powered directly from one power supply channel. The power supplies are located on balconies in the experimental cavern and are connected to the detector via cables of about 50\,m length. The power system for the \cmsph had to respect two important boundary conditions: firstly, the cable plant had to be re-used, as access is difficult and in any case additional cables would not have fit into the existing cable channels; and secondly, the power supply system, including firmware and software, was to be changed as little as possible.

The \cmsph comprises 1856 pixel detector modules with 16 ROCs each, almost doubling the number of ROCs with respect to the original pixel detector. Since the operating voltages of the ROCs stayed the same, the supply currents doubled, leading, in a direct powering scheme, to large voltage drops along the cables. Calculations have indicated that the ohmic losses, which are proportional to the square of the currents, would increase by a factor of three, from about 1.5\,kW to about 4.5\,kW. This corresponds to an average heat load of 70\,W per cable, and was considered unacceptable. Therefore, the powering of the \cmsph is based on the \dcdc conversion technique, with \dcdc converters placed roughly 1\,m away from the detector modules, on the service half-cylinders. Several \dcdc converters are connected in parallel to a power supply channel. The implementation of the power distribution differs between the \bpix and \fpix detectors. The \dcdc converters receive an input voltage of the order of 10\,V (11\,V in 2017 and 9\,V in 2018) and convert this into the voltages that are required by the pixel detector modules, taking into account voltage drops on the supply lines to the pixel detector modules. The output voltages amount to 2.4\,V for the analog part of the ROC, and 3.3\,V (\bpix L1, L3 and L4) and 3.5\,V (\bpix L2 and \fpix) for the TBM and the digital part of the ROC. The input currents are therefore reduced by the conversion ratios of 3-4, being even lower than in the original pixel detector. 

\subsection{\dcdc converters}
Custom \dcdc converter modules of sufficient radiation hardness had to be developed specifically for the \cmsph. They are based on the radiation-tolerant FEAST2 ASIC~\cite{DCDC-CERN} developed by the CERN EP-ESE group. The converters are designed implementing the buck topology, at a switching frequency of 1-2\,MHz. A large external inductor is required as an energy storage element. The \dcdc converters regulate their output voltage in a feedback loop based on pulse-width modulation and act like a negative differential resistance. The FEAST2 ASIC implements several control features (over-temperature, over-current, under-voltage). The delivery of the output voltage (voltage regulation) can be started and stopped remotely via a dedicated line (enable and disable signal). The chip outputs a status bit (``power-good bit''). 

\begin{figure}[tb!]
\begin{center}
        \includegraphics[width=0.38\textwidth]{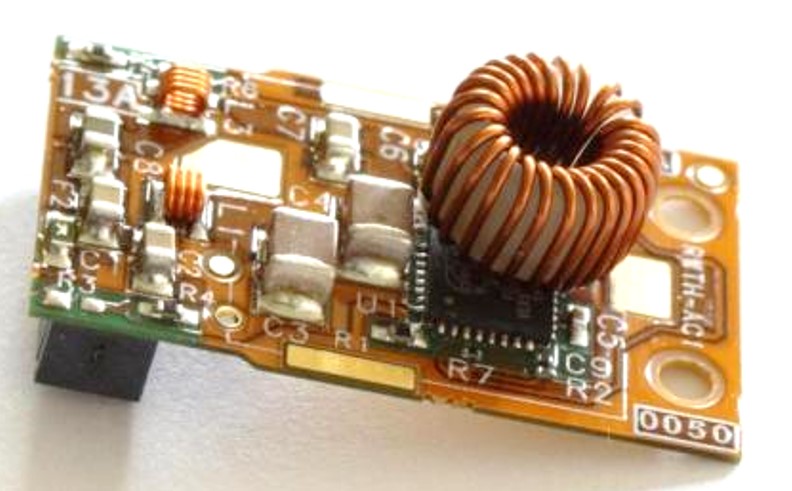}
        \hfill
        \includegraphics[width=0.61\textwidth]{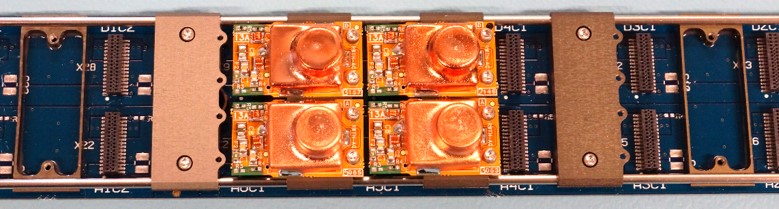}    
\end{center}
\caption{(Left) \dcdc converter with the shield not yet mounted. The black FEAST2 chip is partly covered by the air-core toroid. (Right) part of a \dcdc bus board. Four \dcdc converters with shields are mounted. To the left and the right the cooling bridges are visible, which clamp around the cooling pipes. \label{fig:DCDC_converters}}
\end{figure}

The \dcdc converter modules (Fig.~\ref{fig:DCDC_converters}, left) have a footprint of $28\times 17\mm^2$ and a height of 8\,\mm (without connector), optimized for the very limited space available on the \bpix service half-cylinder. The two-copper-layer PCBs are rigid. The boards carry the FEAST2 ASIC in a QFN32 package along with a number of required passive components, including a voltage divider to set the output voltage, and a resistor that sets the switching frequency to 1.5\,MHz. The main inductor is a custom toroid with a nominal inductance of 430\,nH; the core is made from plastic since ferrite would saturate in the 3.8\,T magnetic field of CMS. Filter networks with pi topology are implemented both at the input and output, to reduce the noise arising from the switching ripple. Lines for the remote control and status bit are implemented. The boards feature a thin film fuse at the input, to prevent a situation where all \dcdc converters on one power line could not be powered anymore because of a short. The connector is located on the back side. Finally, electromagnetic shielding covers the part of the board where large AC components are present (Fig.~\ref{fig:DCDC_converters}, right). The shield is a custom development; a 60\,\mum thick copper layer is galvanically deposited onto the outside of a deep-drawn plastic body. The shield is soldered to the PCB ground. 

The \dcdc converter modules have been optimized in a several year-long R\&D campaign for small size, low material budget, high power efficiency, and low radiated and conducted noise emissions~\cite{TWEPP2013, TWEPP2014}. A total of 1216 \dcdc converters are installed in the \cmsph~\cite{TWEPP2015}.  

\subsection{Low voltage power distribution}

\begin{figure}[tb!]
\begin{center}
        \includegraphics[angle=90,width=0.8\textwidth]{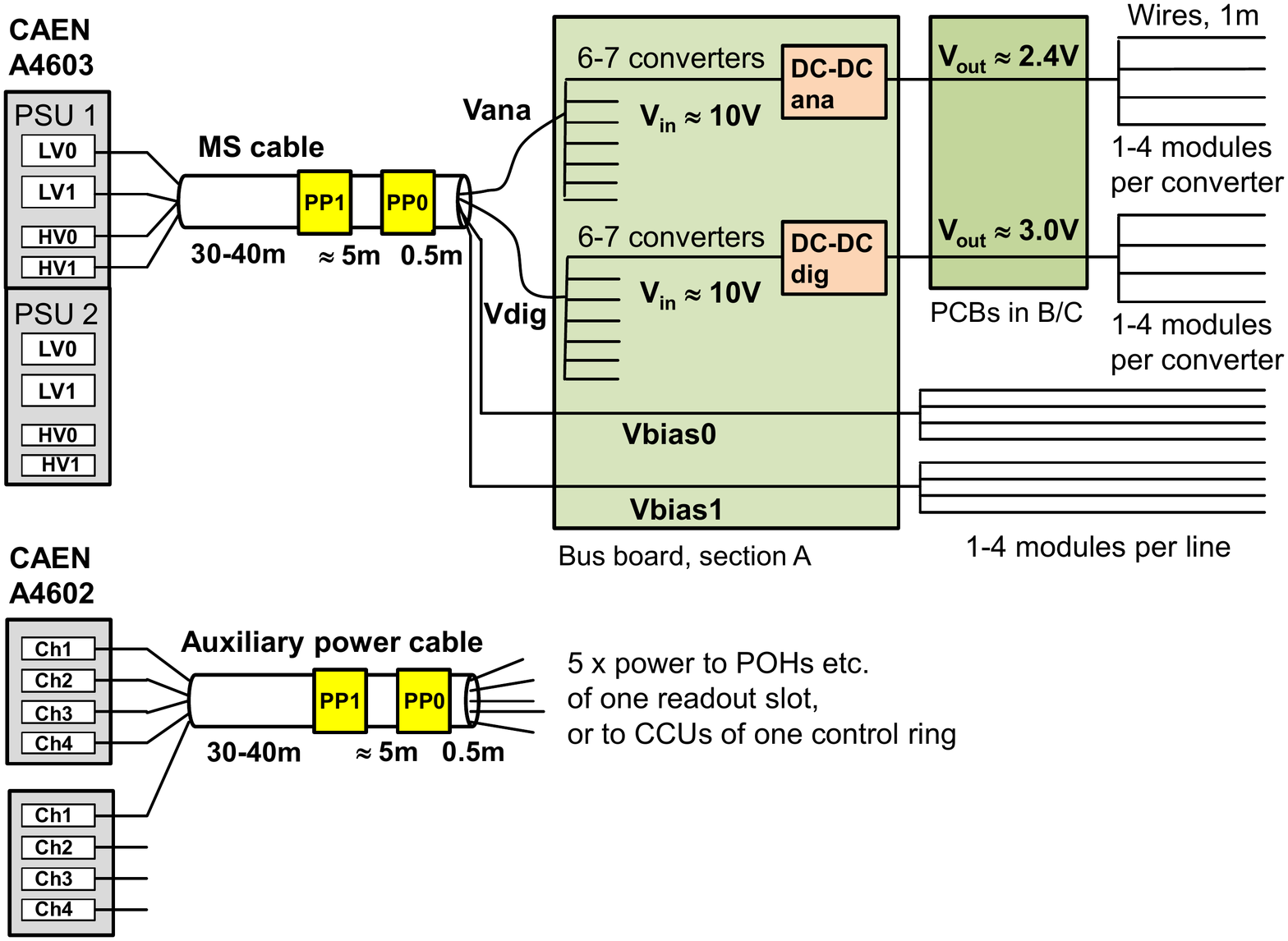}
\end{center}
\caption{Simplified schematic of the pixel power supply system. The details are explained in the main text. \label{fig:poweroverview}}
\end{figure}

\begin{figure}[tb!]
\begin{center}
        \includegraphics[width=0.75\textwidth]{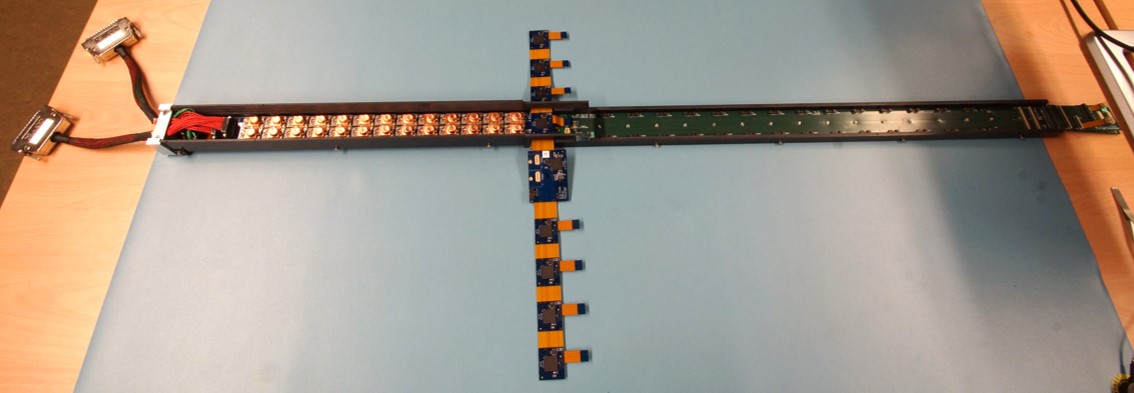}   
\end{center}
\caption{Mockup of one \bpix sector, showing a fully equipped bus board, the (perpendicular) CCU ring board, and the POH motherboard (green). Thirteen pairs of \dcdc converters are shown mounted on the busboard. The extension boards are below the POH motherboard and cannot be seen. \label{fig:PowerDistributionbpix}}
\end{figure}

Figure~\ref{fig:poweroverview} shows a simplified schematic of the pixel power supply system. For the \bpix detector, one pair of \dcdc converters delivers the analog and digital voltage for the following number of pixel detector modules: one module in L1, two or three modules in L2, and four modules in L3 and L4, respectively. The delivered currents vary between 0.4 and 1.7\,A for the analog and between 1.3 and 2.4\,A for the digital \dcdc converters. The power efficiency depends on the load, and amounts to about 80\% and 84\% for the analog and digital \dcdc converters, respectively, which leads to a dissipated power of up to 1.4\,W per \dcdc converter. The \dcdc converters are plugged into a motherboard (Fig.~\ref{fig:DCDC_converters}, right), called \dcdc bus board, that distributes the input voltage and the enable signals to the \dcdc converters, carries the status lines, and forwards the output voltage towards the detector modules. One bus board carries 13 pairs of \dcdc converters. Eight bus boards are mounted in Segment A of each service half-cylinder. The \dcdc converters are screwed to aluminum cooling bridges that clamp around the \coo cooling pipes. In addition to serving as thermal contact, the cooling bridges also keep the cooling loops in place on the service half-cylinder. The low voltages are further distributed to the pixel detector modules via two passive boards in Segment B (``extension boards''), which connect via two adapter boards to the three connector boards, into which the module cables are plugged. The CCU in charge of that sector generates the enable signals and receives the status signals, via its PIA interface, serving one \dcdc bus board. To minimize the number of connections, enabling is done in pairs of \dcdc converters, and the status signal is also combined (logical OR) for one pair. Boards of one \bpix sector can be seen in Fig.~\ref{fig:PowerDistributionbpix}.

\begin{figure}[tb!]
\begin{center}
        \includegraphics[width=0.5\textwidth]{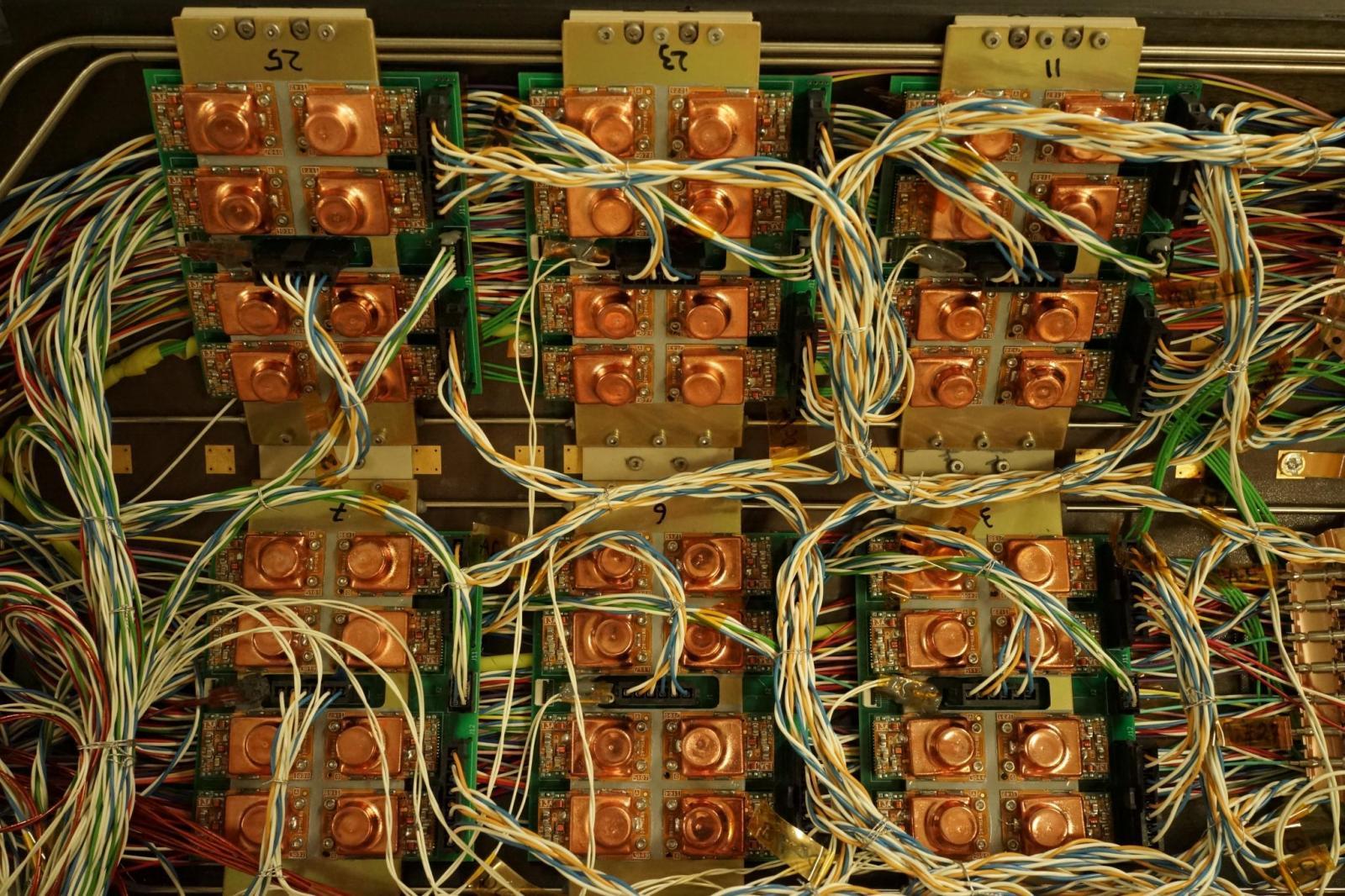}    
\end{center}
\caption{Six \fpix \dcdc motherboards as installed in the service half-cylinder. The cooling blocks that are placed under the \dcdc converters and provide the thermal path to the cooling loops are clearly visible. \label{fig:PowerDistributionfpix}}
\end{figure}

In the \fpix detector, one pair of \dcdc converters serves either three or four pixel detector modules. The input voltage for the \dcdc converters is passed through power-filter boards located at the far end of the service half-cylinder. Analog currents amount to 1.3 and 1.7\,A and digital currents to 2.0 and 2.3\,A for \dcdc converters that serve three and four pixel detector modules, respectively. Four \dcdc converter pairs are mounted on one \dcdc motherboard. A total of twelve such motherboards are installed on the inside of each service half-cylinder. Installed motherboards are shown in Fig.~\ref{fig:PowerDistributionfpix}. Cooling is provided by cooling blocks, with \dcdc converters mounted via a thermal film. Low voltages are supplied to the port card via a wire connection, and from there via the aluminum flex cables to the pixel detector modules. In each \fpix service half-cylinder one CCU board, each carrying four CCUs, is used to provide the enable signals and to receive the status signals; in the \fpix detector, two \dcdc converter pairs are served by one enable and one status line. 

Extensive system tests, in particular of the \bpix power chain, have been conducted to evaluate and optimize the system performance. This included readout tests with pixel detector modules, evaluation of the thermal management, and detailed measurements of the voltage drops along the distribution line, to optimize the supply voltages~\cite{TWEPP2014}. 

\subsection{Cables and power supplies}
Multi-service (MS) cables are used to deliver the high and low voltages (analog and digital) from the power supplies, located on the balconies, to the service half-cylinders. One \bpix bus board is served by two cables, one for L1 and L4 and one for L2 and L3. In the \fpix detector, one multi-service cable serves one \dcdc motherboard. 
 
The A4603 CAEN power supplies that were used for the original pixel detector needed two adaptations in order to be compatible with the \dcdc conversion powering scheme. This affected only the two low-voltage channels (analog and digital), while the high-voltage (HV) channels remained unchanged. The modified version of the power supply is named A4603D. One A4603D power supply module consists of two identical power supply units. Each unit features two low voltage channels and two high voltage channels.

First, the maximum output voltages of both low voltage channels were raised to 12.5\,V. The output voltage is software-programmable and stabilized in the range of 8-12.5\,V, such that the conversion ratio for the \dcdc converters can still be adjusted as desired. The output power is limited to 130-140\,W, with 90\,W available for the digital channel and 45\,W for the analog channel, limiting the maximum current one channel can deliver. Secondly, the fast remote sensing of the original system was abandoned, so as not to interfere with the regulation of the \dcdc converter itself. The sense wires, however, are used to measure the input voltage of the \dcdc converters. This information is used in a slow control loop, which can be enabled or disabled, to adjust the output voltage such that the voltage drop along the cables is compensated. The adjustment is very slow by construction, taking 1-2 seconds. The slow control loop was shown to work well and was therefore always enabled during the operation of the detector. 

One A4603D power supply module features four independent high voltage channels, each able to deliver a maximum voltage of 600\,V and a current of 20\,mA. Depending on the position, three to ten detector modules are supplied together from one HV channel. During LS2, the power supplies will be upgraded so that a maximum voltage of 800\,V can be delivered. 

The electronics components on the \bpix and \fpix service half-cylinders, including POHs, DOHs, and CCUs, require a supply voltage of 2.5\,V. This supply voltage is delivered by A4602 CAEN power supply modules in a direct powering scheme.

\subsection{Issues with \dcdc converters during 2017 operation}
\label{sec:powerops}

In general, the power system has worked according to its specification. However, from the beginning of October 2017 onwards, after about six months of flawless operation, some \dcdc converters started to fail, not delivering any output voltage. At the end of the 2017 run, about 5\% of \dcdc converters had failed. The failures were typically happening after a cycle of disabling and enabling the respective \dcdc converter, performed in order to recover modules affected by an SEU in the TBM. In the 2017/2018 LHC year-end technical stop the pixel detector was extracted, in order to replace defective \dcdc converters and to investigate the failures. An extensive investigation campaign was launched, involving the ASIC designers, and including several new irradiation campaigns. Failures similar to those seen in the \cmsph were observed when disable/enable cycles were performed on the \dcdc converters during irradiation, with a minimal dose of about 1\,\mrad necessary for failure. This is consistent with the dose accumulated by the \dcdc converters in the \cmsph by autumn 2017.

The problem was determined~\cite{Faccio} to be due to a transistor in the FEAST ASIC that develops a radiation-induced leakage current, amplified by a current mirror by a factor of about 500. The resulting current can flow to ground when the ASIC is enabled, but when it is disabled this is not the case anymore and a voltage exceeding specifications of transistors in the chip builds up on a certain (external) capacitor. This causes damage and makes the ASIC unusable. 

All \dcdc converters were exchanged with fresh but identical ones in the 2017/2018 LHC year-end technical stop. The input voltage was reduced from 11 to 9\,V, disabling was abandoned, and replaced by switching off and on the power supply. No failures occurred in the 2018 run. In LS2, the \dcdc converters will be replaced once again, using an improved version of the ASIC that features a path to ground for the leakage current. This is expected to cure the problem and to allow for operation of the power system without further interventions until the end of Run 3.

\section{Cooling}
\label{s:cooling}

The \cmsph uses evaporative \coo cooling to extract the heat dissipated by the detector modules and auxiliary electronics inside the detector, and to keep the silicon modules at low temperatures to mitigate effects of radiation damage. The evaporative \coo cooling technology has been chosen because of the low density, low viscosity and high heat transfer capacity of \coo that allow the use of small-diameter, thin-walled pipes inside the detector volume. The thermodynamic cycle developed for particle physics detectors is simple and the use of any active components near the particle interaction region can be avoided. 

\subsection{Cooling concept}

The \cmsph cooling system~\cite{ref:jerome, ref:paola} features the two-phase accumulator controlled loop (2-PACL)~\cite{ref:verlaat} approach which has been developed at Nikhef laboratories for the AMS Tracker~\cite{Haino:2010yu} and LHCb VELO~\cite{Alves:2008zz} detectors. 

\begin{figure}[tb!]
  \centering
      \includegraphics[trim=0 0 0 0,clip,width=0.7\textwidth]{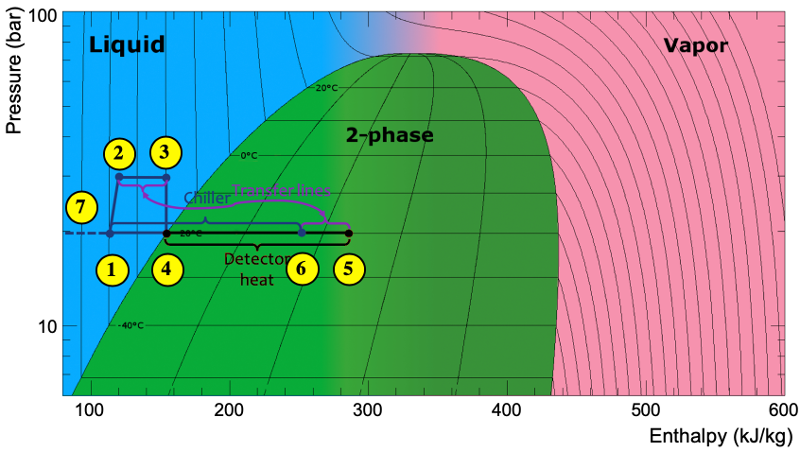}
       \includegraphics[trim=0 0 0 0,clip,width=1\textwidth]{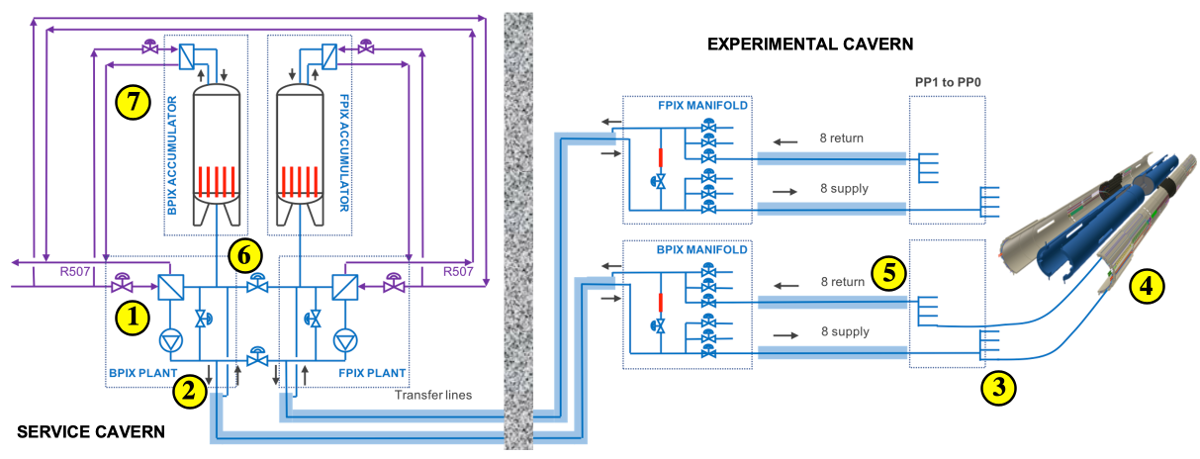}	\\
    \caption{Illustration of the 2-PACL concept. The upper figure is an illustration of the thermodynamic cycle, while the lower figure shows a schematic drawing of the \cmsph cooling system.}
    \label{fig:pacl}
\end{figure}

The thermodynamic cycle is illustrated in Fig.~\ref{fig:pacl}. The sub-cooled liquid \coo is pumped from the cooling plant to the detector (1-2). The liquid \coo flows towards the detector in concentric transfer lines, and it is warmed up by the two-phase returning fluid (2-3). The passage of the fluid through capillaries, at the entrance of the detector, brings the liquid to saturation pressure (3-4). Once inside the detector volume, the \coo partially evaporates (4-5) up to a targeted vapor quality of 33\%, because it extracts the heat produced by the sensor modules and auxiliary electronics. The vapor-liquid mixture returns to the cooling plant in the same transfer lines, where it starts to condensate due to the heat exchange with the incoming cold liquid (5-6). The \coo is then completely condensed and sub-cooled in the condenser (6-1) in the cooling plant (the condenser is connected to a primary cold source). A pressure vessel (known as accumulator), connected to the return line, regulates the return pressure (7) and thus the evaporating temperature. The temperature control is achieved by means of a double regulation loop acting alternatively on heating or cooling elements, depending on if the system needs to increase or decrease its evaporation temperature. All the active components of the system are placed at the cooling plant, which is located in an accessible area in the service cavern.

\subsection{Cooling requirements}

The cooling system is designed to remove the thermal load from the detectors as well as the heat spreading from the ambient environment to the cold parts of the detector system. The estimated maximum power at the design stage was 6\,kW for the \bpix detector, 3\,kW for the \fpix detector, and about 2\,kW for the heat leakage from the ambient environment~\cite{Dominguez:1481838}. The cooling system has been designed to cope with a total power of 15\,kW, providing an ample safety margin.

The range of operational temperature requirements spans from the needs of the commissioning phase to the ones of long-term operation. In the commissioning phase, when the detector volume is not sealed, the operating temperature must remain above the ambient dew point of the CMS cavern, \ie above +13\,\cel, to avoid any condensation. Therefore, +15\,\cel is chosen as the maximum coolant temperature. During operation, the silicon sensors need to be kept at a temperature below 0\,\cel to mitigate radiation damage effects. To fulfill this requirement, a coolant temperature of $-$23\,\cel is chosen as the lower limit of the operation range, while the design of the detector cooling loops and interfaces provides a temperature difference of less than 10\,\cel between the sensors and the coolant\footnote{Assuming a module power dissipation of 3.6-6\,W, as expected for operation of an irradiated detector at an instantaneous luminosity of $2\times10^{34}\,\rm{cm^{-2}s^{-1}}$.}. The pressure range for operation of the \coo cooling system is 10--70\,bar, while the system has been tested to a maximum pressure of 157\,bar.


To ensure continuous operation of the cooling system, even during maintenance periods or failure of one component, the cooling system has been designed such that two full power pumping modules are operating in parallel, as shown in Fig.~\ref{fig:pacl} (bottom). During normal operation, each cooling plant serves the manifold distributing cooling to one subdetector part (\bpix or \fpix), typically working at half of its maximum cooling capacity. In case of maintenance or failure of one of the two plants, the other one can take over and feed both subdetectors (backup mode). Furthermore, the use of two plants has the advantage that different temperatures for the two subdetectors could be chosen, if needed.


\subsection{Cooling system}

Both cooling plants are composed of three main modules: the accumulators and the pumping modules, located in the service cavern, and the distribution manifolds, located in the experimental cavern, as shown in Fig.~\ref{fig:pacl} (bottom). The connections between the pumping modules and the manifold and between the manifolds and detector are ensured by vacuum insulated transfer lines.

The cooling plants include a membrane pump, a condenser, a discharge filter, a suction strainer, a mass flowmeter, and valves and bypass valves for the supply and return lines. On the accumulator, five 1\,kW heaters are located at the bottom of the vessel, whilst a condenser, a level transmitter, and a backup chiller are mounted on top of the vessel. The manifolds distribute the liquid to the detector through eight cooling loops per subdetector. Each cooling loop is equipped with a supply and a return valve, a filter, a differential pressure flowmeter, and a needle valve. Each manifold hosts a ninth loop which is equipped with supply and return valves, a regulation valve and a 15\,kW dummy load for testing purposes. 

The main transfer line between the accumulator and the distribution manifold is designed with three coaxial pipes serving, from inside to outside, as liquid supply pipe, two-phase return pipe, and vacuum insulation, respectively. For this first section, industrial cryogenics solutions have been applied and standard dimensions were used. Because of space constraints, the transfer lines connecting the manifold to the detector are designed with a different geometry. The second section goes from the distribution manifold to a patch panel and it is designed with four concentric process lines insulated by a single vacuum pipe. The third section goes from the patch panel to the detector and is designed with three coaxial pipes. The first and second sections are insulated with static vacuum, while the third section is equipped with getter pumps to keep the vacuum level low enough to ensure proper insulation. The nominal mass flow in the transfer lines is 10.0\,$\rm{g/s}$, corresponding to a mass flow of 2.5\,$\rm{g/s}$ in the cooling lines inside the detector volume. 


\subsection{Detector thermal mockup}
\label{mockup}
In order to test and characterize the thermal behavior of the \cmsph, a mockup with mechanics identical to the \bpix L2 (two cooling loops) was set up in a clean room at the surface of the CMS experimental area. The mockup is equipped with dummy modules with heating resistors and temperature sensors as well as a preheating system to emulate the heat load of the electronics located in the service half-cylinder~\cite{thesis:renner}. 

\begin{figure}[tb!]
  \centering
     \includegraphics[trim=0 0 0 0,clip,width=0.9\textwidth]{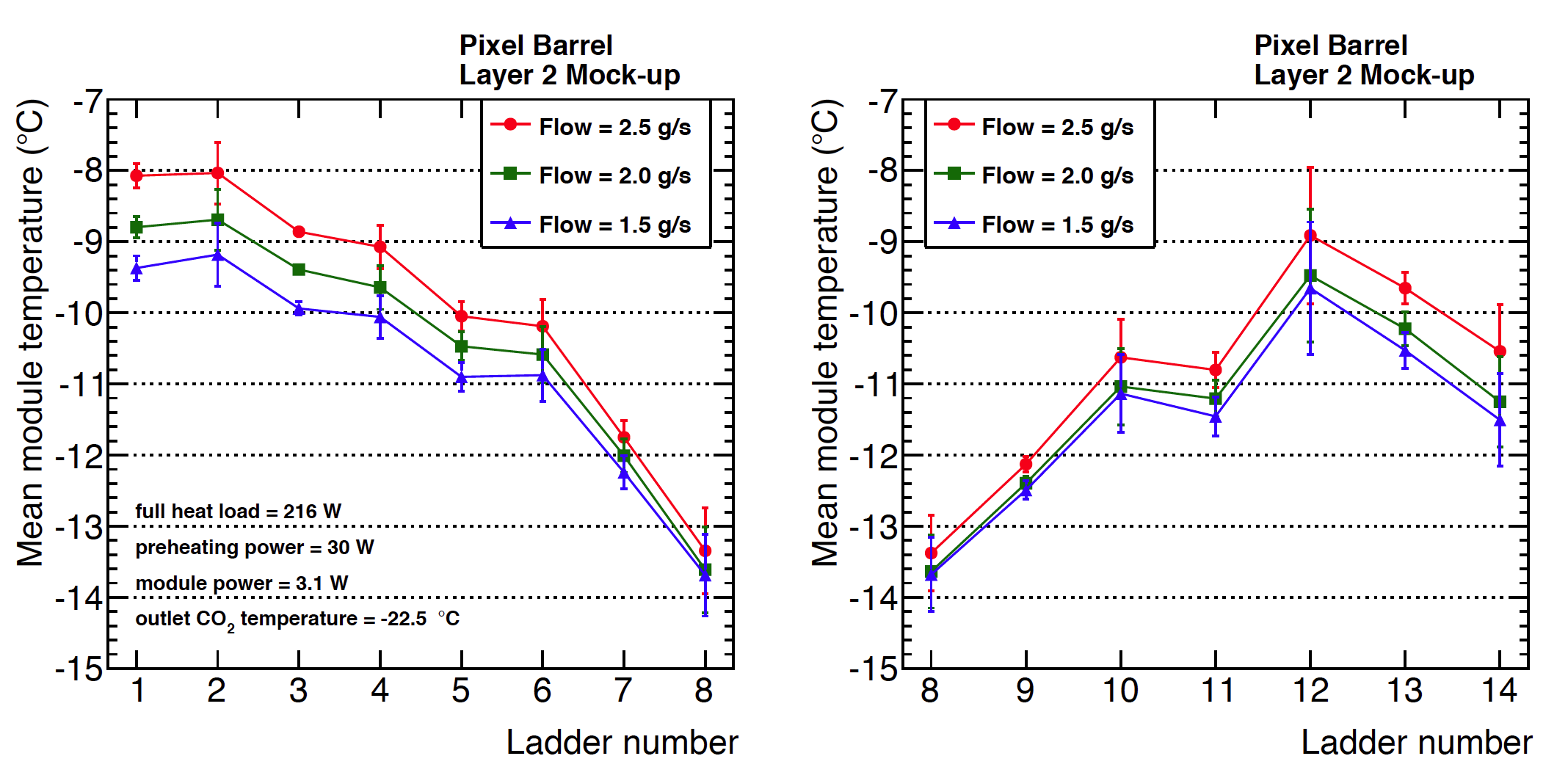}
    \caption{Module temperatures, measured with the thermal mockup and averaged over all modules on a ladder, as a function of the module position along the cooling loop. The module position is defined by the ladder number, which is counted starting from top to bottom. The two plots show the temperature measurements for the loop entering the detector from the $+z$ (left) and $-z$ (right) side. The loops entering the detector from opposite sides ($\pm z$) differ in their geometry~\cite{Dominguez:1481838}. The direction of flow of the coolant is left to right for the $+z$ side and right to left for the $-z$ side. The measurements are performed for different \coo mass flows and a nominal power load, expected for detector operation after having collected data corresponding to an integrated luminosity of 300\,\fbinv. The temperature measurements for ladder number 12 are affected by a bad thermal contact between the temperature sensor and the cooling loop.}
    \label{fig:bpixmockup}
\end{figure}

The mockup is inserted in a commercial refrigerator and connected to an evaporative \coo cooling plant. With this setup, measurements of the thermal behavior in the active detector region are performed, and the dependence on the operating parameters such as \coo mass flow, temperature, or heat load is determined. The temperature distribution of the modules along the cooling loops for nominal power load is shown in Fig.~\ref{fig:bpixmockup}. The temperature along the loops decreases by up to 5\,\cel from the inlet to the outlet of the loops because of the pressure drop. The overall temperature difference in the mockup is about 1.5\,\cel smaller when the \coo mass flow is lowered from 2.5\,$\rm{g/s}$ to 1.5\,$\rm{g/s}$. Based on these results, the mass flow in the pixel detector cooling loops has been lowered by 0.7\,$\rm{g/s}$ to a value of 1.8\,$\rm{g/s}$ during 2018 operation, which improved the temperature uniformity along the loops. A detailed understanding of the temperature distribution along the cooling loops is crucial for the prediction of the leakage current of the sensor modules, as it depends strongly on the sensor temperature.

\subsection{Cooling system performance and operation}

The cooling system was installed and commissioned in CMS in 2015/2016, well ahead of the detector installation, allowing for an extensive test campaign before the detector operation.
During this test campaign, the plants were successfully operated, also beyond specifications. In normal operation, the lowest temperature reached with the nominal load of 7\,kW (for each of the two subdetectors) was $-$32\,\cel at the accumulator and the maximum cooling power extracted at a temperature of $-$23\,\cel was 11\,kW on each system. In backup mode, the minimum temperature reached with the nominal load of 15\,kW was $-$27\,\cel and the maximum cooling power extracted at a temperature of $-$23\,\cel was 18\,kW.
This allowed for a smooth operation of the detector with a safe margin from the system limits. Since its start-up in March 2017, the cooling system has been continuously operating at $-$22\,\cel, excluding the standard maintenance periods during the LHC year-end technical stops.


\section{Pilot system}
\label{s:pilot}

\begin{figure}[tb!]
\centering 
\includegraphics[height=4.9cm]{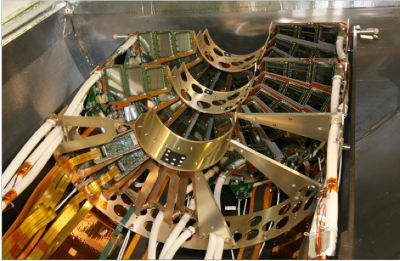}
\qquad
\includegraphics[height=4.9cm]{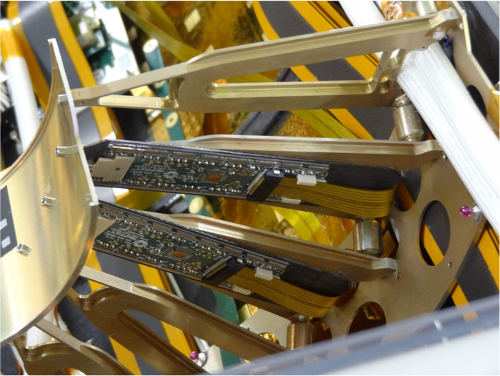}
\caption{\label{fig:pilot} Picture of one of the original \fpix service half-cylinders with two modules of the \phone pixel detector pilot system installed on the added third half-disk mechanics (left) and close up of two modules of the \phone pixel detector pilot system (right).}
\end{figure}

A pilot system~\cite{Akgun:2015afw} for the \cmsph was operated from 2015 to 2016 to be prepared for the short commissioning period during the 2016/2017 LHC extended year-end technical stop. The system consisted of eight prototype \fpix sensor modules and a complete readout chain. It was installed during LS1 in 2014 on a third half-disk mechanical structure, added in the available space in the original \fpix service half-cylinders, as shown in Fig.~\ref{fig:pilot}. A detailed description of the pilot system can be found in Ref.~\cite{ref:bora}. A prototype microTCA DAQ system was used to read out the pilot system. The goal of the installed pilot system was to learn how the power, control, readout, and DAQ systems perform in the CMS environment. Before installation, the pilot system was commissioned in a test stand at CERN using a prototype power and DAQ system developed for the \phone upgrade. Operating the pixel detector pilot system within CMS during proton-proton collisions in 2015-2016 provided valuable experience and enabled an early start for the modifications of the readout system that were required in view of the final system. 

The commissioning and operation of the pilot system helped to improve the design of the TBM and readout system and to complete the development of the FED firmware. In particular, an issue with decoding data at high trigger rates, observed in the pilot system test stand, was resolved. Two separate sources were identified as the causes of the problem: an asymmetric eye diagram due to the TBM design and jitter on the port card. While an asymmetric eye diagram could be accepted for the pilot system, a new version of the TBM was designed for the final sensor modules for the \cmsph. For the jitter on the port card, an external QPLL chip was put in between the TPLL and DELAY25 chip on the pilot port card. The final detector system incorporated the QPLL chip directly on the PCB (on the DOH motherboard for the \bpix detector and on the port card for the \fpix detector). 

The pilot system was also used to gain first experience with the new \dcdc conversion powering system. Six prototype \dcdc converters were installed. One pair of \dcdc converters was used to power four pixel detector modules, while the other four pixel detector modules were powered directly from the power supplies. No differences in performance were seen. The four remaining \dcdc converters were connected to external, passive loads, which were located at the power supply racks. The output voltages of these four \dcdc converters were monitored for single event effects, such as power glitches or spikes. No such events were observed. Furthermore, it was checked that the \dcdc converters did not cause any additional noise in the surrounding tracker detector. The problem with the \dcdc converter chips described in Sec.~\ref{sec:powerops} was not found, as the critical cycles of disabling and enabling were not regularly performed. Also the number of installed \dcdc  converters was very small (six devices), while the fraction of failed \dcdc converters in 2017 was 5\%.

\section{Integration, testing, and installation}
\label{s:integration}

In this section, the integration and testing of the \cmsph system is reviewed. This includes the assembly of the modules on the mechanical support structures as well as the integration of the detector and the service half-cylinders. Furthermore, the procedure of installing the \cmsph into the volume in the middle of the CMS detector is described. 

\subsection {Integration of the \bpix detector}

The integration of the \bpix detector took place between September 2016 and February 2017. The modules were mounted on the support structure at PSI, while the assembly of the
service half-cylinders was done at the University of Zurich. The final system was assembled and fully commissioned at PSI and then transported to CERN. Before the installation, further testing was carried out in a cleanroom at the surface of the CMS experimental site. 



\begin{figure}[tb!]
  \centering
    \includegraphics[trim=0 0 0 0,clip,height=5cm]{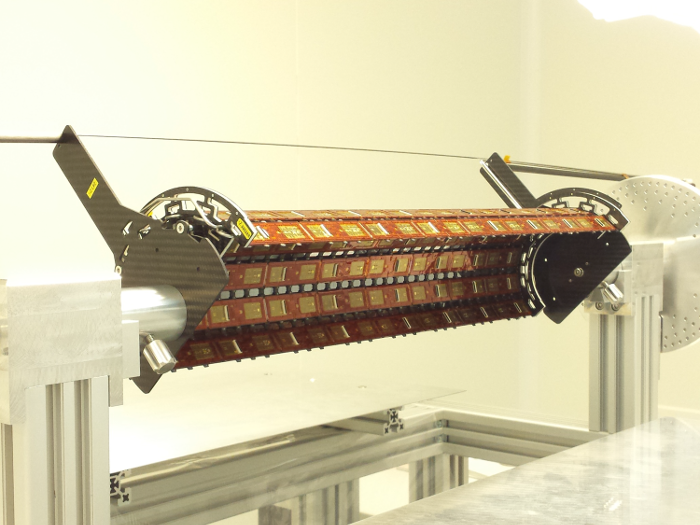}
    \includegraphics[trim=0 0 0 0,clip,height=5cm]{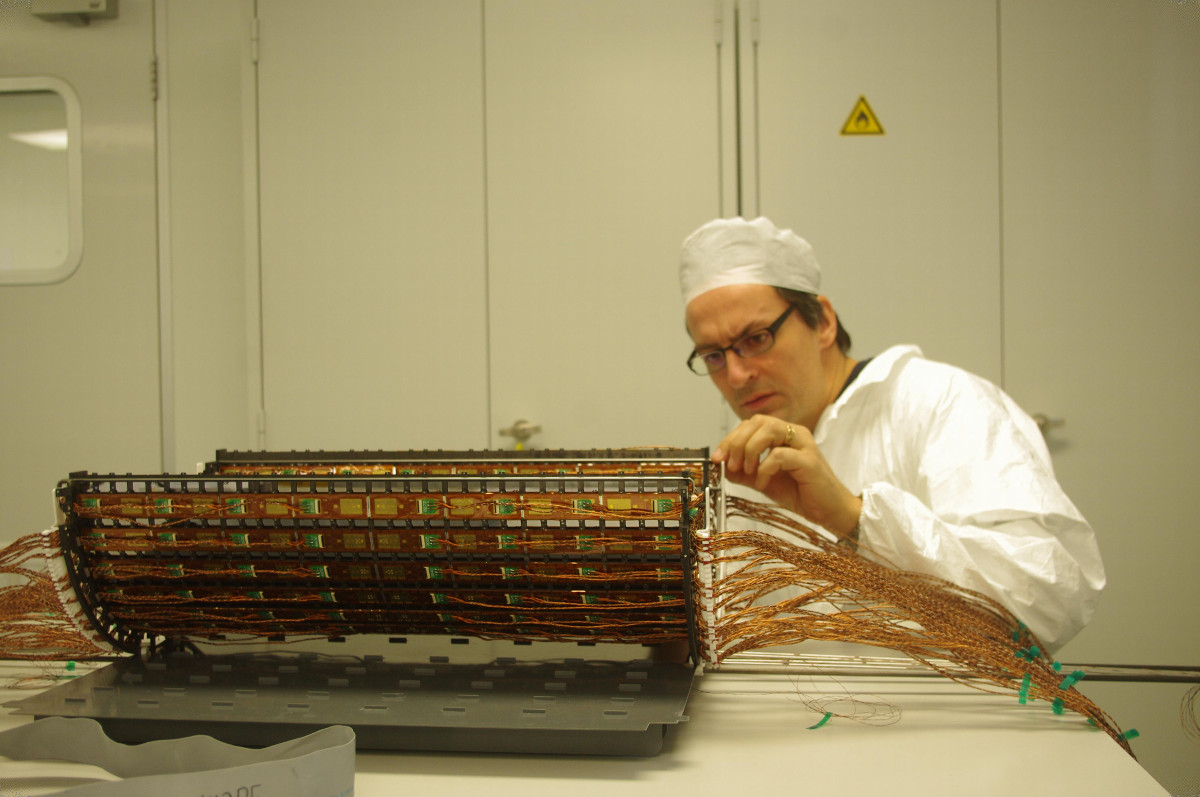}
    \caption{\bpix L2 half-shell after module mounting (left), and \bpix L4 half-shell with modules and cables (right).}
    \label{fig:modulemounting}
\end{figure}

The 1184 \bpix modules were mounted manually and fixed with screws on the carbon fiber support structure. Thermal grease was used in the innermost layer to improve the thermal contact between the modules and the cooling structure. The mounting procedure proved to be very efficient and up to 80 modules were mounted in a day. Only 8 modules were damaged during the mounting procedure, most likely by excessive mechanical pressure or electrostatic discharges. They were all replaced before proceeding with the detector integration. Figure~\ref{fig:modulemounting} (left) shows one \bpix L2 half-shell with modules mounted. In the next step, the micro-twisted-pair cables were connected to the modules and routed to the detector end-flange, as shown in Fig.~\ref{fig:modulemounting} (right).

\begin{figure}[tb!]
  \centering
    \includegraphics[trim=0 0 0 0,clip,width=0.7\textwidth]{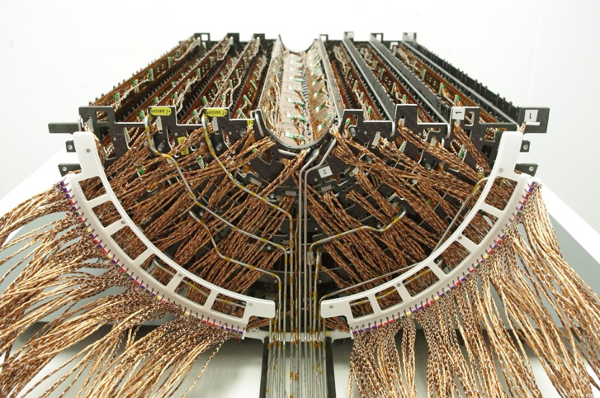}
    \caption{One half of the \bpix detector.}
    \label{fig:bpixfinal}
\end{figure}

In the final step, the four layers of half-shells were merged and the other side of the cables connected to the connector boards mounted on a temporary support structure. The modules were tested individually after the mounting and groups of modules connected to a connector board (11-16 modules) were retested after the merging of the half-shells. The module mounting was completed with only four non-working modules, which corresponds to less than 0.4\% of the \bpix detector channels. Figure~\ref{fig:bpixfinal} shows a picture of one half of the fully assembled \bpix detector.

\begin{figure}[tb!]
  \centering
    \includegraphics[trim=0 0 0 0,clip,width=0.44\textwidth]{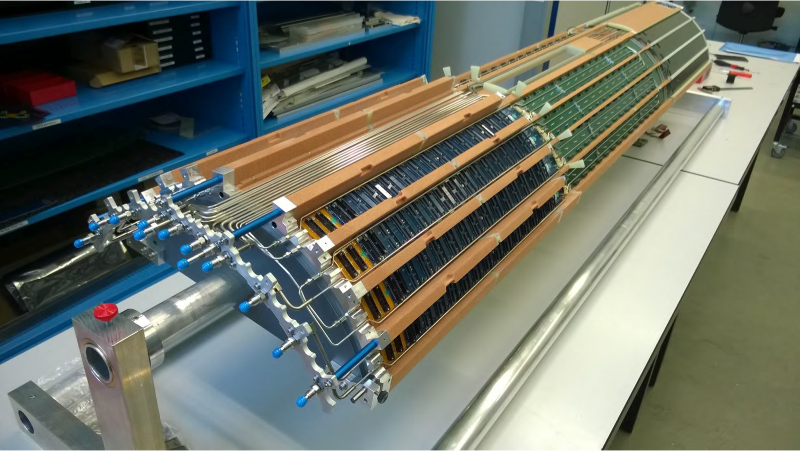}
    \includegraphics[trim=0 0 0 0,clip,width=0.44\textwidth]{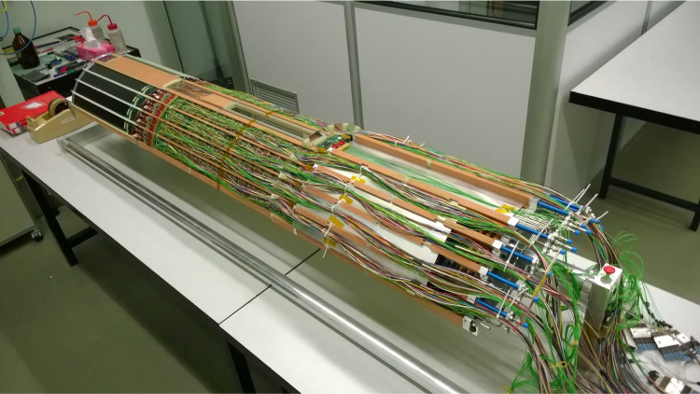}
    \caption{\bpix service half-cylinder during assembly: mechanics with power boards and cooling loops installed (left), and fully equipped \bpix service half-cylinder (right).}
    \label{fig:bpixstelectronics}
\end{figure}

The integration of the four \bpix service half-cylinders consisted of the assembly and testing of the components for power, cooling, and readout. In order to facilitate the mounting procedure, the service half-cylinders were placed on a rotatable mandril (Fig.~\ref{fig:bpixstelectronics}). In the first step, the \dcdc bus board, the extension boards, the POH motherboards, and the adapter boards were put onto the mechanical support structure and fixed by screws. The cooling tubes were then laid in and fixed by mounting the cooling bridges. The \dcdc converters were subsequently screwed on top of the cooling bridges. In the next step, the CCU ring was placed, before the optical components (POHs, the DOH motherboard and the DOHs) and flex cables were attached. Groups of three POHs were connected to a bundle of 12 optical fibers before mounting them onto the service half-cylinder. The 1984 optical fibers that run along the service half-cylinders were arranged within the slots to fit into the tight space available and to allow for future simple replacement of faulty components. Pictures of the \bpix service half-cylinders during the assembly are shown in Fig.~\ref{fig:bpixstelectronics}. 

All electronics parts of the \bpix readout and power system were tested for functionality before and during mounting. With the components of the pixel detector DAQ and powering system available for the testing, one sector could be tested at a time. The functionality of the CCU ring architecture and the \itwoc communication with the readout electronics on the \bpix service half-cylinder was checked. Furthermore,  a set of detector modules was connected to each sector to verify programming, clock, and trigger distribution, and to ensure the quality of the signal transmission. The power distribution was tested by measuring the low voltage at the module connectors while applying a constant load corresponding to the expected maximum power consumption of the modules. The high-voltage power distribution was checked up to voltages of 1000\,V. In addition, the \bpix service half-cylinders underwent a series of thermal cycles between $+20$\cel and $-20$\cel.

The few components that were damaged during the mounting procedure (mainly individual opto hybrids and optical fibers) were replaced on the spot and the \bpix service half-cylinders were sent for the integration with the \bpix detector in a fully functional state.

\begin{figure}[tb!]
  \centering
    \includegraphics[trim=0 0 0 0,clip,width=0.7\textwidth]{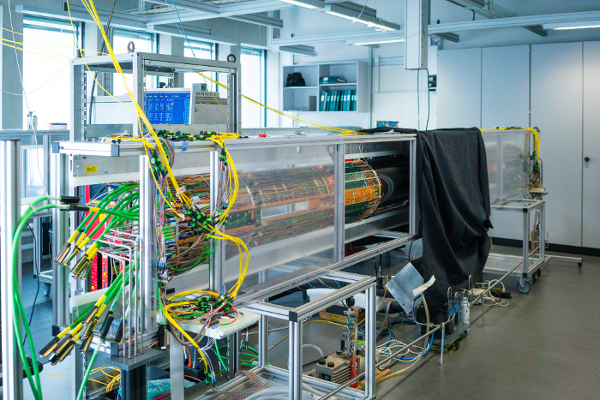}
    \caption{One half of the \bpix detector with the service half-cylinders connected. The detector system is placed on a rail system inside a 5\,m long custom-built transport box.}
    \label{fig:bpixbox}
\end{figure}

The \bpix detector and service half-cylinders were assembled and commissioned at PSI in January 2017. The detector halves together with the service half-cylinders were integrated in two 5\,m long boxes custom-built for the installation into CMS. Within the transport boxes, the pixel detector was placed on wheels on a rail system, which could later be used to slide the detector into position in the CMS detector. A picture of one half of the \bpix system inside the transport box is shown in Fig.~\ref{fig:bpixbox}.

After the assembly, the complete system was fully commissioned by performing a detailed functionality test connecting one sector at a time to the DAQ system. A custom-made mobile two-phase \coo cooling system was used to keep the temperature of the detector constant at +15\cel. The tests included the verification of the functionality of all electrical components on the service half-cylinders, checks of the low-voltage and high-voltage power distribution, as well as module performance tests based on charge injection to all pixels. If accessible, the few individual components that did not pass the tests were replaced. After all tests, the number of non-functional channels was found to be less than 0.7\%. In addition to the four problematic modules present after detector construction, four modules were found with either one or more ROCs not responding to programming commands or no data transmitted by the TBM. 

On February 7, 2017, the two halves of the \bpix detector system were transported to CERN. After the transportation, the system was tested again in a cleanroom at the surface of the CMS experimental area. In addition to functionality tests at room temperature, detailed tests and calibrations were also performed when operating the detector at $-20$\cel (cold test). 

An issue with some groups of modules drawing too much current was observed during the cold test. After detailed investigations, the issue was traced to module cables with damaged isolation leading to short circuits among cables. The damage happened due to mechanical friction at the aluminum end flange of the service half-cylinder. After replacing individual cables and adding additional isolation in the affected area, the issue was resolved and the detector was ready for installation.

\subsection {Integration of the \fpix detector}

The integration of the \fpix detector took place between late spring of 2016 and January
2017. The modules were mounted on half-disks at FNAL, while the integration of the
completed half-disks and support electronics into service half-cylinders was done both at FNAL and CERN. 
As with the original detector, shipment from FNAL to CERN was done with the half-disk assemblies
hand-carried separately, on a commercial airplane, from the rest of the service half-cylinder components,
which were shipped conventionally.


The integration began with Laird TPCM-583 thermal film being applied to the carbon fiber half-ring 
blade structures in the module mounting locations. The Laird film provides an adhesive as well as a thermal connection between the modules and the blades. To laminate the Laird film to the blade properly, half-rings were baked at 55\,\cel for 30 minutes in a vacuum oven. After setting the hub addresses for the TBMs, the modules were
placed on half-rings using vacuum holding jigs and a clamping
fixture for convenient module mounting. The modules were then screwed to the blades, before the aluminum flex cables were attached and clamped.
Fully loaded half-rings were baked a second time to set up the Laird film connection to the
modules. Modules were tested before and after the baking and, throughout the whole mounting process, only four modules needed replacement. All modules were working prior to integration in the half cylinder.

The service half-cylinders were prepared for half-ring installation by first mounting the filter boards at the end of the service half-cylinder and then laying in the cooling tubes. 
In the next step, the port-card cooling structures, the \dcdc motherboards and the port card pairs were installed. The port-card power cables, CCU boards, and optical fibers were added and the functionality of the port cards and the optical power output were tested using spare
sensor modules. The \dcdc converters were installed, cabled, and tested.
Cooling blocks for the \dcdc converters and the port cards were thermally joined to the
cooling pipes with thermal grease, and all parts were fixed to the service half-cylinders with screws.

\begin{figure}[tb!]
  \centering
    \includegraphics[trim=250 0 200 0,clip,height=7cm]{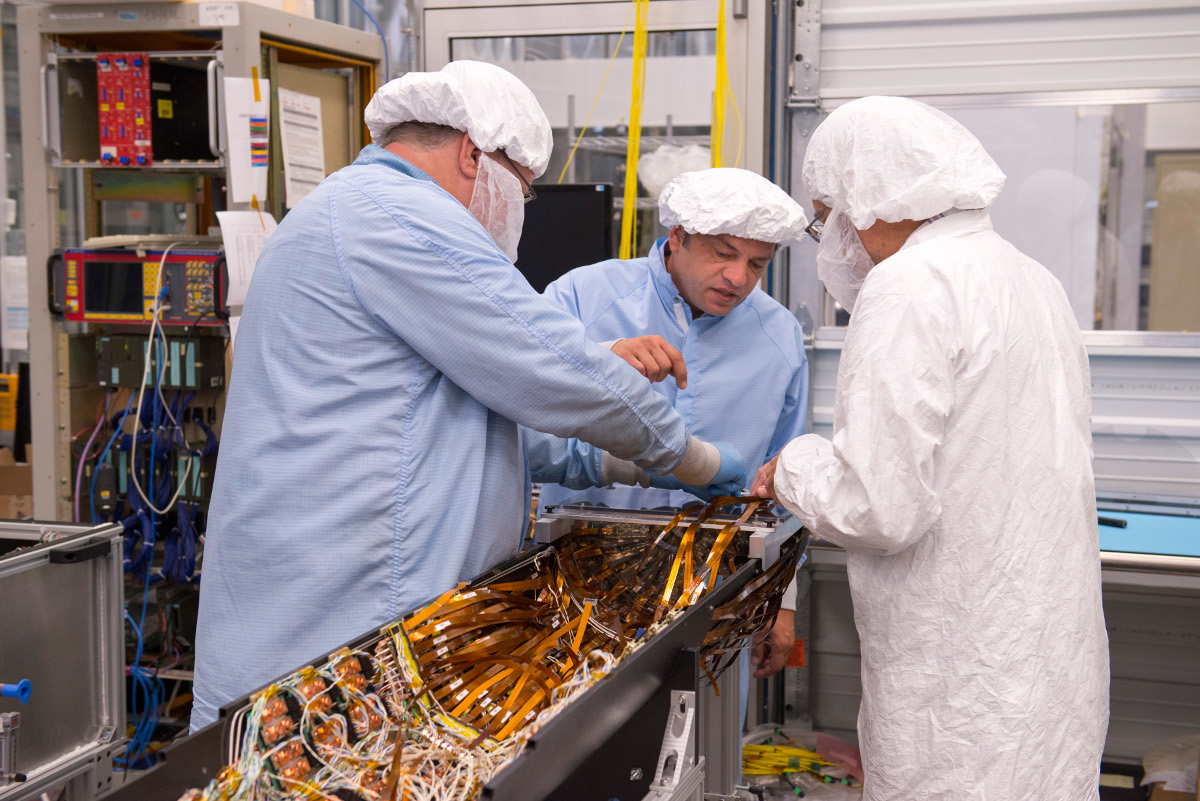}
    \includegraphics[trim=0 0 0 0,clip,height=7cm]{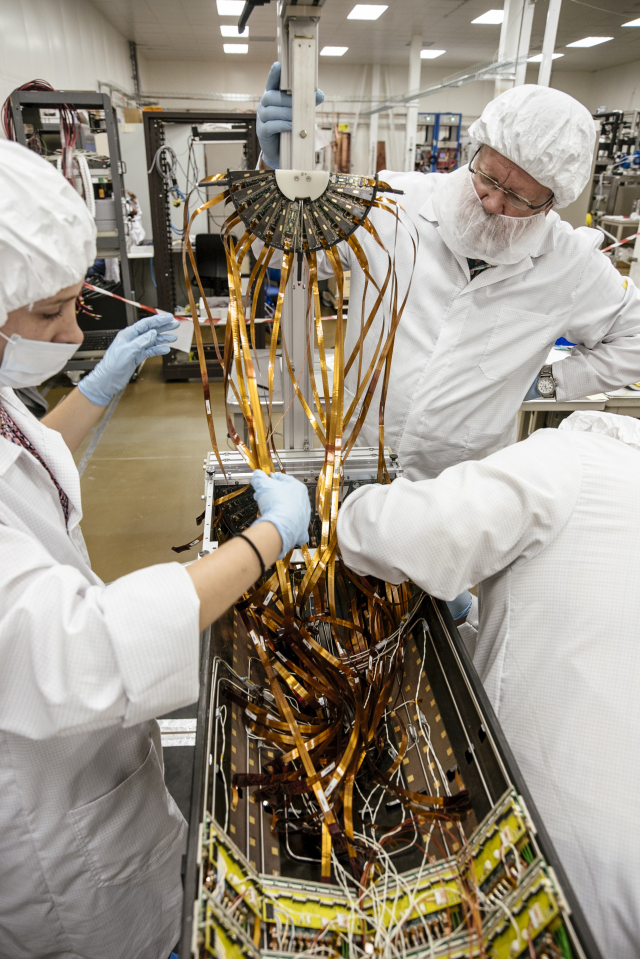}
    \caption{Pictures of the installation of the \fpix half-disks into the service half-cylinder during the assembly at FNAL (left) and during the reassembly at CERN (right). Pictures used by permission from FNAL and CERN.}
    \label{fig:fpixmodulemounting}
\end{figure}

Completed half-rings were lowered in place using a special manual crane and holding fixture.
Installation of half-disks proceeded in order from the outer half-ring closest 
to the interaction point followed by the inner half-ring, until all three half-disks were installed and cabled to port cards. Pictures of the installation procedures of the half-disks are shown in Figure~\ref{fig:fpixmodulemounting}. After a half-disk was in place and the cooling pipes were attached, a leak check was performed. When testing the module cables, it was found that the flex cable ends could be easily damaged if
forced to bend in a small radius, either in storage or during installation. This led to an additional disassembly, inspection, reassembly and test at CERN. A total of 44 out of 672 flex cables were replaced. An in-situ test of the ROC functionality was used to assess any damage to the final assembled detector. Nine out of 672 modules (1.3\%) of the \fpix detector were not working after all practical recovery procedures were exhausted. Out of these nine non-working modules, seven were due to flex cable issues, one had an issue with a broken TBM, and one had broken wire bonds due to improper handling.

\begin{figure}[tb!]
  \centering
    \includegraphics[trim=0 0 0 0,clip,width=7cm]{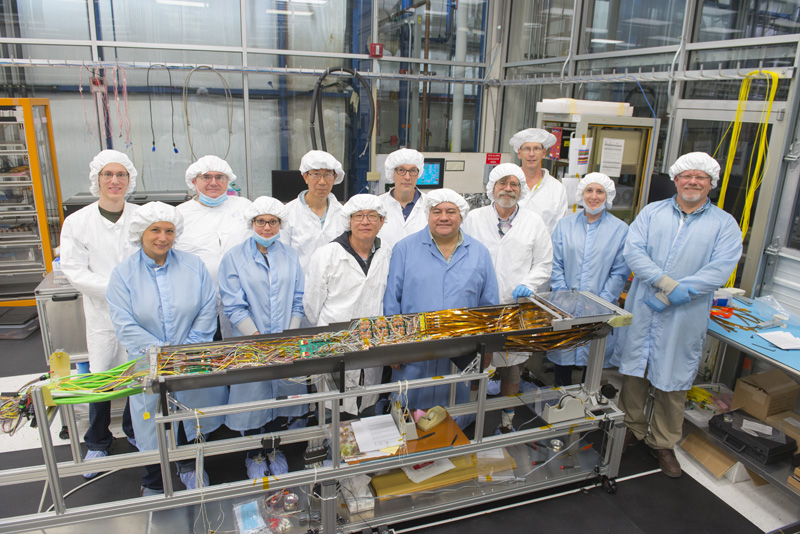}
    \includegraphics[trim=0 0 0 0,clip,width=7cm]{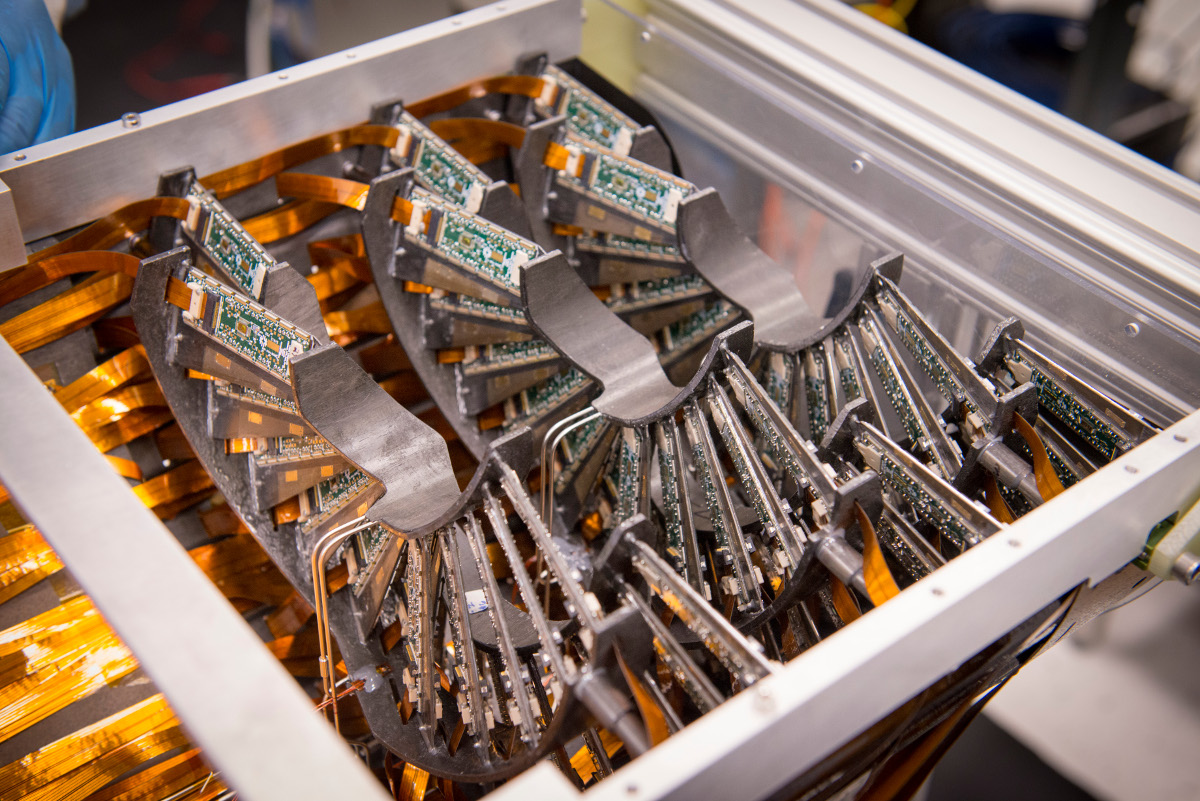}
    \caption{Picture of a fully assembled \fpix half-cylinder (left) and close-up of three \fpix half-disks (right). Pictures used by permission from FNAL.}
    \label{fig:fpixstelectronics}
\end{figure}

Pictures of a fully assembled \fpix half-cylinder are shown in Fig.~\ref{fig:fpixstelectronics}. Assembled half-cylinders were transferred to a sealed box, flushed with dry air, for full electrical/optical testing, and final dressing, e.g. for adding additional temperature probes. 
For the first half-cylinder tested, 0.15\% of pixels were lost in patterns at the corners of modules due to mechanical stresses. In order to minimize further potential damage, the second half-cylinder was tested by leaving the detector on and cooled, while the \coo system provided coolant at an adjustable temperature. The temperature of the coolant was lowered in steps of 5\,\cel. As a result, only 0.02\% of pixels were lost. The \coo system at CERN provided cool-down rates smaller than 1\,\cel/min and no increase in bump-bond loss after reassembly and testing at CERN was noted.


\subsection{Installation into CMS}

\begin{figure}[tb!]
  \centering
    \includegraphics[trim=0 0 0 0,clip,width=0.8\textwidth]{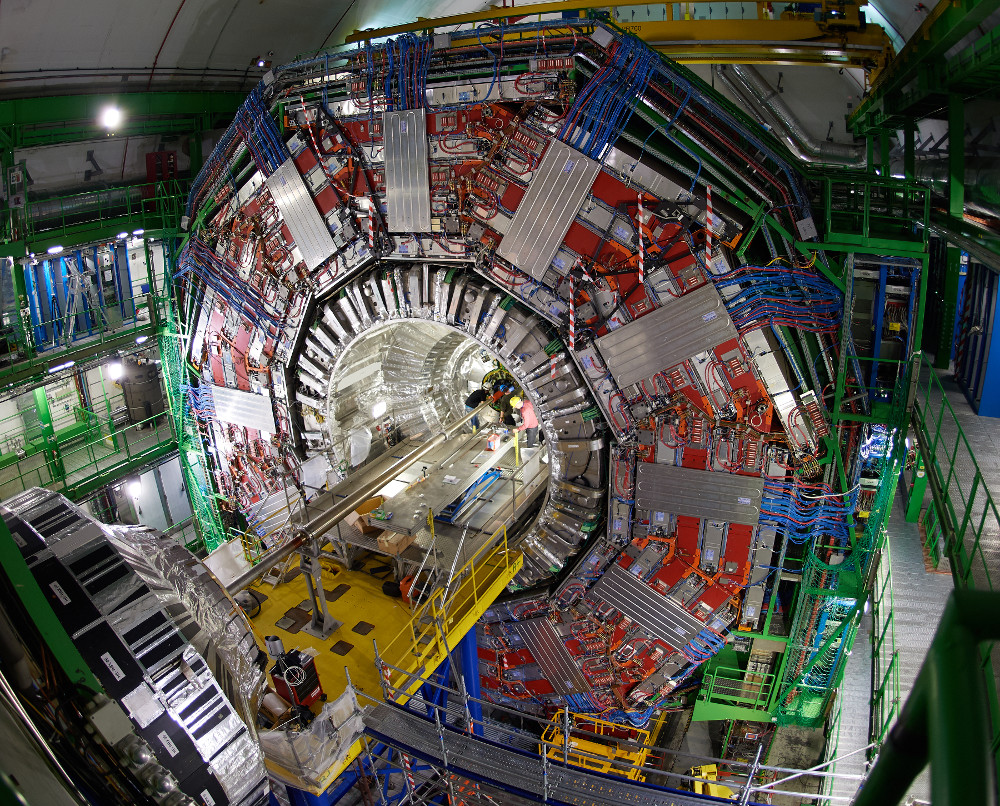}
    \caption{Picture of one side of the CMS detector with the endcaps opened and the temporary platform for the pixel detector installation set up. Pictures used by permission from CERN.}
    \label{fig:installationtable}
\end{figure}

The pixel detector support tube inside the CMS detector has curved rails on the top and bottom, that were designed to insert the pixel detector parts and service half-cylinders along the beam pipe. A clearance of 5\,mm with respect to the beam pipe is ensured during any time of the installation procedure. For the installation of the pixel detector, the CMS endcaps had to be opened to get access to the central part of the detector. Temporary platforms holding the installation tables were then installed on both ends of CMS (Fig.~\ref{fig:installationtable}). 

After the extraction of the original pixel detector in December 2016, the detector services were adapted for the \cmsph system. The cooling tubes arriving at \mbox{patch panel 0} (PP0), located at both ends of the pixel detector system, were replaced with capillaries and transfer lines (between PP0 and the patch panel at the edge of the magnet solenoid, PP1) for the \coo cooling system. Furthermore, work was carried out on the optical fiber plant. Given that the number of fiber ribbons needed to read out and control the \cmsph increased by almost 30\% with respect to the number previously installed, it was necessary to install new optical fiber cables arriving at PP0. This brought the opportunity to change the optical connector type used at PP0, from the Multi-fiber System (MFS) connector, used previously, to the Multi-fiber Push On (MPO) type. MPO connectors do not require the use of a tool to (dis-)connect, making the installation and removal time shorter, which is important to maintain the serviceability of the detector in a radiation environment.

In order to verify the installation tools and procedures, the installation was rehearsed at the surface of the CMS experimental area using a mockup of the volume where the pixel detector was to be installed.  

The installation of the \phone \bpix and \fpix detectors took place in the first week of March 2017. The \bpix system was installed first. The installation boxes were craned down to the experimental cavern through the main shaft and lifted onto the installation table. Both halves of the \bpix detector were inserted from the same end of the CMS detector (as for the original detector). The rail system inside the box was joined with the rail system inside CMS using temporary extension rails. This way, the pixel detector could slide out of the transport box along the rails inside the CMS detector. The \bpix detector features adjustable wheels such that the detector half-shells stay away from the beam pipe during insertion and are only closed when the detector reaches its final position. 

The pixel detector has to be placed with an accuracy of 0.2\,\mm in the transverse plane due to the tight tolerance with respect to the beam pipe. In the longitudinal direction there is less need for precision and the position accuracy is of the order of 1\,\mm. The detector position is then determined through the alignment process using tracks of charged particles from collisions.

After both halves of the \bpix detector system were installed, the power cables, capillaries for cooling lines and optical fibers were connected at PP0. A fast check-out of the \bpix connections and functionality was performed within a few days before proceeding with the installation of the \fpix detector. 


The \fpix half-cylinders were craned into the experimental cavern inside installation boxes, which have an internal rail system. The \fpix half-cylinders were installed from both sides of the CMS detector. The two halves that were inserted from the same $z$ end were installed simultaneously. This allows modules in the two different half-cylinders to overlap in the direction transverse to the beam during the installation. While pushing the half-cylinder pair into the bore, the force needed to advance each half-cylinder was monitored. A sudden rise in force indicated that the half-cylinders had reached the physical stops at the end of the rails, 
and the half-cylinders were fixed in place. At this point, the capillaries were connected at PP0 and the cooling pressure integrity was checked. After installation in 2017, a small leak in one \fpix half-cylinder at 50\,bar, well above the standard operating pressure, was noted. As the leak was small, 
and not present at lower pressures, the installation proceeded with the connection of power cables and control and readout fibers, and was followed by a checkout. No channels were lost during the installation and no effect on operations was seen due to the leak at 50\,bar, nor was a leak present when the detector was re-installed in 2018, leading to the conclusion that the high pressure leak was due to a connection made 
during installation rather than an issue inside the half-cylinder.

\begin{figure}[tb!]
  \centering
     \includegraphics[trim=0 0 0 0,clip,width=0.5\textwidth]{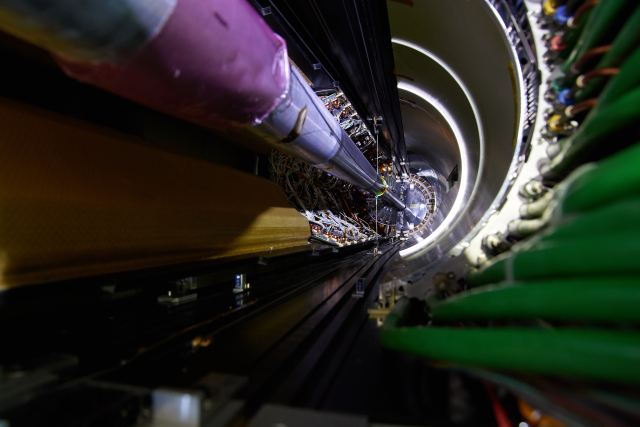}
      \includegraphics[trim=0 0 0 0,clip,width=0.44\textwidth]{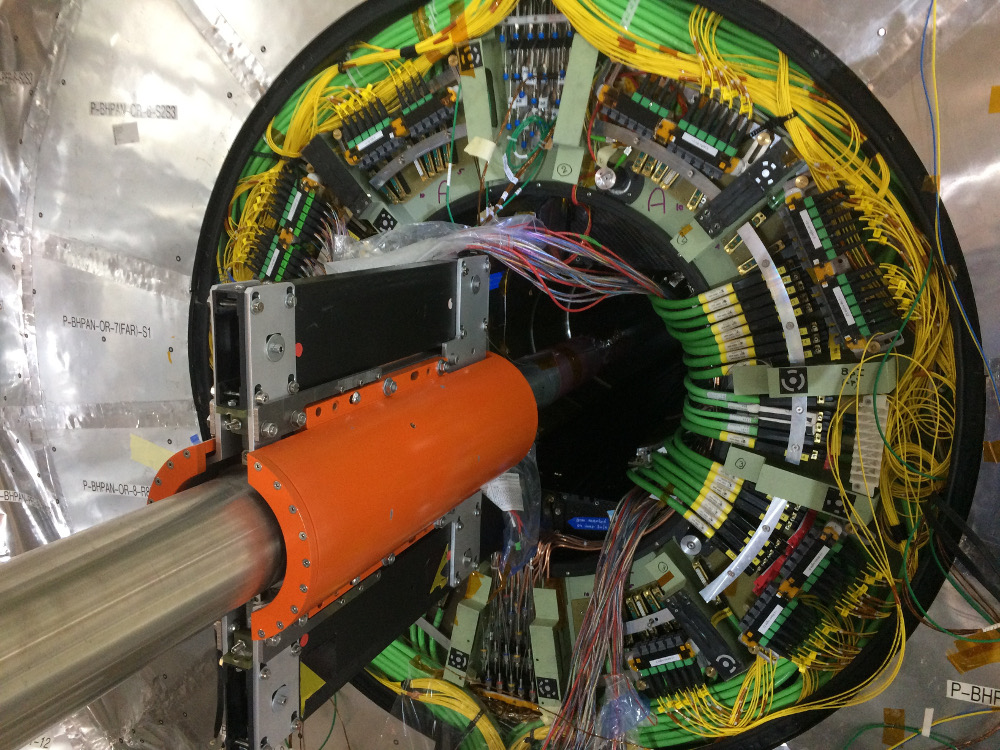}
    \caption{ Picture of the \bpix detector and one quarter of the \fpix detector (left of the beam pipe) during installation in CMS (left). Picture of power cable, optical fiber and cooling tube connections at PP0 (right).}
    \label{fig:pp0}
\end{figure}

A picture of the \bpix and \fpix detectors during installation inside CMS is shown in Fig.~\ref{fig:pp0} (left), while Fig.~\ref{fig:pp0} (right) shows a picture of the connections of power cables, cooling lines and optical fibers at PP0. 

The same installation procedure has been used for the re-installation of the \bpix and \fpix detectors after the consolidation work performed during the 2017/2018 LHC year-end technical stop. No issues were experienced, confirming that the procedure is well established, reliable and reproducible.

\begin{figure}[tb!]
\centering
\includegraphics[height=6.3cm]{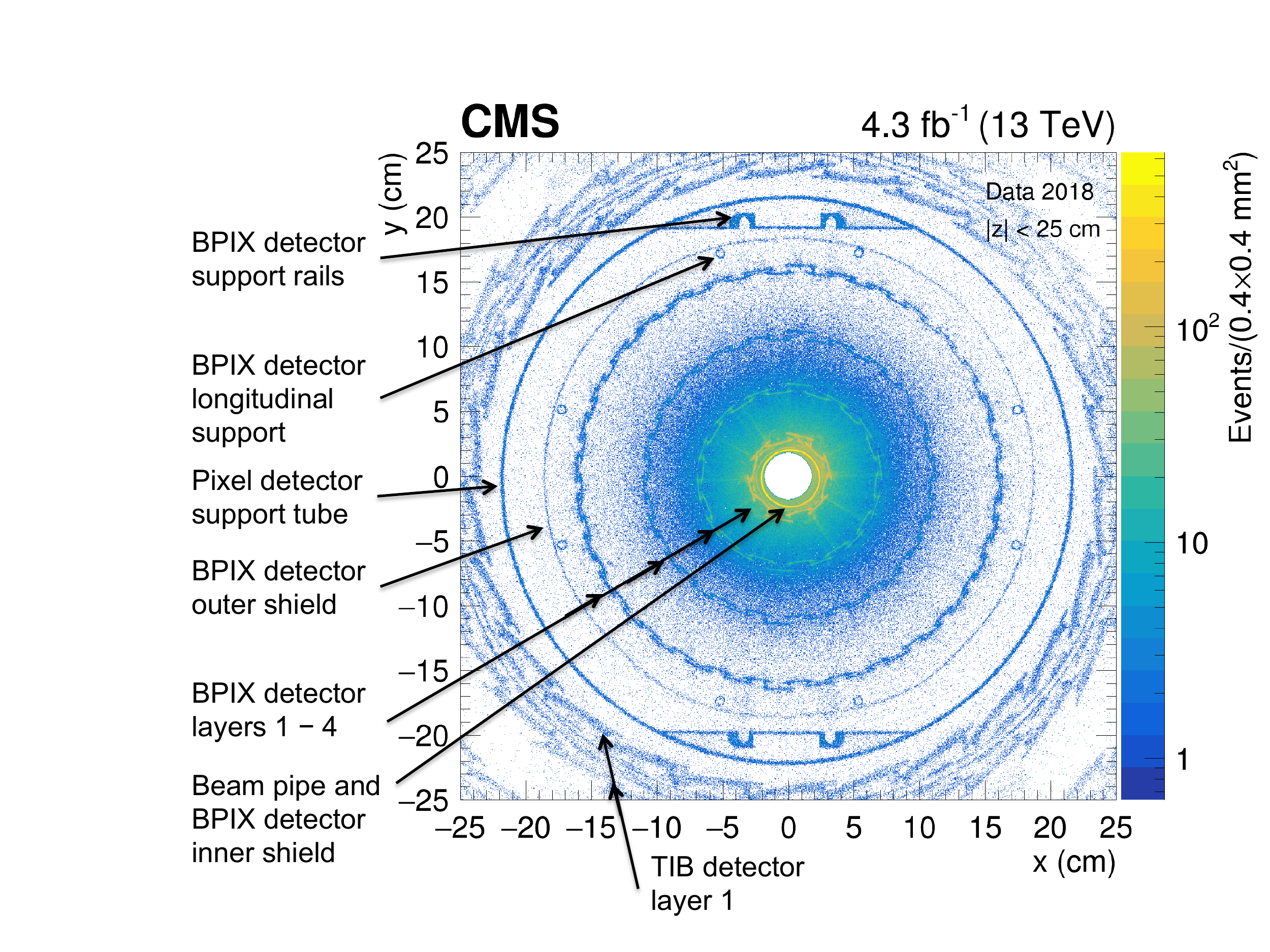}
\includegraphics[height=6.3cm]{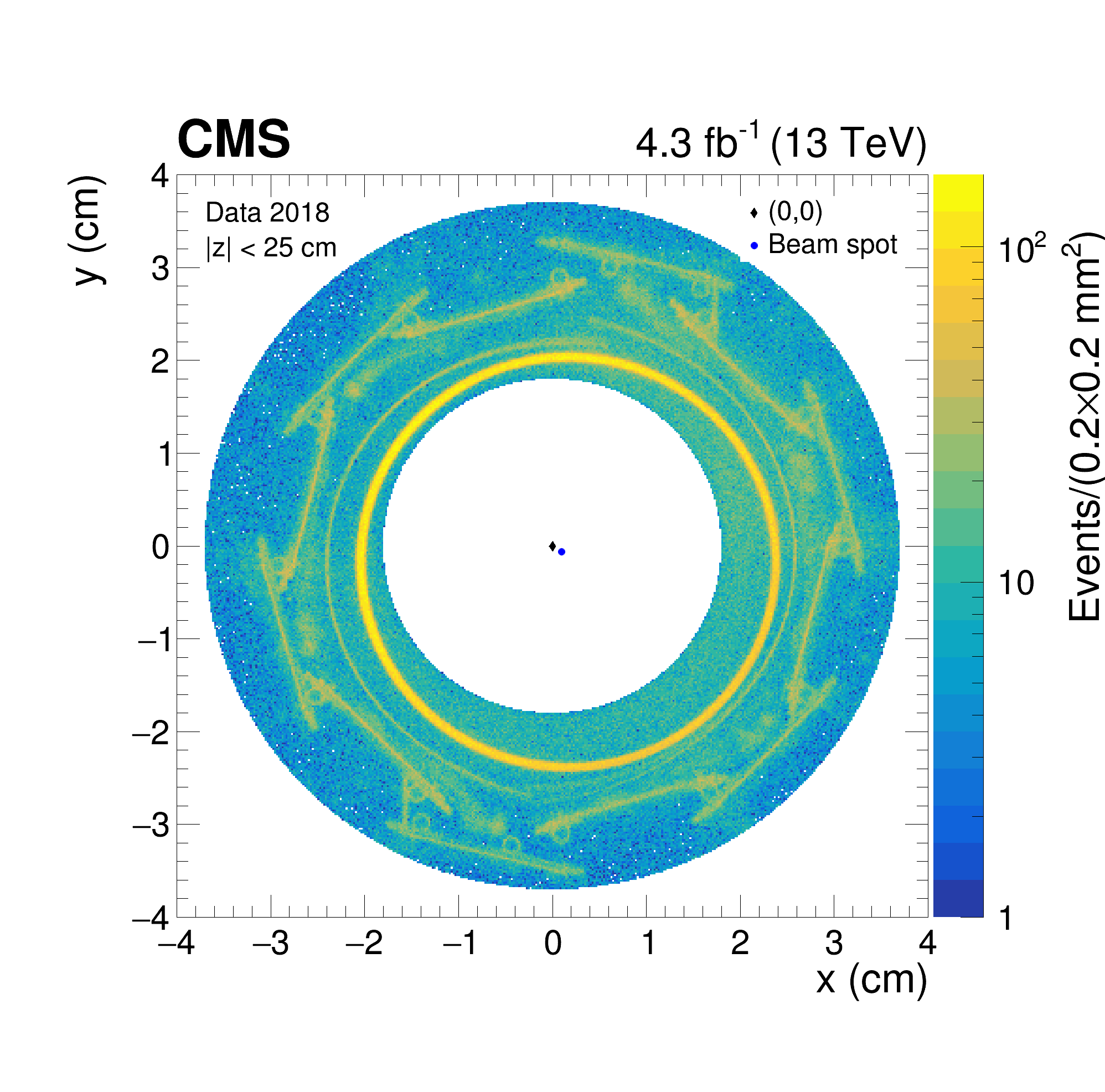}
\caption{
    (Left) vertices from nuclear interactions of hadrons reconstructed in the innermost part of the CMS tracking system in the $x$-$y$ plane in the barrel region (\AbsZbarrel). The density of vertices is indicated by the color scale. The signatures of the beam pipe, the \bpix detector with its support, and the first barrel layer of the silicon strip tracking detector (Tracker Inner Barrel, TIB)~\cite{cmspaper} can be seen. (Right) close up of the innermost region, showing the beam pipe, the inner shield and \bpix L1.}
\label{fig:TrackerHadrography}
\end{figure}

The position of the detector layers has been measured, relative to the inactive parts, using reconstructed vertices of nuclear interactions of hadrons from proton-proton collisions with the detector material. The analysis uses 4.3\,\fbinv of proton-proton collision data recorded in 2018 at a center-of-mass energy of 13\,\TeV. It follows the procedure described in Ref.~\cite{Sirunyan:2018icq}. The density of vertices from nuclear interactions, reconstructed in the \bpix detector (\AbsZbarrel) and projected onto the $x$-$y$ plane, is shown in Fig.~\ref{fig:TrackerHadrography}. The data were also used to determine the position of the beam pipe. The center of the beam pipe was found to be at $x=1.71\pm0.03$\,\mm and $y=-1.76\pm0.03$\,\mm, consistent with expectations from a geometrical survey. Figure~\ref{fig:TrackerHadrography} (right) illustrates the tight space between the beam pipe and \bpix L1 as well as the accurately fitting closure between the two detector halves in the overlap region.

\section{Detector calibration}
\label{s:calib}

The testing and calibration of the \cmsph after installation started as soon as the power cables, optical fibers, and cooling lines were connected and the cooling was available. The detector calibrations have been performed at the nominal operating temperature of $-22$\cel in March and April 2017, before the pixel detector took cosmic-ray data together with all other CMS subsystems. During the course of collision data-taking, the detector is usually recalibrated twice a year during technical stops of the LHC, to mitigate radiation effects and to optimize the detector performance. 


The calibration procedure includes the adjustment of phases for programming and readout signals, the determination and equalization of pixel charge thresholds, the calibration of pixel charge measurements, as well as the masking of non-working channels. The procedure is similar to the one used for the original detector~\cite{Chatrchyan:2009aa}, except for the phase tuning of the newly introduced digital readout. The results of the module testing during production (Sec.~\ref{s:modules}) have been used as a starting point in the calibration procedure.

\subsection{Adjustment of programming phase}

%

In order to establish communication with the detector modules, the delays in the programming lines have to be adjusted. The phase of each transmission line can be independently programmed in steps of 0.5\,ns from 0 to 25\,ns using the corresponding register in the DELAY25 chip. Programming signals can only be decoded by the TBM if they are in phase with the clock signal, so that a start/stop signal is recognized. The working region for sending programming data is common to all modules in a readout group and is determined by performing a delay scan, changing the phase of the programming data with respect to the clock. The region of correct phase is about 6-8\,ns wide and a working point in the middle of the valid region is chosen.

\subsection{Adjustment of POH laser bias current}

\begin{figure}[tb!]
  \centering
    \includegraphics[trim=0 0 0 0,clip,width=0.6\textwidth]{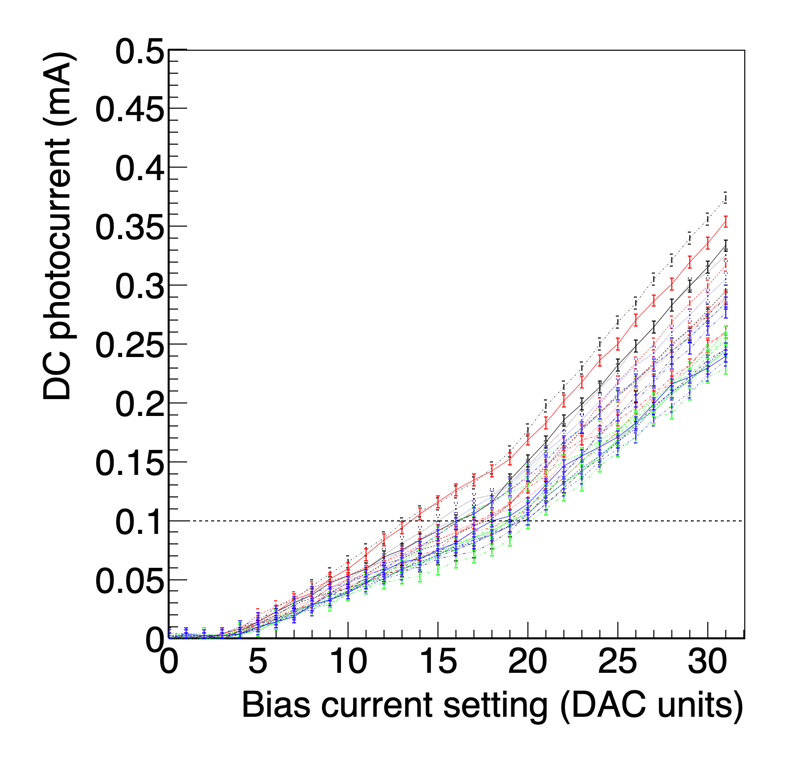}
    \caption{DC photocurrent measured at the FED receiver as a function of the setting of the laser bias current on the POH. The measured values for one FED (24 optical links) are shown. Each curve corresponds to one optical link. The bias current is programmable through the \itwoc interface in steps of 0.45\,mA per setting. The setting is adjusted such that the output current is above 0.1\,mA (indicated by the horizontal line). The spread between optical links is caused by differences in the quality of the optical connection.}
    \label{fig:pohbias}
\end{figure}

The bias current and the gain of the four or seven lasers of each POH can be adjusted individually. The setting of the bias current is scanned, keeping the gain at a constant value. A higher setting leads to a higher current, which translates into an increased optical power received at the FED. The receiver modules at the FED have a diagnostic feature which allows the DC photocurrent to be measured on each optical link individually. This is used to adjust the setting for each laser such that the output current is above 0.1\,mA. An example of a scan of the POH bias current for one FED is shown in Fig.~\ref{fig:pohbias}. 

\subsection{TBM delay adjustment}

The TBM delays have to be adjusted to synchronize the ROC readout data at 160\,\mbs with the signal transmission speed of 400\,\mbs. The programming is done via 8-bit registers internal to the TBM. The following delay settings are implemented:
\begin{itemize}
\item the phases of the 160 and 400\,MHz clocks generated by the TBM PLL can be adjusted in steps of 1 and 0.4\,ns, respectively; 
\item the timing of each group of ROCs that is multiplexed on the output data stream can be adjusted in a range of 0-7\,ns in steps of 1\,ns; 
\item the token passed from the TBM to a group of ROCs can be delayed by 6.25\,ns;
\item the header/trailer sent by the TBM can be delayed by 6.25\,ns.
\end{itemize}

The latter two delay settings are set to the values obtained from module testing. The setting for the TBM PLL phases and the timing of the groups of ROCs are adjusted for each module. The optimal values are obtained by scanning the delay setting and verifying the integrity of the  readout data package received by the FED. The set value is chosen in the center of the working region and is found to be stable over the course of the data-taking period.

\subsection{Threshold adjustment and noise measurement}
\label{sec:threshold}

The lowest possible value that can be chosen for the threshold of a given pixel is an important performance parameter, since thresholds directly influence the position resolution. The lower the threshold, the more the effects of charge sharing can be used, resulting in a better position resolution. However, if thresholds are too low, pixel noise starts saturating the double-column readout and the ROCs stop acquiring data. 

The strategy adopted is to adjust each ROC to its lowest possible threshold, which results in a non-uniform threshold distribution
between different ROCs (with a standard deviation of about 150-200\,\e). Within ROCs, pixels are trimmed to a common threshold using
the available 3-bit adjustment in each pixel (resulting in an average standard deviation per ROC of about 75-100\,\e).  
More details are given in Ref.~\cite{Chatrchyan:2009aa}. Note that the ROCs in \bpix \lone had to be adjusted to thresholds higher than
the possible minimum due to the larger cross-talk noise; this will be discussed below.

 For the pixel threshold adjustment, the values obtained during module testing for the DAC settings of the ROCs and trim bits for the pixels were used as initial values. The timing of the charge injection is adjusted by performing a scan over the \caldel DAC.

The threshold and the noise are determined by recording the S-curves (as described in Sec.~\ref{s:testing}) for a number of pixels per ROC. A subset of 81 pixels per ROC was found to be sufficient to determine the average threshold and noise values per ROC. After the initial adjustment of the pixel thresholds, periodic re-adjustments are needed. Radiation influences various parameters of the ROC, which results in threshold shifts. The pixel thresholds are closely monitored over the course of the data-taking period, and re-adjusted if necessary.
During the technical stops in 2017 and at the beginning of 2018, the thresholds have been re-optimized.
The \bpix L2-4 thresholds stayed roughly constant at about 1400\,\e, similarly for the \fpix detector at about 1500\,\e.
For \bpix \lone, because of the higher noise mentioned in Sec.~\ref{s:roc}, the thresholds had to be higher,
about 2600\,\e in 2017. After optimization in 2018, the L1 thresholds were about 2200\,\e.

With these pixel thresholds the number of noise hits was very low, below 10 pixels per bunch crossing per layer, resulting in a per pixel noise hit probability of less than $10^{-6}$. Individual pixels that showed a hit probability exceeding 0.1\% were masked during operation. The total fraction of masked pixels was less than 0.01\%. 

The time-walk effect is small for the \psidig ROC, about 300\,\e (see Sec.~\ref{s:roc}). Therefore, the in-time thresholds of
the pixels in the outer \bpix layers and the \fpix detector is about 1700\,\e and 1800\,\e, respectively. For the \proc,
the time walk is larger, about 1300\,\e, therefore the in-time threshold for \lone is about 3500\,\e.


A better understanding of the detector and software improvements developed during the two-year long data-taking period eventually showed that \bpix L1 could be operated at lower thresholds. The main improvement was the implementation of a modified programming sequence of the \proc, which left analog busses inside double-columns in a low voltage state, making them less sensitive to cross-talk during data taking. A test was made toward the end of the collision data-taking period in 2018, where the thresholds in \bpix L1 were reduced by 700\,\e without any negative effect. 
In addition, the revised ROC to be used in the new \lone will have significantly less time-walk and lower cross-talk noise (as discussed in Sec.~\ref{s:roc}). The expected threshold, noise and time-walk behavior of the revised \proc version is similar to the one of the \psidig. This improvement suggests that for Run~3 the \bpix L1 can be operated with significantly lower thresholds than before.

\subsection{Pixel pulse height calibration}
\label{gaincalibration}

The most probable value of energy deposition for normally incident minimum-ionizing particles (MIPs) in a silicon sensor with a thickness of 285\,\mum corresponds to approximately 21000\,electrons. This charge is frequently spread over more than one pixel due to Lorentz drift and diffusion of collected electrons, leading to charge clusters. An exact inter-calibration of the pixel charge measurement is crucial for a precise position resolution since the hit position is interpolated from the charge information of all pixels in a cluster. 

\begin{figure}[tb!]
  \centering
   \begin{tabular}{cc}
   \includegraphics[height=7cm]{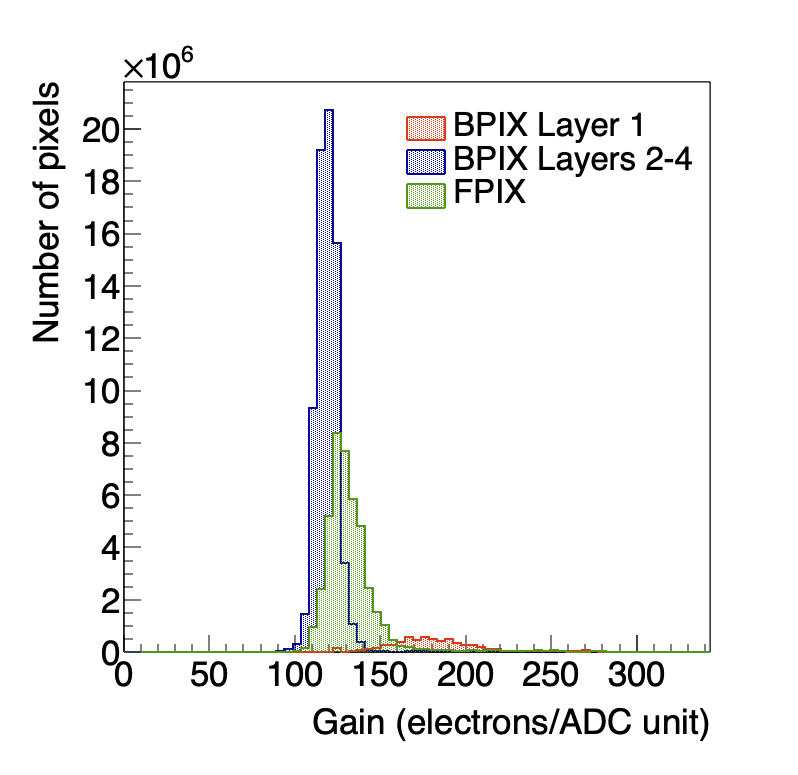} 
   \includegraphics[height=7cm]{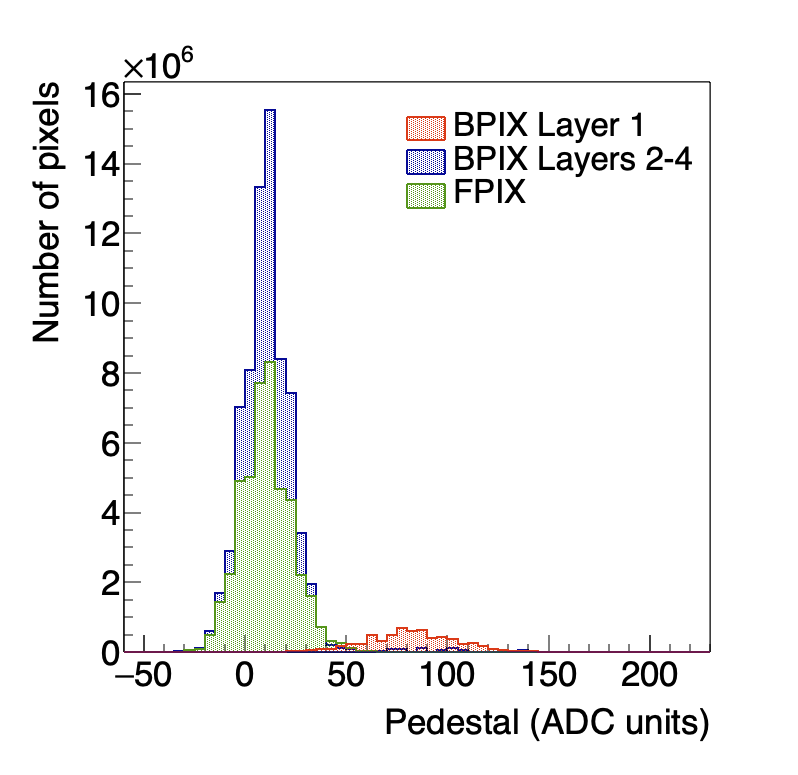} 
    \end{tabular}
    \caption{Distribution of gain (left) and pedestal (right) for all pixels. The distributions for L1 differ from the other layers because of the different design of the double column in the \proc.}
    \label{fig:adctoe} 
\end{figure}

The conversion of pixel charge measurements from units of ADC counts to units of electrons requires the calibration of the pixel response function.  The calibration is performed by injecting signals with increasing amplitude to each pixel and measuring the recorded pulse height, as described in Sec.~\ref{s:testing}. The pulse height curve is approximately linear below saturation, which happens at about 45000 electrons, and is parameterized by the slope (gain) and offset (pedestal) of a linear fit. The initial values of the \phoffset and \phscale DACs, which adjust the pulse height gain and pedestal for each ROC, were obtained from module testing and retuned at the beginning of data-taking to optimally center the pulse height within the ADC range.
Due to radiation, the optimal signal settings change with time and need to be re-adjusted about once a year.
The gain and pedestal distributions measured in the detector are shown in Fig.~\ref{fig:adctoe}. The pedestal variations have a stronger influence on the hit resolution than the gain variations. This is due to the fact that they can vary by hundreds of electrons while the slope variation is smaller and is also smeared out by the Landau fluctuations. Therefore, for hit charge reconstruction, the individual value of the pedestal per pixel is used, while only an average gain is used for all pixels in a  double column. In the offline reconstruction, the conversion of the internally injected charge from \vcal DAC units to electrons is done using values averaged over all ROCs obtained during module testing, as discussed in Sec.~\ref{s:modules}.

\section{Operation and performance}
\label{s:ops}

\subsection{Detector working fraction}

The \cmsph was exposed to the first stable proton-proton collisions in May 2017. Initially the detector working fraction, i.e. the fraction of working ROCs, was 98.4\% for the \bpix detector and 96.1\% for the \fpix detector, where the main cause of failures in the \fpix detector was due to an issue with the clock distribution in one sector. Further modules had to be masked in the \bpix detector due to issues in individual readout or power groups caused by connector failures; this decreased the \bpix working fraction to 95.0\%. 
However, towards the end of 2017, the fraction of non-working modules was dominated by the failures of \dcdc converters described in Sec.~\ref{sec:powerops},
resulting in a working fraction of 90.9\% and 85.0\% for the \bpix and \fpix detectors, respectively. 

During the consolidation work in the 2017/2018 LHC year-end technical stop, all infrastructure failures were repaired and all \dcdc converters were replaced. Subsequently, the functionality of all detector components was verified and it was found that damage had occurred in modules connected to broken \dcdc converters during the operation in 2017. In the pixel detector system, the granularity of the low voltage supply for the ROCs is finer than the granularity of the high voltage supply for the silicon sensor bias. Therefore, modules connected to broken \dcdc converters were operated with high voltage on, but low voltage off. When the low voltage is off, the pixel leakage current can no longer flow through the preamplifier structure of the ROC. The current will instead flow through the sensor bias grid, building up a voltage difference at the input of the pixel preamplifier which can cause damage. The extent of damage depends on the amount of leakage current (and thus the pixel detector module position) and the time operated without low voltage. All affected modules that were accessible without disassembling the detector layers were replaced (6 out of 8 damaged modules in \bpix L1). The modules that were severely damaged and could not be replaced stayed disabled for data taking during 2018. This led to a hit inefficiency of the pixel detector of about 1\%.

The 2018 data taking started with a working \bpix detector fraction of about 98\%, but decreased to 93.5\%
due to faulty connections between PCBs (adapter board and extension board) in two sectors, leading to three \dcdc converters being disconnected. The \fpix detector working fraction was stable through the whole year
at 96.7\%.


\subsection{Detector time and space alignment with first collisions}
\label{s:phase}

\begin{figure}[!tb]
  \begin{centering}
  \includegraphics[height=6.5cm,angle=0]{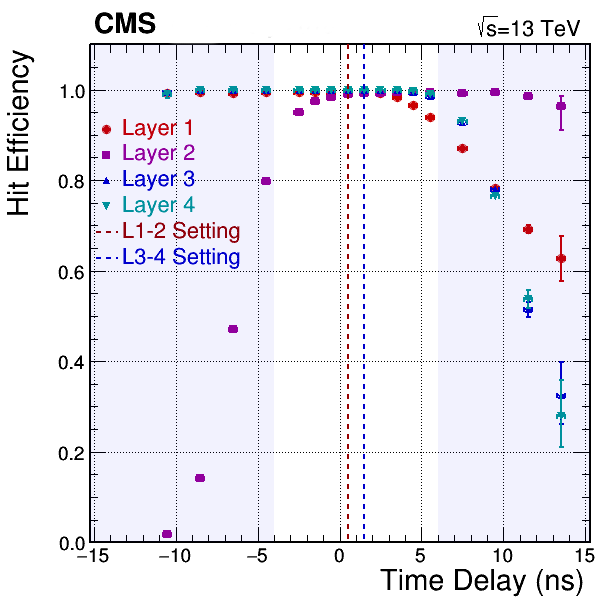}
  \includegraphics[height=6.5cm,angle=0]{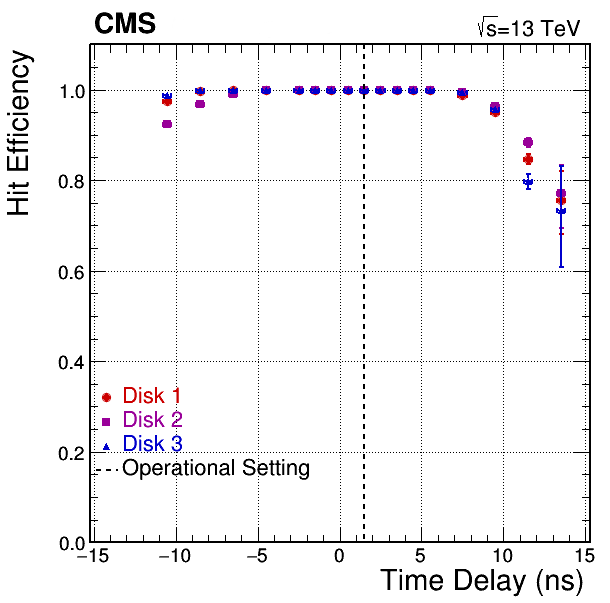}
   \caption{Efficiency of \bpix layers (left) and \fpix disks (right) as a function of the time delay with respect to the LHC clock. The vertical dashed lines indicate the chosen settings. The timing of all \bpix layers is changed simultaneously. If one layer is inefficient, the measurement of the hit efficiency is affected by a large systematic uncertainty. The interval of validity is indicated by the white region between the shaded bands.}
   \label{fig:Fig8-1}
  \end{centering}
\end{figure}

Before high-quality data for physics analyses could be taken, additional calibrations, using tracks from proton-proton collisions, had to
be carried out. First, a timing scan was performed. This procedure adjusts the
phase of the LHC clock to obtain the maximum hit recording efficiency in the pixel detector ROCs.
It consists of varying the clock phase and finding the range of the full efficiency plateau (Fig.~\ref{fig:Fig8-1}).
A  working point is then established within this plateau but close to the edge (the one which corresponds to low signals)
in order to optimize the readout for the small-amplitude signals affected by time walk.
This procedure has worked very well for the \fpix detector and \bpix \lthree and \lfour. Problems were encountered however for \bpix L1 and L2. It was found that the two versions of the ROCs (\proc and \psidig) have an internal timing difference of about 12\,ns.
This difference cannot be compensated for because both layers receive the clock through the
same control links, and thus use the same phase adjustment. Despite the shift, a
common working point could be found at the overlap of the two timing distributions. This is
illustrated in Fig.~\ref{fig:Fig8-1} (left), where the plateaus for \lone and \ltwo are shown together with the plateaus for L3 and L4.
The functionality of the phase adjustment added in the TBM to be used for the \lone replacement in Run~3 (described in Sec.~\ref{s:tbm}) will allow to correct this feature.  

For a proper measurement of the trajectory and the momentum of charged particles, the actual position of the sensors has to be known with an accuracy of 10\,\mum or better. The procedure to determine the sensor positions, orientations, and surface deformations (referred to as "alignment") relies on cosmic-ray and proton-proton collision data and is described in Ref.~\cite{Chatrchyan:2014wfa}. While an initial alignment of the large detector structures was obtained using tracks of cosmic muons collected before the start of collision data-taking in 2017, the final, module level, alignment required the use of collision tracks originating from the interaction point and reached a statistical accuracy of the module positions of a few micrometers~\cite{ref:alignment}.
A complete module-level alignment for the pixel detector requires of the order of 10 million tracks which corresponds to about one week of collision data.
The alignment procedure was repeated several times during the data taking periods in 2017 and 2018. 



%

\subsection{SEU recovery mechanisms during operation}

SEU cross sections have been measured for the 250\,nm CMOS process used for the pixel detector
ROCs and TBMs~\cite{Kastli:2005jj}. The effect of SEUs on the operation of the pixel detector has been evaluated
and the needed recovery procedures have been put in place~\cite{Kastli:2005jj}. Most of the SEU effects go unnoticed since they
alter the detector response in a very minor way. However, with some probability, vital circuits
will be affected, which causes interruption of the data readout of the corresponding channel. In such cases one of two recovery procedures is initiated. In the first procedure, which requires less time, a reset is issued to all
readout chips, which clears and re-enables the affected circuits. If a memory cell is corrupted,
a reset is not sufficient and instead a full reprogramming of all cells has to be performed. This
recovery procedure takes more time (about 30 seconds) and is performed less frequently. Typically, at the full LHC luminosity,
such a recovery is performed about once per hour.

In addition, in order to recover modules affected by SEUs in the TBMs that cannot be reset (Sec.~\ref{s:tbm}), parts of the pixel detector have to be periodically power-cycled. In the 2017 data taking period the power cycles were mainly done during collisions. Once the number of TBM cores getting stuck reached a programmable threshold, the data taking was interrupted, the power cycling was performed by disabling and enabling the \dcdc converters corresponding to the power group with the blocked module, and the affected components were reconfigured. To avoid any damage to the \dcdc converters (Sec.~\ref{sec:powerops}), power cycles were done only during the LHC non-collision periods in 2018. This was achieved by switching off and on the corresponding power supply channels, which took approximately one minute. This simplified the procedure, but also meant that a larger fraction of the detector was affected during data taking (up to 10\% of modules in L1 during an LHC fill).

\subsection{Detector response and performance monitoring}

\subsubsection{Charge measurement} 

In the reconstruction step, hit pixels are combined to form clusters from neighboring pixels~\cite{Chatrchyan:2014fea}. The charge measured within the cluster corresponds to the charge deposited by a single charged particle.
Assuming that most clusters are due to MIPs, the charge is distributed according to a Landau-like function. The shape of the Landau distribution is modified by the presence of noise, out-of-time
hits, hit inefficiencies, and beam-background events. In addition, the cluster charge is affected
by radiation effects in the sensors (such as charge trapping and changes of the
depletion voltage or LA) and in the ROCs (such as voltage drifts).
Periodic recalibration and an increase of the sensor bias voltages can correct for some of these effects, but
not all. Eventually, especially at the end of the lifetime of the detector, the measured charge will
be strongly affected by the reduced charge collection efficiency after irradiation.

\begin{figure}[!tb]
  \begin{centering}
\includegraphics[trim=0 2 0 0,clip,height=7cm]{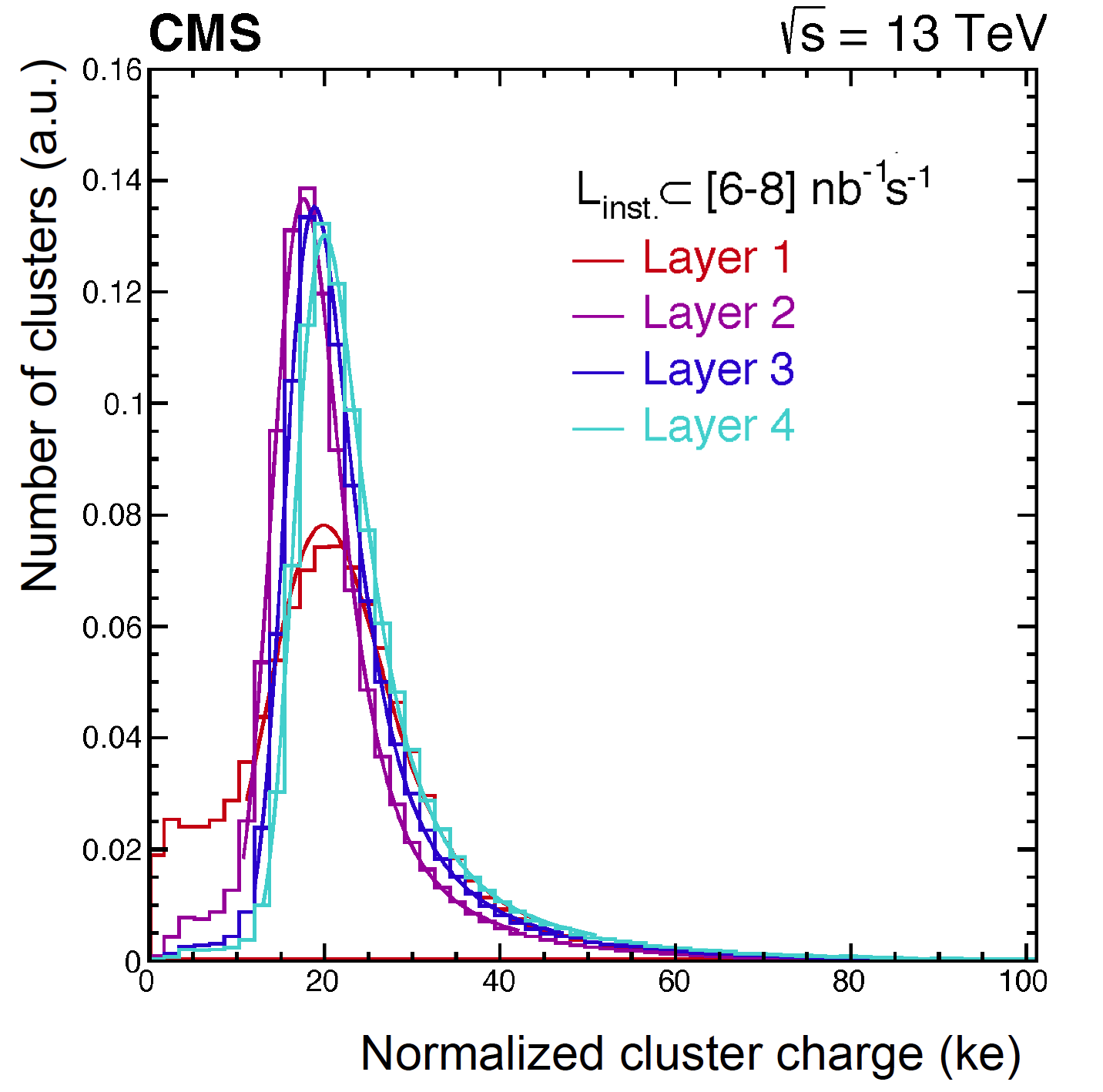} 
\includegraphics[trim=0 2 0 0,clip,height=7cm]{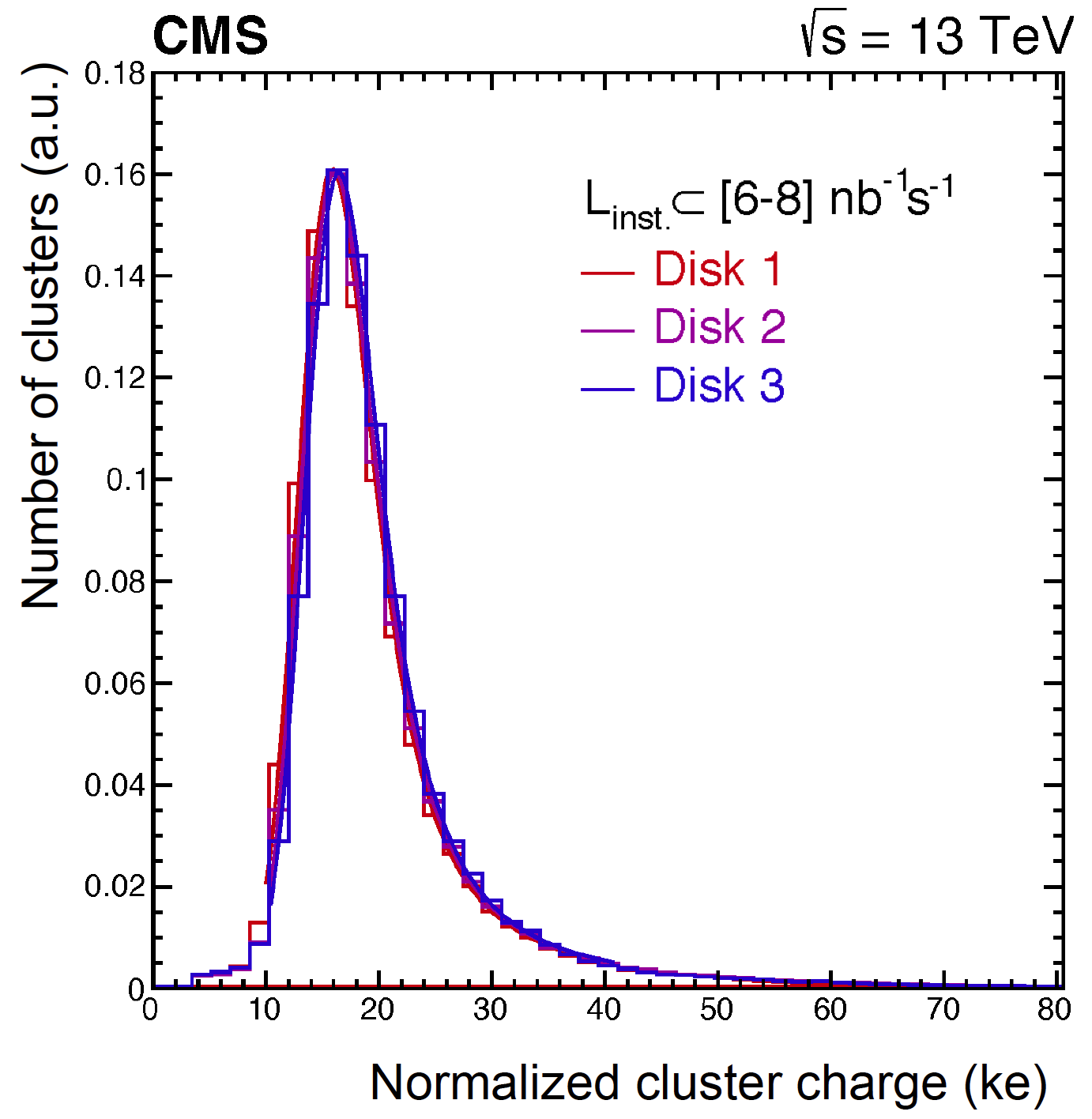} 
    \caption{Cluster charge for the four \bpix layers (left) and the three \fpix disks (right). The data correspond to an integrated luminosity of about 75\,\fbinv. The distributions are fit with a Landau function convolved with a Gaussian function. The increased number of clusters with low charge in \bpix L1 is due to split clusters caused by inefficiencies of the \proc. The difference in the most probable value of the Landau distributions for the four \bpix layers are explained by radiation effects.}
   \label{fig:Fig8-3}
  \end{centering}
\end{figure}

Figure~\ref{fig:Fig8-3} shows the measured charge distributions for the \bpix and \fpix detectors for hits associated with reconstructed tracks.
For \lone the shape of the charge distribution differs from the other layers. This is due to the higher thresholds and the additional contribution of low charge clusters resulting from split clusters due to the \proc inefficiencies.
The charge loss caused by irradiation can be partly recovered by increasing the bias voltage. The bias voltages during the 2017-2018 data taking were first set to 100\,V and were later increased to 450\,V for L1, 300\,V for L2, 250\,V for L3 and L4, and 350\,V and 300\,V for \fpix ring 1 and ring 2 modules, respectively. 

Beyond a certain irradiation level, full charge collection cannot be recovered anymore, leading to a degraded position resolution. This is partly recovered by a proper modeling of the radiation effects in the offline reconstruction~\cite{Chatrchyan:2014fea}.

\subsubsection{Lorentz angle} 
\label{s:la}


The value of the LA (introduced in Sec.~\ref{s:sensor}) affects the position resolution of the pixel hits.
The LA depends strongly on the bias voltage of the sensor and is also affected by radiation damage. Thus, the LA has to be regularly monitored.

The measured LA is used in the data reconstruction process to obtain an optimal position resolution and to minimize a potential bias in the hit reconstruction. Tracks with shallow incident angles to the sensor surface are used  in order to measure the charge drift
in large pixel clusters. The method to measure the LA is described in Ref.~\cite{lorentzangle}. Some residual effects caused by the radiation induced changes of the LA are absorbed by the detector alignment procedure.

\begin{figure}[!tb]
  \begin{centering}
  \includegraphics[height=7cm]{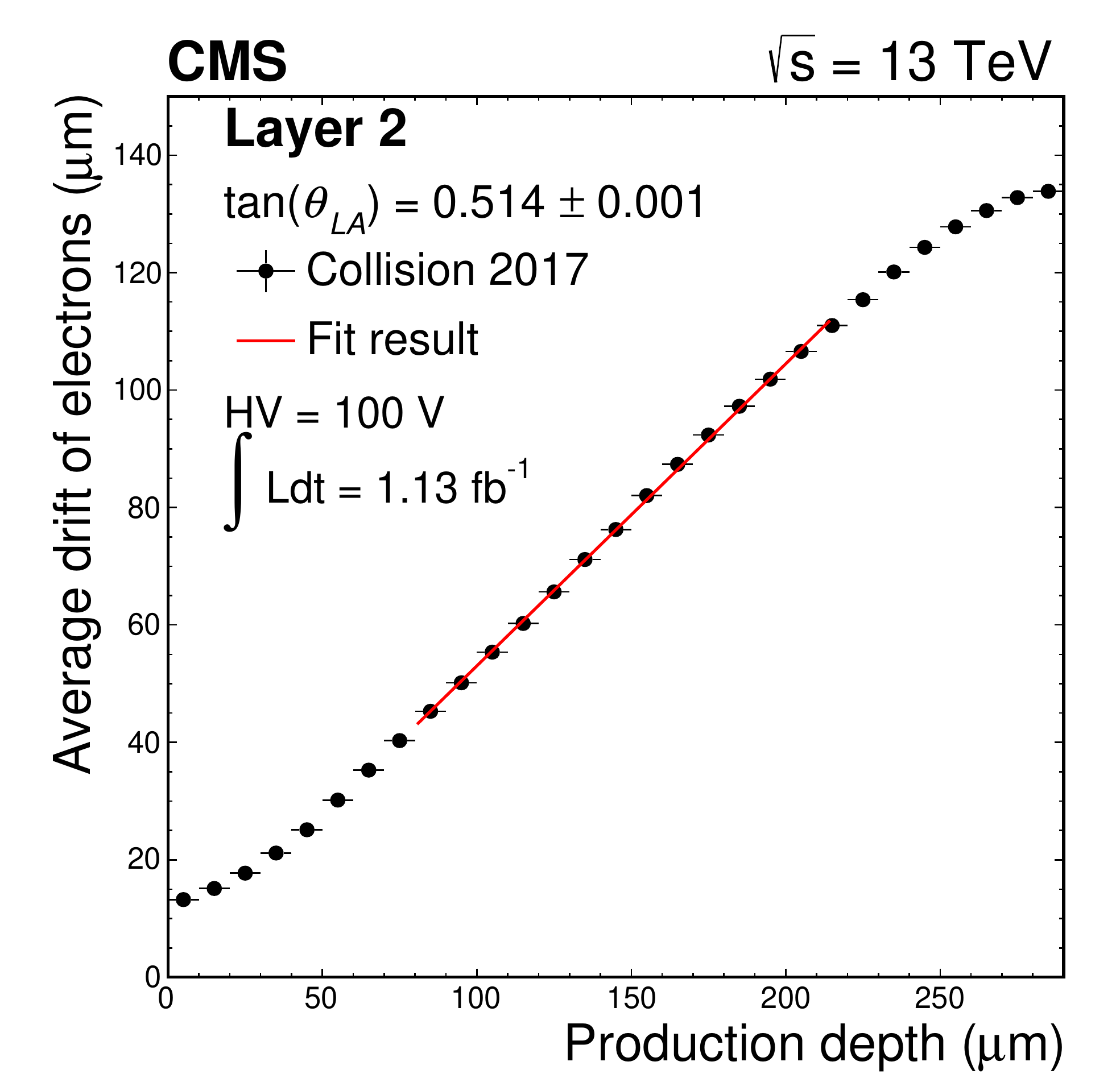}
   \caption{Charge drift as a function of the depth at which the charge was generated. The plot is for modules in \bpix L2 at the nominal CMS B-field of 3.8\,T. The slope of the linear fit is interpreted as the tangent of the Lorentz angle.}
   \label{fig:Fig8-5}
  \end{centering}
\end{figure}

Figure~\ref{fig:Fig8-5} illustrates the method to measure the LA in the \bpix detector. It shows the average charge
drift as a function of its production depth\footnote{The production depth is defined as the depth in the sensor at which the charge was generated by the traversing charged particle.} and reflects the value of the LA (see Ref.~\cite{lorentzangle} for details).
A linear function is fitted to the central, linear,  part of the distribution where the LA is constant.
The average slope of the charge drift versus generation depth is interpreted as the tangent of the LA. The LA values measured at the beginning of the 2017 data taking are shown in Tab.~\ref{tab:LA}. The larger value of the LA for \bpix L1 is due to the radiation effects accumulated during the first two months of data taking.

\begin{table}
  \begin{center}
    \caption{Values for the tangent of the Lorentz angle, $\tan(\theta_{LA})$, at the beginning of 2017 for the \bpix layers. All layers were operated at a bias voltage of 100\,V at this point. The uncertainties shown in the table are only statistical. The systematic uncertainties are larger and are estimated to be about 1\%.}
    \label{tab:LA}
   \vspace{0.1in}
   \begin{tabular}{l | c |   c |   c | c}
     \hline
   & {\bpix L1}  & {\bpix L2} & {\bpix L3} & {\bpix L4}  \\
   \hline
   \hline
   $\tan(\theta_{LA})$ &   $0.532\pm0.001$    &  $0.514\pm0.001$     &  $0.520\pm0.001$    &  $0.508\pm0.002$  \\
      
    \hline
    \hline
      \end{tabular}
  \end{center}
\end{table}


\subsubsection{Detector hit efficiency} 
\label{sec:hiteff}

The hit efficiency is one of the two most important performance parameters for the pixel detector.
High cluster hit-detection efficiency is mandatory in order to maintain good track pattern recognition
and b-tagging performance~\cite{Chatrchyan:2014fea, Sirunyan:2017ezt}. Therefore, the hit efficiency is carefully monitored by measuring the probability that a cluster is found
in a pixel detector layer to which a reconstructed track has been extrapolated. The search window for matching hits
is set to a rather large value of 1\,mm in order to include hits with poor position resolution. Poor position resolution
can be due to various effects, like the presence of delta-rays, single dead or inefficient pixels, or
sensor edge effects. 

One of the dominant effects that affects the cluster hit efficiency measurement is the ROC readout
inefficiency that is explained in Sec.~\ref{s:roc}.
This effect can lead to a temporary inefficiency due to loss of data in a single pixel or in the whole double column.
The loss mechanism depends strongly on the data rate, which in turn depends on the instantaneous luminosity and the
pixel detector layer position (mainly the radius). There is also a weak dependence on the trigger rate.

\begin{figure}[!tb]
  \begin{centering}  
 \includegraphics[width=.7\linewidth]{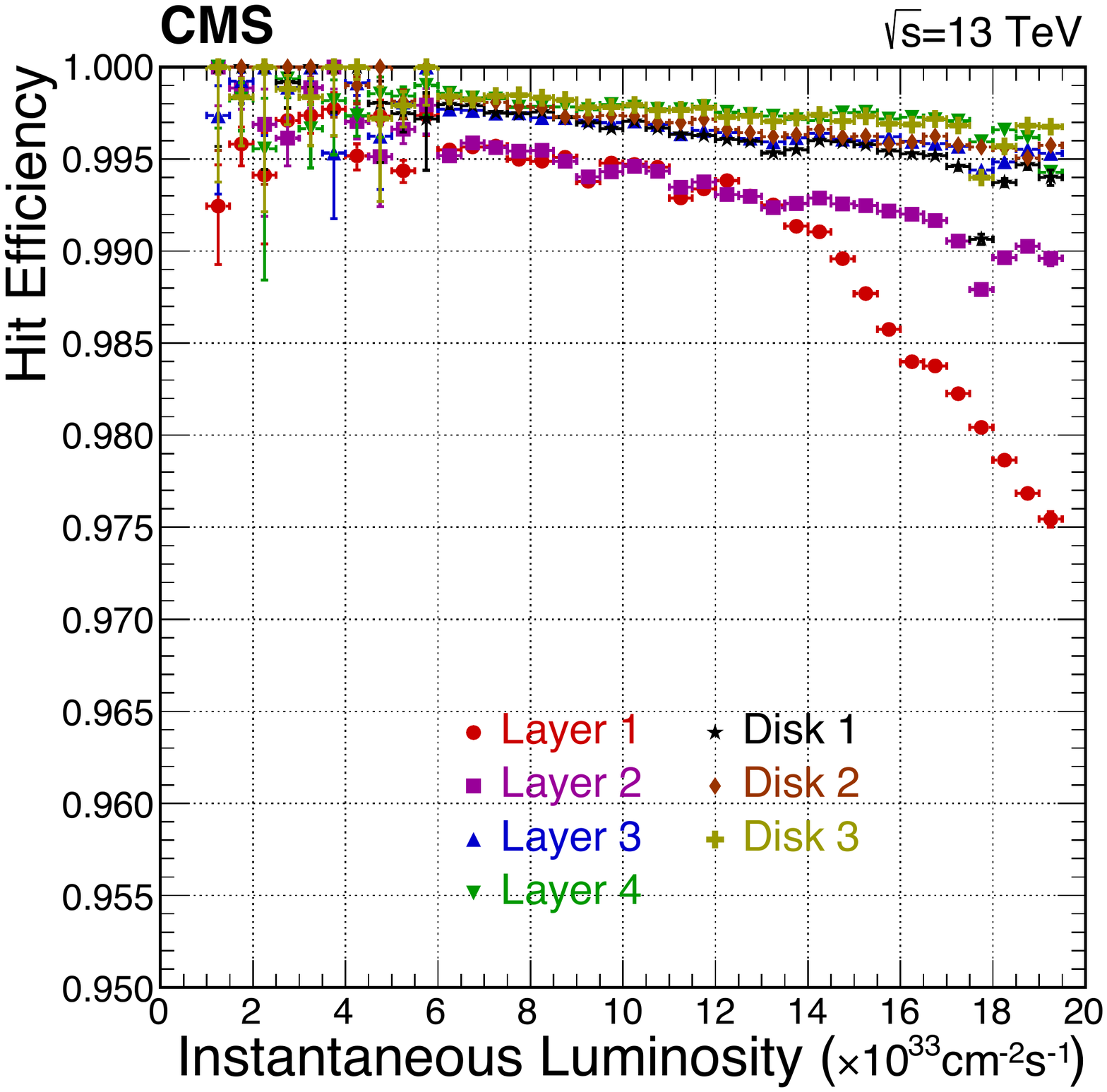}
   \caption{Cluster hit efficiency as a function of instantaneous luminosity measured using tracks from collisions in 2018.}
   \label{fig:Fig8-6}
  \end{centering}
\end{figure}

The cluster hit efficiency is measured by propagating  a track to the detector plane and checking if a hit is found on it. Figure~\ref{fig:Fig8-6} shows the cluster hit efficiency measured as a function of instantaneous luminosity, split into different \bpix layers and \fpix disks. Even at an instantaneous luminosity of $\rm 2\times10^{34}$\,\percms, the cluster hit efficiency for \bpix L2-L4 and the \fpix detector is larger than 99\%. Despite the low-rate inefficiency discussed in Sec.~\ref{sec:roc}, the cluster hit efficiency in L1 is well above 99\% for low instantaneous luminosities due to the fact that periodic resets are sent during operation. The hit efficiency in L1 drops at luminosities above $\rm1.4\times10^{34}$\,\percms, reaching a value of 97.5\% at $\rm 2\times10^{34}$\,\percms. Note that the efficiency using the \proc is still far higher than what it would have been when using the \psidig chip also in L1, and it will be further improved when using the revised version of the \proc in the replacement of L1 to be installed during LS2. 

\subsubsection{Position resolution} 
\label{s:pos}

Together with the  efficiency, the position resolution is the most important performance pa\-ra\-me\-ter of the
detector. The position resolution is sensitive to many effects and must be regularly measured. Several aspects of the CMS data
analysis depend directly on the position resolution, notably b-tagging, secondary vertex resolution and
primary vertex identification~\cite{Chatrchyan:2014fea, Sirunyan:2017ezt}.

The position resolution is monitored using the ``triplet'' method~\cite{triplet}. In this method, the tracks measured by the whole tracker detector are used together with the hits in two pixel detector layers to predict the hit position in a third layer. Whenever possible, this third layer is located in between the other two and the hit position is interpolated. For outer layers, extrapolation from neighboring layers is used. The interpolated/extrapolated position is then compared with the measured hit position and residuals are calculated. The residual distribution is fitted using a Student's t function in order to estimate its width.
The width of the residual distribution is closely related to the hit position resolution (sometimes called "point resolution")
of the layer under study, but also includes contributions from detector alignment and the accuracy of the measurement of the hits in other layers used for the track interpolation.

\begin{figure}[!tb]
  \begin{centering}
  \includegraphics[height=7cm]{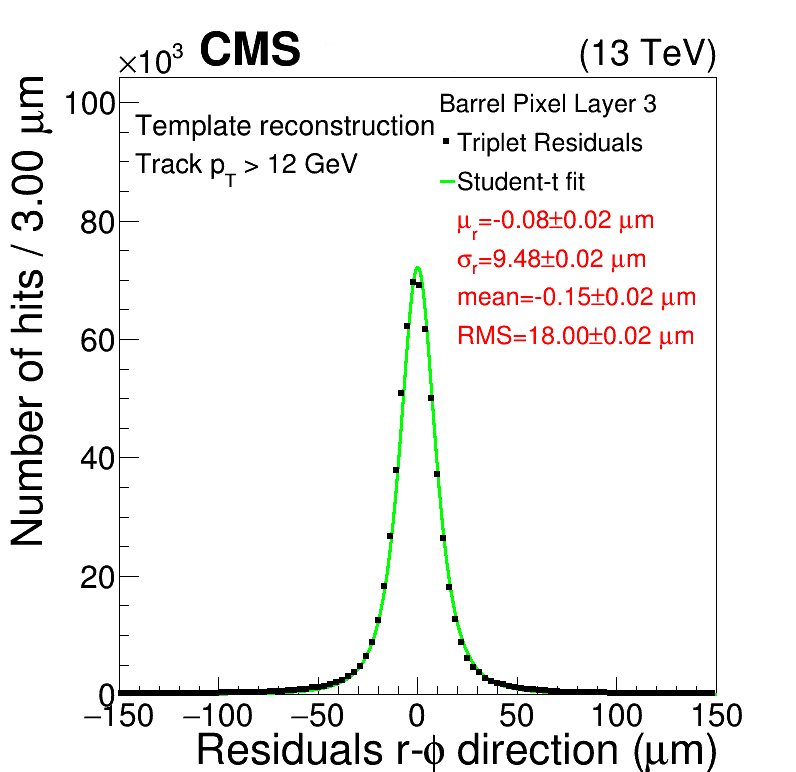}
  \includegraphics[height=7cm]{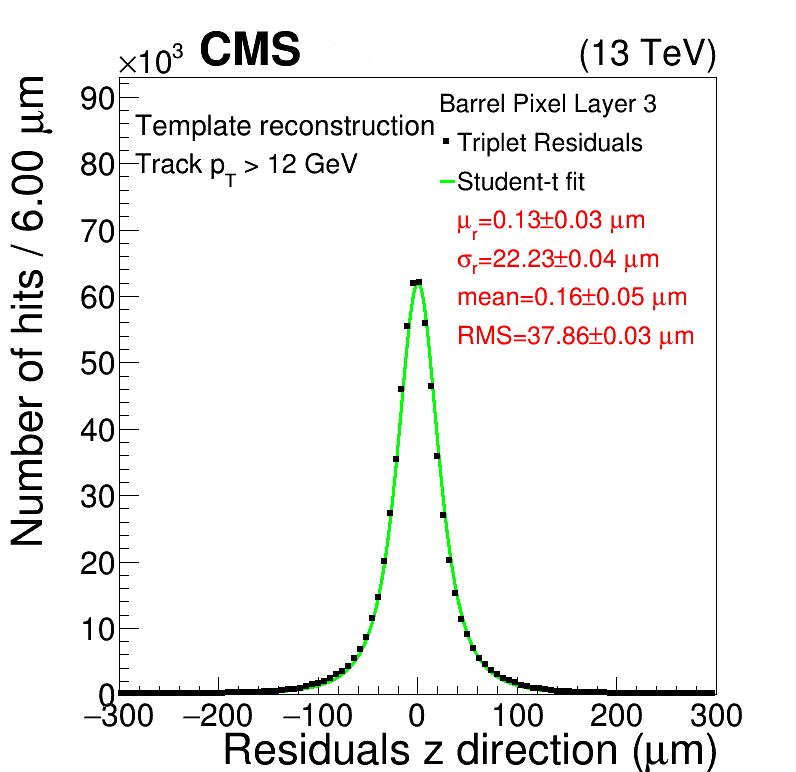}
  \caption{Pixel hit residuals for \bpix L3 in $r\phi$ (left) and $z$ direction (right). The data correspond to an integrated luminosity of about 75\,\fbinv.}
   \label{fig:Fig8-7}
  \end{centering}
\end{figure}

The resolution is measured separately for the two directions along the pixels, $r\phi$ and $z$ in the \bpix detector and $r$ and $\phi$ in the \fpix detector.
The measured value depends on the track angle and on the radial position of the layer.
Figure~\ref{fig:Fig8-7} shows an example of the fitted residuals for \bpix \lthree. 
For \bpix \lthree the fitted width of the residual distribution is 9.5\,\mum in the $r\phi$ direction and 22.2\,\mum in the $z$ direction.
This is close to the result obtained from simulation, 12.2\,\mum and 24.1\,\mum, respectively, which shows that the position
resolution is correctly modeled.
The width of the residual distribution is used to monitor the position resolution over time.


\section{Summary}
\label{s:summary}
The original silicon pixel detector of the CMS experiment at the CERN LHC has been replaced with an upgraded pixel detector during the extended year-end technical stop of the LHC in 2016/2017. The \cmsph features an additional barrel layer and additional forward disks. It has four-hit coverage up to an absolute value of pseudorapidity of $|\eta|$=2.5, improved pattern recognition and track reconstruction, but also added redundancy to mitigate hit losses. The upgraded system is able to cope with significantly higher data rates because of the new readout chip design and the increased bandwidth of the digital data transmission via optical links. The \cmsph has been successfully taking data in 2017 and 2018 and, together with the other CMS subsystems, has delivered high-quality physics data corresponding to an integrated luminosity of more than 100\,\fbinv. During LS2 the innermost barrel layer of the \cmsph will be replaced in order to maintain efficient and robust tracking in CMS until the end of Run~3.

\newpage
\section{Glossary of special terms and acronyms}
\label{chap:glossary}

\begin{footnotesize}%

\begin{tabbing}
\hspace*{2.2cm}\=\hspace*{5cm} \kill
ADC:\>Analog-to-digital converter \\
 AMS:\> Alpha Magnetic Spectrometer \\
 \bpix:\> Barrel pixel detector\\
 \caldel:\> DAC setting to adjust the delay of the calibration signal in the ROC\\
 CCU:\> Communication and control unit chip, used in the auxiliary electronics\\
                       \> of the \cmsph\\
 CFK:\> Carbon fiber reinforced polymer\\
 CMS:\> Compact Muon Solenoid\\
 CNC:\>Computer numerical control\\
 D1:\> First disk of the forward pixel detector\\
 D2:\> Second disk of the forward pixel detector\\
 D3:\> Third disk of the forward pixel detector\\
 DAC:\> Digital-to-analog converter\\
 DAQ:\> Data acquisition\\
 DC:\> Direct current\\
 DELAY25:\> Delay chip, used in the auxiliary electronics of the \cmsph\\
 DLT:\> Digital level-translator chip, used in the auxiliary electronics\\
                       \> of the \cmsph\\
 DOH:\> Digital opto-hybrid, used in the auxiliary electronics of the \cmsph\\
 DTB:\> Digital test board\\
 \fpix:\> Forward pixel detector\\
 FEC:\> Front-end controller, used in the DAQ system of the \cmsph\\
 FED:\> Front-end driver, used in the DAQ system of the \cmsph\\
 FZ:\> Float zone\\
 HDI:\> High-density interconnect \\
 HL-LHC:\> High-luminosity LHC\\
 HV:\> High voltage\\
 IV:\> Current-voltage\\
 L1:\> First layer of the barrel pixel detector\\
 L2:\> Second layer of the barrel pixel detector\\
 L3:\> Third layer of the barrel pixel detector\\
 L4:\> Fourth layer of the barrel pixel detector\\
 L1A:\> CMS level-1 trigger accept signal\\
 LA:\> Lorentz angle\\
 LCDS:\> Low-current differential signal\\
 LHC:\> Large Hadron Collider\\
 LHCb VELO:\> Vertex locator subdetector of the LHCb detector at the LHC\\
 LLD:\> Linear laser driver chip, used in the auxiliary electronics of the \cmsph\\
 LS1:\> Long Shutdown 1 (2013-2014)\\
 LS2:\> Long Shutdown 2 (2019-2020)\\
 MIP:\> Minimum-ionizing particle\\
 MPO:\> Multi-fiber Push On connector for optical fibers\\
 MS cable:\> Multi-service cable\\
 NRZI:\> Non-Return to Zero Inverted, encoding scheme\\
 \phscale:\>  DAC setting to adjust the gain of the ADC in the pulse height measurement  in the ROC\\
 \phoffset:\>  DAC setting to adjust the offset of the ADC in the pulse height measurement  in the ROC\\
 2-PACL:\> Two-phase accumulator controlled loop\\
 PCB:\> Printed circuit board\\
 PEEK:\> Polyether ether ketone\\
 PIA:\> Peripheral interface adapter\\
 PxFEC:\> Pixel front-end controller, used in the DAQ system of the \cmsph\\
 PLL:\> Phase-locked loop\\
 POH:\> Pixel opto-hybrid, used in the auxiliary electronics of the \cmsph\\
 PP0:\> Patch panel 0, used for the CMS tracker detector, located at the far end of the detector\\
 PP1:\> Patch panel 1, used for the CMS tracker detector, located at the edge of the solenoid\\
 \proc:\> Name of the readout chip used in \bpix L1 of the \cmsph\\
 \psidig:\> Name of the readout chip used in \bpix L2-L4 and \fpix of the \cmsph\\
 \psiana:\> Name of the readout chip used in the original CMS pixel detector\\
 PUC:\> Pixel unit cell\\
 ROC:\> Readout chip\\
 Run 1:\> First data-taking period at the LHC (2009-2012)\\
 Run 2:\> Second data-taking period at the LHC (2015-2018)\\
 Run 3:\> Third data-taking period at the LHC (expected for 2021-2024)\\
 SEU:\> Single-event upset\\
 TBM:\> Token bit manager chip\\
 TBM08:\> Name of the TBM used in \bpix L3-L4 and \fpix of the \cmsph\\
 TBM09:\> Name of the TBM used in \bpix L2 of the \cmsph\\
 TBM10:\> Name of the TBM used in \bpix L1 of the \cmsph\\
 TCDS:\> CMS trigger control and distribution system\\
 TkFEC:\> Tracker front-end controller, used in the DAQ system of the \cmsph\\
 TPLL:\> Tracker PLL chip, used in the auxiliary electronics of the \cmsph\\
 TPG:\> Thermal pyrolytic graphite\\
 TOSA:\> Transmitter optical subassemblies\\
 QPLL:\> Quartz-based PLL chip, used in the auxiliary electronics of the \cmsph\\
 UBM:\> Under-bump metallization\\
 \vana:\> DAC setting to adjust the analog voltage regulator  in the ROC\\
 \vcal:\> DAC setting to adjust the amplitude of the calibration signal in the ROC\\
 \vthrcomp:\> DAC setting to adjust the pixel threshold globally in the ROC\\
 \vtrim:\> DAC setting to adjust the range of the trim bits in the ROC\\
\end{tabbing}
 
\end{footnotesize}


\acknowledgments
The tracker groups gratefully acknowledge financial support from the following funding agencies: BMWFW and FWF (Austria); FNRS and FWO (Belgium); CERN; MSE and CSF (Croatia); Academy of Finland, MEC, and HIP (Finland); CEA and CNRS/IN2P3 (France); BMBF, DFG, and HGF (Germany); GSRT (Greece); NKFIA K124850, and Bolyai Fellowship of the Hungarian Academy of Sciences (Hungary); DAE and DST (India); IPM (Iran); INFN (Italy); PAEC (Pakistan); SEIDI, CPAN, PCTI and FEDER (Spain); Swiss Funding Agencies (Switzerland); MST (Taipei); STFC (United Kingdom); DOE and NSF (U.S.A.). Individuals have received support from HFRI (Greece). The irradiation campaign has received funding from the European Union's Horizon 2020 Research and Innovation programme under Grant Agreement no. 654168.

\newpage
\bibliographystyle{JHEP}
\bibliography{leac_001.bib}

\newpage
\section*{Tracker Group of the CMS Collaboration}

\vspace{0.3cm}
\noindent
\textcolor{black}{\textbf{Institut~f\"{u}r~Hochenergiephysik, Wien, Austria}\\*[0pt]
W.~Adam, T.~Bergauer, D.~Bl\"{o}ch, M.~Dragicevic, R.~Fr\"{u}hwirth\cmsAuthorMark{1}, V.~Hinger, H.~Steininger}

\vspace{0.2cm}
\noindent
\textbf{Universiteit~Antwerpen, Antwerpen, Belgium}\\*[0pt]
W.~Beaumont, D.~Di~Croce, X.~Janssen, J.~Lauwers, P.~Van~Mechelen, N.~Van~Remortel

\vspace{0.2cm}
\noindent
\textcolor{black}{\textbf{Vrije~Universiteit~Brussel, Brussel, Belgium}\\*[0pt]
F.~Blekman, S.S.~Chhibra, J.~De~Clercq, J.~D'Hondt, S.~Lowette,  S.~Moortgat, A.~Morton, K.~Skovpen, A.R.~Sahasransu, E.~S{\o}rensen~Bols, P.~Van~Mulders}

\vspace{0.2cm}
\noindent
\textcolor{black}{\textbf{Universit\'{e}~Libre~de~Bruxelles, Bruxelles, Belgium}\\*[0pt]
Y.~Allard, D.~Beghin, B.~Bilin, B.~Clerbaux, G.~De~Lentdecker, H.~Delannoy, W.~Deng, L.~Favart, R.~Goldouzian, A.~Grebenyuk, A.~Kalsi, M.~Mahdavikhorrami, I.~Makarenko, L.~Moureaux, A.~Popov, N.~Postiau, F.~Robert, Z.~Song, L.~Thomas, P.~Vanlaer, D.~Vannerom, Q.~Wang, H. Wang, Y.~Yang}

\vspace{0.2cm}
\noindent
\textcolor{black}{\textbf{Universit\'{e}~Catholique~de~Louvain,~Louvain-la-Neuve,~Belgium}\\*[0pt]
O.~Bondu, G.~Bruno, C.~Caputo, P.~David, C.~Delaere, M.~Delcourt, A.~Giammanco, V.~Lemaitre, J.~Prisciandaro, A.~Saggio, N.~Szilasi, M.~Vidal~Marono, P.~Vischia, J.~Zobec}

\vspace{0.2cm}
\noindent
\textcolor{black}{\textbf{Institut Ru{\dj}er Bo\v{s}kovi\'{c}, Zagreb, Croatia}\\*[0pt]
V.~Brigljevi\'{c}, D.~Feren\v{c}ek, D.~Majumder, M.~Rogulji\'{c}, A.~Starodumov\cmsAuthorMark{2}, T.~\v{S}u\v{s}a}

\vspace{0.2cm}
\noindent
\textcolor{black}{\textbf{Department~of~Physics, University~of~Helsinki, Helsinki, Finland}\\*[0pt]
P.~Eerola}

\vspace{0.2cm}
\noindent
\textcolor{black}{
\textbf{Helsinki~Institute~of~Physics, Helsinki, Finland}\\*[0pt]
E.~Br\"{u}cken, A.~G\"adda, T.~Lamp\'{e}n, L.~Martikainen, E.~Tuominen, R.~Turpeinen, A.~Winkler}

\vspace{0.2cm}
\noindent
\textcolor{black}{\textbf{Lappeenranta-Lahti~University~of~Technology, Lappeenranta, Finland}\\*[0pt]
P.~Luukka, T.~Tuuva}

\vspace{0.2cm}
\noindent
\textcolor{black}{\textbf{Universit\'{e}~de~Strasbourg, CNRS, IPHC~UMR~7178, Strasbourg, France}\\*[0pt]
J.-L.~Agram\cmsAuthorMark{3}, J.~Andrea, D.~Bloch, C.~Bonnin, G.~Bourgatte, J.-M.~Brom, E.~Chabert, L.~Charles, C.~Collard, E.~Dangelser, U.~Goerlach, C.~Grimault, L.~Gross, M.~Krauth, N.~Ollivier-Henry, E.~Silva~Jim\'{e}nez}

\vspace{0.2cm}
\noindent
\textcolor{black}{\textbf{Universit\'{e}~de~Lyon, Universit\'{e}~Claude~Bernard~Lyon~1, CNRS/IN2P3, IP2I Lyon, UMR 5822, Villeurbanne, France}\\*[0pt]
G.~Baulieu, G.~Boudoul, L.~Caponetto, N.~Chanon, D.~Contardo,  P.~Den\'{e}, T.~Dupasquier, G.~Galbit, M.~Gouzevitch, N.~Lumb, L.~Mirabito, B.~Nodari, S.~Perries, M.~Vander~Donckt, S.~Viret}

\vspace{0.2cm}
\noindent
\textcolor{black}{\textbf{RWTH~Aachen~University, I.~Physikalisches~Institut, Aachen, Germany}\\*[0pt]
L.~Feld, C.~Fimmers, M.~Fleck, M.~Friedrichs, L.~Gromann, W.~Karpinski, K.~Klein, M.~Lipinski, P.~Malek, D.~Meuser, R.~Narr, I.~Oezen, A.~Pauls, G.~Pierschel, M.~Preuten, M.~Rauch, D.~Rittich, N.~R\"{o}wert, J.~Sammet, S.~Schmitz, J.~Schulz, M.~Teroerde, M.~Wlochal}

\vspace{0.2cm}
\noindent
\textcolor{black}{\textbf{RWTH~Aachen~University, III.~Physikalisches~Institut~B, Aachen, Germany}\\*[0pt]
C.~Dziwok, G.~Fluegge, O.~Pooth, A.~Stahl, T.~Ziemons}

\vspace{0.2cm}
\noindent
\textcolor{black}{\textbf{Deutsches~Elektronen-Synchrotron, Hamburg, Germany}\\*[0pt]
S.~Arab, C.~Asawatangtrakuldee\cmsAuthorMark{4}, V.~Botta, A.~Burgmeier,  L.~Calligaris\cmsAuthorMark{5}, C.~Cheng, S.~Choudhury\cmsAuthorMark{6}, P.~Connor, C.~Contreras-Campana, A.~De~Wit, G.~Dolinska, G.~Eckerlin, D.~Eckstein, T.~Eichhorn, E.~Gallo, M.~Guthoff, K.~Hansen, M.~Haranko, A.~Harb, C.~Kleinwort, I.~Korol, S.~Kousar, R.~Mankel, H.~Maser, M.~Meyer, M.~Missiroli, C.~Muhl, A.~Mussgiller, A.~Petrukhin, D.~Pitzl, O.~Reichelt, M.~Savitskyi, P.~Schuetze, C.~Seitz, S.~Spannagel, R.~Stever, A.~Vargas, R.~Walsh, A.~Zuber}

\vspace{0.2cm}
\noindent
\textcolor{black}{\textbf{University~of~Hamburg,~Hamburg,~Germany}\\*[0pt]
A.~Benecke, H.~Biskop, P.~Buhmann, M.~Centis Vignali, A.~Ebrahimi, M.~Eich, J.~Erfle, F.~Feindt, A.~Froehlich, E.~Garutti, P.~Gunnellini, J.~Haller, T.~Hermanns, A.~Hinzmann, G.~Kasieczka, R.~Klanner, V.~Kutzner, T.~Lange, T.~Lapsien, S.~M\"attig, Ch.~Matthies, M.~Matysek, M.~Mrowietz, C.~Niemeyer, Y.~Nissan, K.~Pena, A.~Perieanu, J.~Poehlsen, T.~Poehlsen, O.~Rieger, P.~Schleper, J.~Schwandt, D.~Schwarz, V.~Sola, J.~Sonneveld, G.~Steinbr\"{u}ck, A.~Tews, B.~Vormwald, J.~Wellhausen, I.~Zoi}

\vspace{0.2cm}
\noindent
\textcolor{black}{\textbf{Institut~f\"{u}r~Experimentelle Teilchenphysik, KIT, Karlsruhe, Germany}\\*[0pt]
M.~Abbas, L.~Ardila, M.~Balzer, T.~Barvich, M.~Baselga, T.~Blank, F.~B\"ogelspacher, E.~Butz, M.~Caselle, F.~Colombo, W.~De~Boer, A.~Dierlamm, K.~El~Morabit, B.~Freund, J.-O.~Gosewisch, F.~Hartmann, S.~Heindl, J.~Hunt, U.~Husemann, R.~Koppenh\"ofer, S.~Kudella, S.~Maier, S.~Mallows, M.~Metzler, Th.~Muller, M.~Musich, A.~N\"urnberg, O.~Sander, D.~Schell, M.~Schr\"oder, T.~Schuh, I.~Shvetsov, H.-J.~Simonis, P.~Steck, M.~Wassmer, M.~Weber, A.~Weddigen, T.~Weiler}

\vspace{0.2cm}
\noindent
\textcolor{black}{\textbf{Institute~of~Nuclear~and~Particle~Physics~(INPP), NCSR~Demokritos, Aghia~Paraskevi, Greece}\\*[0pt]
G.~Anagnostou, P.~Asenov, P.~Assiouras, G.~Daskalakis, I.~Kazas, A.~Kyriakis, D.~Loukas, L.~Paspalaki}

\vspace{0.2cm}
\noindent
\textbf{MTA-ELTE Lend\"ulet CMS Particle and Nuclear Physics Group, E\"otv\"os Lor\'{a}nd University, Budapest, Hungary}\\*[0pt]
M.~Bart\'{o}k\cmsAuthorMark{7}

\vspace{0.2cm}
\noindent
\textcolor{black}{\textbf{Wigner~Research~Centre~for~Physics, Budapest, Hungary}\\*[0pt]
T.~Bal\'{a}zs, K.~M\'{a}rton, F.~Sikl\'{e}r, V.~Veszpr\'{e}mi}

\vspace{0.2cm}
\noindent
\textbf{Institute for Nuclear Research ATOMKI, Debrecen, Hungary}\\*[0pt]
J.~Karancsi

\vspace{0.2cm}
\noindent
\textbf{National Institute of Science Education and Research, HBNI, Bhubaneswar, India}\\*[0pt]
P.~Mal, S.~Swain

\vspace{0.2cm}
\noindent
\textcolor{black}{\textbf{University~of~Delhi,~Delhi,~India}\\*[0pt]
A.~Bhardwaj, C.~Jain, G.~Jain, A.~Kumar, K.~Ranjan, S.~Saumya}

\vspace{0.2cm}
\noindent
\textcolor{black}{\textbf{Saha Institute of Nuclear Physics, Kolkata, India}\\*[0pt]
R.~Bhattacharya, S.~Dutta, P.~Palit, G.~Saha, S.~Sarkar}

\vspace{0.2cm}
\noindent
\textcolor{black}{\textbf{INFN~Sezione~di~Bari$^{a}$, Universit\`{a}~di~Bari$^{b}$, Politecnico~di~Bari$^{c}$, Bari, Italy}\\*[0pt]
P.~Cariola$^{a}$, D.~Creanza$^{a}$$^{,}$$^{c}$, M.~de~Palma$^{a}$$^{,}$$^{b}$, G.~De~Robertis$^{a}$, A.~Di~Florio$^{a}$$^{,}$$^{b}$, L.~Fiore$^{a}$, M.~Ince$^{a}$$^{,}$$^{b}$, F.~Loddo$^{a}$, G.~Maggi$^{a}$$^{,}$$^{c}$, S.~Martiradonna$^{a}$,  M.~Mongelli$^{a}$, S.~My$^{a}$$^{,}$$^{b}$, G.~Selvaggi$^{a}$$^{,}$$^{b}$, L.~Silvestris$^{a}$}

\vspace{0.2cm}
\noindent
\textcolor{black}{\textbf{INFN~Sezione~di~Catania$^{a}$, Universit\`{a}~di~Catania$^{b}$, Catania, Italy}\\*[0pt]
S.~Albergo$^{a}$$^{,}$$^{b}$, S.~Costa$^{a}$$^{,}$$^{b}$, A.~Di~Mattia$^{a}$, R.~Potenza$^{a}$$^{,}$$^{b}$, M.A.~Saizu$^{a,}$\cmsAuthorMark{8}, A.~Tricomi$^{a}$$^{,}$$^{b}$, C.~Tuve$^{a}$$^{,}$$^{b}$}

\vspace{0.2cm}
\noindent
\textcolor{black}{
\textbf{INFN~Sezione~di~Firenze$^{a}$, Universit\`{a}~di~Firenze$^{b}$, Firenze, Italy}\\*[0pt]
G.~Barbagli$^{a}$, M.~Brianzi$^{a}$, A.~Cassese$^{a}$, R.~Ceccarelli$^{a}$$^{,}$$^{b}$, R.~Ciaranfi$^{a}$, V.~Ciulli$^{a}$$^{,}$$^{b}$, C.~Civinini$^{a}$, R.~D'Alessandro$^{a}$$^{,}$$^{b}$, F.~Fiori$^{a}$, E.~Focardi$^{a}$$^{,}$$^{b}$, G.~Latino$^{a}$$^{,}$$^{b}$, P.~Lenzi$^{a}$$^{,}$$^{b}$, M.~Lizzo$^{a}$$^{,}$$^{b}$, M.~Meschini$^{a}$, S.~Paoletti$^{a}$, E.~Scarlini$^{a}$$^{,}$$^{b}$, R.~Seidita$^{a}$$^{,}$$^{b}$, G.~Sguazzoni$^{a}$, L.~Viliani$^{a}$}

\vspace{0.2cm}
\noindent
\textcolor{black}{\textbf{INFN~Sezione~di~Genova$^{a}$, Universit\`{a}~di~Genova$^{b}$, Genova, Italy}\\*[0pt]
F.~Ferro$^{a}$, R.~Mulargia$^{a}$$^{,}$$^{b}$, E.~Robutti$^{a}$}

\vspace{0.2cm}
\noindent
\textcolor{black}{\textbf{INFN~Sezione~di~Milano-Bicocca$^{a}$, Universit\`{a}~di~Milano-Bicocca$^{b}$, Milano, Italy}\\*[0pt]
F.~Brivio$^{a}$, M.E.~Dinardo$^{a}$$^{,}$$^{b}$, P.~Dini$^{a}$, S.~Gennai$^{a}$, L.~Guzzi$^{a}$$^{,}$$^{b}$, S.~Malvezzi$^{a}$, D.~Menasce$^{a}$, L.~Moroni$^{a}$, D.~Pedrini$^{a}$, D.~Zuolo$^{a}$$^{,}$$^{b}$}

\vspace{0.2cm}
\noindent
\textcolor{black}{\textbf{INFN~Sezione~di~Padova$^{a}$, Universit\`{a}~di~Padova$^{b}$, Padova, Italy}\\*[0pt]
P.~Azzi$^{a}$, N.~Bacchetta$^{a}$, P.~Bortignon$^{a,}$\cmsAuthorMark{9}, D.~Bisello$^{a}$, T.Dorigo$^{a}$, M.~Tosi$^{a}$$^{,}$$^{b}$, H.~Yarar$^{a}$$^{,}$$^{b}$}

\vspace{0.2cm}
\noindent
\textcolor{black}{\textbf{INFN~Sezione~di~Pavia$^{a}$, Universit\`{a}~di~Bergamo$^{b}$, Bergamo, Universit\`{a}~di Pavia$^{c}$, Pavia, Italy}\\*[0pt]
L.~Gaioni$^{a}$$^{,}$$^{b}$, M.~Manghisoni$^{a}$$^{,}$$^{b}$, L.~Ratti$^{a}$$^{,}$$^{c}$, V.~Re$^{a}$$^{,}$$^{b}$, E.~Riceputi$^{a}$$^{,}$$^{b}$, G.~Traversi$^{a}$$^{,}$$^{b}$}

\vspace{0.2cm}
\noindent
\textcolor{black}{\textbf{INFN~Sezione~di~Perugia$^{a}$, Universit\`{a}~di~Perugia$^{b}$, CNR-IOM Perugia$^{c}$, Perugia, Italy}\\*[0pt]
G.~Baldinelli$^{a}$$^{,}$$^{b}$, F.~Bianchi$^{a}$$^{,}$$^{b}$, G.M.~Bilei$^{a}$, S.~Bizzaglia$^{a}$, M.~Caprai$^{a}$, B.~Checcucci$^{a}$, D.~Ciangottini$^{a}$, L.~Fan\`{o}$^{a}$$^{,}$$^{b}$, L.~Farnesini$^{a}$, M.~Ionica$^{a}$, G.~Mantovani$^{a}$$^{,}$$^{b}$, V.~Mariani$^{a}$$^{,}$$^{b}$, M.~Menichelli$^{a}$, A.~Morozzi$^{a}$, F.~Moscatelli$^{a}$$^{,}$$^{c}$, D.~Passeri$^{a}$$^{,}$$^{b}$, A.~Piccinelli$^{a}$$^{,}$$^{b}$, P.~Placidi$^{a}$$^{,}$$^{b}$, A.~Rossi$^{a}$$^{,}$$^{b}$, A.~Saha$^{a,}$\cmsAuthorMark{10}, A.~Santocchia$^{a}$$^{,}$$^{b}$, D.~Spiga$^{a}$, L.~Storchi$^{a}$, T.~Tedeschi$^{a}$$^{,}$$^{b}$, C.~Turrioni$^{a}$$^{,}$$^{b}$}

\vspace{0.2cm}
\noindent
\textcolor{black}{\textbf{INFN~Sezione~di~Pisa$^{a}$, Universit\`{a}~di~Pisa$^{b}$, Scuola~Normale~Superiore~di~Pisa$^{c}$, Pisa, Italy}\\*[0pt]
K.~Androsov$^{a}$, P.~Azzurri$^{a}$, G.~Bagliesi$^{a}$, A.~Basti$^{a}$, R.~Beccherle$^{a}$, V.~Bertacchi$^{a}$$^{,}$$^{c}$, L.~Bianchini$^{a}$, T.~Boccali$^{a}$, L.~Borrello$^{a}$, F.~Bosi$^{a}$, R.~Castaldi$^{a}$, M.A.~Ciocci$^{a}$$^{,}$$^{b}$, R.~Dell'Orso$^{a}$, S.~Donato$^{a}$, L.~Giannini$^{a}$$^{,}$$^{c}$, A.~Giassi$^{a}$, M.T.~Grippo$^{a}$$^{,}$$^{b}$, F.~Ligabue$^{a}$$^{,}$$^{c}$, G.~Magazzu$^{a}$, E.~Manca$^{a}$$^{,}$$^{c}$, G.~Mandorli$^{a}$$^{,}$$^{c}$, E.~Mazzoni$^{a}$, A.~Messineo$^{a}$$^{,}$$^{b}$, A.~Moggi$^{a}$, F.~Morsani$^{a}$, F.~Palla$^{a}$, F.~Palmonari$^{a}$, S.~Parolia$^{a}$$^{,}$$^{b}$, F.~Raffaelli$^{a}$, A.~Rizzi$^{a}$$^{,}$$^{b}$, S.~Roy Chowdhury$^{a}$$^{,}$$^{c}$, P.~Spagnolo$^{a}$, R.~Tenchini$^{a}$, G.~Tonelli$^{a}$$^{,}$$^{b}$, A.~Venturi$^{a}$, P.G.~Verdini$^{a}$}

\vspace{0.2cm}
\noindent
\textcolor{black}{\textbf{INFN~Sezione~di~Torino$^{a}$, Universit\`{a}~di~Torino$^{b}$, Torino, Italy}\\*[0pt]
R.~Bellan$^{a}$$^{,}$$^{b}$, M.~Costa$^{a}$$^{,}$$^{b}$, R.~Covarelli$^{a}$$^{,}$$^{b}$, G.~Dellacasa$^{a}$, N.~Demaria$^{a}$, S.~Garbolino$^{a}$, E.~Migliore$^{a}$$^{,}$$^{b}$, E.~Monteil$^{a}$$^{,}$$^{b}$, M.~Monteno$^{a}$, G.~Ortona$^{a}$, L.~Pacher$^{a}$$^{,}$$^{b}$, A.~Paterno$^{a}$, A.~Rivetti$^{a}$, A.~Solano$^{a}$$^{,}$$^{b}$}

\vspace{0.2cm}
\noindent
\textcolor{black}{\textbf{Instituto~de~F\'{i}sica~de~Cantabria~(IFCA), CSIC-Universidad~de~Cantabria, Santander, Spain}\\*[0pt]
E.~Curras Rivera, J.~Duarte Campderros, M.~Fernandez, A.~Garcia~Alonso, G.~Gomez, F.J.~Gonzalez~Sanchez, R.~Jaramillo~Echeverria, D.~Moya, I.~Vila, A.L.~Virto}

\vspace{0.2cm}
\noindent
\textcolor{black}{\textbf{CERN, European~Organization~for~Nuclear~Research, Geneva, Switzerland}\\*[0pt]
D.~Abbaneo, I.~Ahmed, B.~Akgun\cmsAuthorMark{11}, E.~Albert, G.~Auzinger, J.~Bendotti, G.~Bergamin\cmsAuthorMark{12}, G.~Blanchot, F.~Boyer, A.~Caratelli, R.~Carnesecchi, D.~Ceresa, J.~Christiansen, K.~Cichy, J.~Daguin, S.~Detraz, D.~Deyrail, O.~Dondelewski, N.~Emriskova\cmsAuthorMark{13}, B.~Engegaard, F.~Faccio, A.~Filenius, N.~Frank, T.~French, K.~Gill, G.~Hugo, W.~Hulek\cmsAuthorMark{14}, L.M.~Jara~Casas, J.~Kaplon, K.~Kloukinas, A.~Kornmayer, N.~Koss, L.~Kottelat, D.~Koukola, M.~Kovacs, A.~La Rosa, P.~Lenoir, R.~Loos, A.~Marchioro, S.~Marconi, I.~Mateos Dominguez\cmsAuthorMark{15}, S.~Mersi, S.~Michelis, A.~Onnela, S.~Orfanelli, T.~Pakulski, A.~Papadopoulos\cmsAuthorMark{16}, S.~Pavis, A.~Peisert, A.~Perez, F.~Perez Gomez, J.-F.~Pernot, P.~Petagna, Q.~Piazza, H.~Postema, K.~Rapacz, S.~Scarf\'{i}\cmsAuthorMark{17}, P.~Tropea, J.~Troska, A.~Tsirou, F.~Vasey, B.~Verlaat, P.~Vichoudis, A.~Zografos\cmsAuthorMark{18}, L.~Zwalinski}

\vspace{0.2cm}
\noindent
\textcolor{black}{\textbf{Paul~Scherrer~Institut, Villigen, Switzerland}\\*[0pt]
W.~Bertl$^{\dag}$, S.~Burkhalter\cmsAuthorMark{19}, L.~Caminada\cmsAuthorMark{20}, W.~Erdmann, R.~Horisberger, H.-C.~Kaestli, D.~Kotlinski, U.~Langenegger, B.~Meier, T.~Rohe, S.~Streuli}

\vspace{0.2cm}
\noindent
\textcolor{black}{\textbf{Institute~for~Particle~Physics, ETH~Zurich, Zurich, Switzerland}\\*[0pt]
F.~Bachmair, M.~Backhaus, R.~Becker, P.~Berger, D.~di~Calafiori, A.~Calandri, L.~Djambazov, M.~Donega, C.~Dorfer, P.~Eller, C.~Grab, D.~Hits, J.~Hoss,  W.~Lustermann, M.~Meinhard, V.~Perovic, M.~Reichmann, B.~Ristic, U.~Roeser, M.~Rossini, D.~Ruini, V.~Tavolaro, R.~Wallny, D.~Zhu}

\vspace{0.2cm}
\noindent
\textcolor{black}{
\textbf{Universit\"{a}t~Z\"{u}rich,~Zurich,~Switzerland}\\*[0pt]
T.~Aarrestad\cmsAuthorMark{21}, K.~B\"{o}siger, D.~Brzhechko, F.~Canelli, A.~De~Cosa, R.~Del Burgo, C.~Galloni\cmsAuthorMark{22}, M.~Gienal, D.~Hernandez Garland, A.~Jofrehei, B.~Kilminster, C.~Lange\cmsAuthorMark{21}, S.~Leontsinis, A.~Macchiolo, U.~Molinatti, R.~Maier, V.~Mikuni, I.~Neutelings, J.~Ngadiuba\cmsAuthorMark{23}, D.~Pinna\cmsAuthorMark{22}, G.~Rauco, P.~Robmann, Y.~Takahashi, S.~Wiederkehr\cmsAuthorMark{24}, D.~Wolf, A.~Zucchetta\cmsAuthorMark{25}}

\vspace{0.2cm}
\noindent
\textcolor{black}{\textbf{National~Taiwan~University~(NTU),~Taipei,~Taiwan}\\*[0pt]
P.-H.~Chen, W.-S.~Hou, Y.-W.~Kao, R.-S.~Lu, M.~Moya}

\vspace{0.2cm}
\noindent
\textcolor{black}{\textbf{University~of~Bristol,~Bristol,~United~Kingdom}\\*[0pt]
D.~Burns, E.~Clement, D.~Cussans, J.~Goldstein, S.~Seif~El~Nasr-Storey}

\vspace{0.2cm}
\noindent
\textcolor{black}{\textbf{Rutherford~Appleton~Laboratory, Didcot, United~Kingdom}\\*[0pt]
J.A.~Coughlan, K.~Harder, K.~Manolopoulos, I.R.~Tomalin}

\vspace{0.2cm}
\noindent
\textcolor{black}{\textbf{Imperial~College, London, United~Kingdom}\\*[0pt]
J.~Borg, G.~Fedi, G.~Hall, G.~Iles, M.~Pesaresi, A.~Rose, K.~Uchida}

\vspace{0.2cm}
\noindent
\textcolor{black}{\textbf{Brunel~University, Uxbridge, United~Kingdom}\\*[0pt]
K.~Coldham, J.~Cole, M.~Ghorbani, A.~Khan, P.~Kyberd, I.D.~Reid}

\vspace{0.2cm}
\noindent
\textcolor{black}{\textbf{The Catholic~University~of~America,~Washington~DC,~USA}\\*[0pt]
R.~Bartek, A.~Dominguez, R.~Uniyal, A.M.~Vargas~Hernandez}

\vspace{0.2cm}
\noindent
\textcolor{black}{\textbf{Brown~University, Providence, USA}\\*[0pt]
G.~Altopp, A.~Alzahrani, G.~Benelli, B.~Burkle, X.~Coubez, U.~Heintz, E.~Hinkle, N.~Hinton, J.~Hogan\cmsAuthorMark{26}, A.~Honma, A.~Korotkov, J.~Lee, D.~Li, M.~Lukasik, M.~Narain, S.~Sagir\cmsAuthorMark{27},  F.~Simpson, E.~Spencer, E.~Usai, J.~Voelker, W.Y.~Wong, W.~Zhang}

\vspace{0.2cm}
\noindent
\textcolor{black}{\textbf{University~of~California,~Davis,~Davis,~USA}\\*[0pt]
E.~Cannaert, M.~Chertok, J.~Conway, G.~Funk, G.~Haza, F.~Jensen, D.~Pellett, M.~Shi, J.~Thomson, K.~Tos, T.~Welton, F.~Zhang}

\vspace{0.2cm}
\noindent
\textcolor{black}{\textbf{University~of~California,~Riverside,~Riverside,~USA}\\*[0pt]
G.~Hanson, W.~Si}

\vspace{0.2cm}
\noindent
\textcolor{black}{\textbf{University~of~California, San~Diego, La~Jolla, USA}\\*[0pt]
S.B.~Cooperstein, N.~Deelen, R.~Gerosa, S.~Krutelyov, V.~Sharma, A.~Yagil}

\vspace{0.2cm}
\noindent
\textcolor{black}{\textbf{University~of~California, Santa~Barbara~-~Department~of~Physics, Santa~Barbara, USA}\\*[0pt]
O.~Colegrove, V.~Dutta, L.~Gouskos, J.~Incandela, S.~Kyre, H.~Qu, M.~Quinnan}

\vspace{0.2cm}
\noindent
\textbf{University~of~Colorado~Boulder, Boulder, USA}\\*[0pt]
J.P.~Cumalat, W.T.~Ford, E.~MacDonald, A.~Perloff, K.~Stenson, K.A.~Ulmer, S.R.~Wagner

\vspace{0.2cm}
\noindent
\textcolor{black}{\textbf{Cornell~University, Ithaca, USA}\\*[0pt]
J.~Alexander, Y.~Bordlemay~Padilla, Y.~Cheng, J.~Conway, D.~Cranshaw, A.~Datta, K.~McDermott, J.~Monroy, D.~Quach, J.~Reichert, A.~Ryd, K.~Smolenski, C.~Strohman,  Z.~Tao, J.~Thom, J.M.~Tucker, P.~Wittich, M.~Zientek}

\vspace{0.2cm}
\noindent
\textcolor{black}{
\textbf{Fermi~National~Accelerator~Laboratory, Batavia, USA}\\*[0pt]
M.~Alyari, A.~Bakshi, D.R.~Berry, G.~Bolla$^{\dag}$, K.~Burkett, D.~Butler, A.~Canepa, H.~Cheung, G.~Derylo, A.~Ghosh, C.~Gingu, H.~Gonzalez, S.~Gr\"{u}nendahl, S.~Hasegawa, M.~Johnson, S.~Kwan$^{\dag}$, C.M.~Lei, R.~Lipton, M.~Liu, S.~Los, V.~Martinez Outschoorn, P.~Merkel, S.~Nahn, A.~Prosser, F.~Ravera, R.~Rivera, B.~Schneider, W.J.~Spalding, L.~Spiegel, S.~Timpone, L.~Uplegger, M.~Verzocchi, E.~Voirin, H.A.~Weber}

\vspace{0.2cm}
\noindent
\textbf{Florida State University, Tallahassee, USA}\\*[0pt]
R.~Habibullah, R.~Yohay

\vspace{0.2cm}
\noindent
\textcolor{black}{\textbf{University~of~Illinois~at~Chicago~(UIC), Chicago, USA}\\*[0pt]
H.~Becerril Gonzalez, X.~Chen, S.~Dittmer, A.~Evdokimov, O.~Evdokimov, C.E.~Gerber, D.J.~Hofman, C.~Mills, T.~Roy, S.~Rudrabhatla, J.~Yoo}

\vspace{0.2cm}
\noindent
\textcolor{black}{\textbf{The~University~of~Iowa, Iowa~City, USA}\\*[0pt]
M.~Alhusseini, S.~Durgut, J.~Nachtman, Y.~Onel, C.~Rude, C.~Snyder, K.~Yi\cmsAuthorMark{28}}

\vspace{0.2cm}
\noindent
\textcolor{black}{\textbf{Johns~Hopkins~University,~Baltimore,~USA}\\*[0pt]
N.~Eminizer, A.~Gritsan, S.~Kyriacou, P.~Maksimovic, C.~Mantilla Suarez, J.~Roskes, M.~Swartz, T.~Vami}

\vspace{0.2cm}
\noindent
\textcolor{black}{\textbf{The~University~of~Kansas, Lawrence, USA}\\*[0pt]
P.~Baringer, A.~Bean, Z.~Flowers, S.~Khalil, J.W.~King, A.~Kropivnitskaya, E.~Schmitz, G.~Wilson}

\vspace{0.2cm}
\noindent
\textcolor{black}{\textbf{Kansas~State~University, Manhattan, USA}\\*[0pt]
A.~Ivanov, T.~Mitchell, A.~Modak, R.~Taylor}

\vspace{0.2cm}
\noindent
\textcolor{black}{\textbf{University~of~Mississippi,~Oxford,~USA}\\*[0pt]
J.G.~Acosta, L.M.~Cremaldi, S.~Oliveros, L.~Perera, D.~Summers}

\vspace{0.2cm}
\noindent
\textbf{University~of~Nebraska-Lincoln, Lincoln, USA}\\*[0pt]
K.~Bloom, D.R.~Claes, C.~Fangmeier, F.~Golf, I.~Kravchenko, J.~Siado

\vspace{0.2cm}
\noindent
\textcolor{black}{\textbf{State~University~of~New~York~at~Buffalo, Buffalo, USA}\\*[0pt]
I.~Iashvili, A.~Kharchilava, C.~McLean, D.~Nguyen, A.~Parker, J.~Pekkanen, S.~Rappoccio}

\vspace{0.2cm}
\noindent
\textcolor{black}{\textbf{Northeastern~University,~Boston,~USA}\\*[0pt]
J.~Li, A.~Parker, L.~Skinnari}

\vspace{0.2cm}
\noindent
\textcolor{black}{\textbf{Northwestern~University,~Evanston,~USA}\\*[0pt]
K.~Hahn, Y.~Liu, K.~Sung}

\vspace{0.2cm}
\noindent
\textcolor{black}{\textbf{The~Ohio~State~University, Columbus, USA}\\*[0pt]
J.~Alimena, B.~Cardwell, B.~Francis, C.S.~Hill}

\vspace{0.2cm}
\noindent
\textcolor{black}{\textbf{University~of~Puerto~Rico,~Mayaguez,~USA}\\*[0pt]
S.~Malik, S.~Norberg, J.E.~Ramirez Vargas}

\vspace{0.2cm}
\noindent
\textcolor{black}{\textbf{Purdue~University, West Lafayette, USA}\\*[0pt]
R.~Chawla, S.~Das, S.~Gurdasani, M.~Jones, A.~Jung, A.~Koshy, G.~Negro, J.~Thieman}

\vspace{0.2cm}
\noindent
\textcolor{black}{\textbf{Purdue~University~Northwest,~Hammond,~USA}\\*[0pt]
T.~Cheng, J.~Dolen, N.~Parashar}

\vspace{0.2cm}
\noindent
\textcolor{black}{\textbf{Rice~University, Houston, USA}\\*[0pt]
K.M.~Ecklund, S.~Freed, M.~Kilpatrick, A.~Kumar, T.~Nussbaum}

\vspace{0.2cm}
\noindent
\textcolor{black}{\textbf{University~of~Rochester,~Rochester,~USA}\\*[0pt]
R.~Demina, J.~Dulemba, O.~Hindrichs}

\vspace{0.2cm}
\noindent
\textcolor{black}{\textbf{Rutgers, The~State~University~of~New~Jersey, Piscataway, USA}\\*[0pt]
E.~Bartz, A.~Gandrakotra, Y.~Gershtein, E.~Halkiadakis, A.~Hart, A.~Lath, K.~Nash, M.~Osherson, S.~Schnetzer, R.~Stone}

\vspace{0.2cm}
\noindent
\textbf{Texas~A\&M~University, College~Station, USA}\\*[0pt]
R.~Eusebi

\vspace{0.2cm}
\noindent
\textcolor{black}{\textbf{Vanderbilt~University, Nashville, USA}\\*[0pt]
P.~D'Angelo, W.~Johns, K.O.~Padeken}

\vspace{1cm}
\noindent
\dag: Deceased\\
1: Also at Vienna University of Technology, Vienna, Austria \\
2: Also at Institute for Theoretical and Experimental Physics, Moscow, Russia \\
3: Also at Universit\'{e} de Haute-Alsace, Mulhouse, France \\
4: Now at Chulalongkorn University, Faculty of Science, Department of Physics, Bangkok, Thailand \\
5: Now at Universidade Estadual Paulista, Sao Paulo, Brazil \\
6: Now at Indian Institute of Science (IISc), Bangalore, India\\
7: Also at Institute of Physics, University of Debrecen, Debrecen, Hungary\\
8: Also at Horia Hulubei National Institute of Physics and Nuclear Engineering~(IFIN-HH), Bucharest, Romania \\
9: Also at University of Cagliari, Cagliari, Italy \\
10: Also at Faculty of Sciences, Shoolini University, Solan, Himachal Pradesh, India \\
11: Now at Bogazici University, Istanbul, Turkey \\
12: Also at Institut Polytechnique de Grenoble, Grenoble, France \\
13: Also at Universit\'{e}~de~Strasbourg, CNRS, IPHC~UMR~7178, Strasbourg, France \\
14: Also at Tadeusz Kosciuszko Cracow University of Technology, Cracow, Poland \\
15: Also at Universidad de Castilla-La-Mancha, Ciudad Real, Spain \\
16: Also at University of Patras, Patras, Greece \\
17: Also at \'{E}cole Polytechnique F\'{e}d\'{e}rale de Lausanne, Lausanne, Switzerland \\
18: Also at National Technical University of Athens, Athens, Greece \\
19: Also at ETH~Zurich, Zurich, Switzerland \\
20: Also at Universit\"{a}t~Z\"{u}rich,~Zurich,~Switzerland \\
21: Now at CERN, European~Organization~for~Nuclear~Research, Geneva, Switzerland \\
22: Now at University of Wisconsin - Madison, Madison, WI, USA \\
23: Now at California Institute of Technology, Pasadena, USA \\
24: Also at Paul Scherrer Institut, Villigen, Switzerland \\
25: Now at INFN~Sezione~di~Padova and Universit\`{a}~di~Padova, Padova, Italy \\
26: Now at Bethel University, St. Paul, Minnesota, USA \\
27: Now at Karamanoglu Mehmetbey University, Karaman, Turkey \\
28: Also at Nanjing Normal University, Nanjing, China

\end{document}